\documentclass[11pt]{article}

\usepackage{relsize}
\usepackage{amsmath,amssymb,latexsym}
\usepackage[T1]{fontenc}
\usepackage{epsfig}
\usepackage{fullpage}
\usepackage{graphicx}
\usepackage{makeidx}
\usepackage{slashed}
\usepackage{color}
\usepackage[normalem]{ulem}
\usepackage{subcaption}
\usepackage{feynmp}
\usepackage{cite}
\usepackage{hyperref}

\long\def\symbolfootnote[#1]#2{\begingroup%
\def\thefootnote{\fnsymbol{footnote}}\footnote[#1]{#2}\endgroup}

\begin{document}

\vspace*{1.5cm}

\begin{center}
{\Large\sc The role of the threshold variable in soft-gluon resummation of the $t\bar{t}h$ production process}\\[10ex] 
 
 { 
 Melissa van Beekveld$^{a,b,c}$,
 Wim Beenakker$^{b,d}$}\\[1cm]
 {\it 
$^a$ {Rudolf Peierls Centre for Theoretical Physics, 20 Parks Road, Oxford OX1
3PU, United Kingdom}\\
$^b$ {THEP, Radboud University,
 Heyendaalseweg 135, 6525~AJ Nijmegen, the Netherlands}\\
$^c$ {Nikhef, Science Park 105, 1098 XG Amsterdam, the Netherlands}\\
$^d$ {Institute of Physics, University of Amsterdam, Science Park 904, 1018 XE Amsterdam, the Netherlands}
 }
\end{center}

\vspace{1.5cm}

\begin{abstract}
\noindent{We study the role of the threshold variable in soft-gluon threshold resummation. We focus on the computation of the resummed total cross section, the final-state invariant-mass distribution, and transverse-momentum distribution of the Higgs boson when produced in association with a top--anti-top quark pair for the Large Hadron Collider operating at $13$~TeV. We show that different choices for the threshold variable result in differences at next-to-leading power, i.e.~contributions that are down by one power of the threshold variable. These contributions are noticeable numerically, although their effect on the resummed observables lies within the scale uncertainty of those observables. The average central results, obtained after combining several central-scale choices, agree remarkably well for different choices of the threshold variable. However, different threshold choices do affect the resulting scale uncertainty. To compute our results, we introduce a novel numerical method that we call the deformation method, which aids the stabilization of the inverse Mellin transform in cases where the analytical Mellin transform of the partonic cross section is unknown. We show that this method leads to a factor of $10$ less function evaluations, while gaining a factor of $4-5$ in numerical precision when compared to the standard method.}
\end{abstract}

\vspace*{\fill}

\newpage
\reversemarginpar

\section{Introduction}
The top-quark Yukawa coupling is the largest one in the Standard Model (SM). A precise measurement of its properties is one of the prime goals of the experiments at the Large Hadron Collider (LHC), as its measurement can place constraints on beyond-the-SM (BSM) scenarios that modify the SM Yukawa coupling.  Therefore, it is important to have a precise and accurate prediction for the $t\bar{t}h$ cross section. At present, the experimental uncertainties dominate the theoretical ones. However, the experimental uncertainties rapidly decrease with the collection of more data. The top-quark Yukawa coupling has recently been measured by the ATLAS~\cite{Aaboud:2018urx} and CMS~\cite{Sirunyan:2018hoz} experiments, constraining its value to lie within $20-30\%$ of the SM value. The total cross section for the $t\bar{t}h$ production process as reported by ATLAS is $670\pm 90$~(stat.)~$_{-100}^{+110}$~(syst.)~fb at the hadronic center-of-mass (CM) energy of $\sqrt{S} = 13$~TeV, while CMS reports $639^{+157}_{-132}$~fb. The prediction of the SM for the $t\bar{t}h$ cross section (given below) lies within the $2\sigma$-interval of these measurements. \\
The next-to-leading order (NLO) QCD corrections to the $t\bar{t}h$ process were calculated already some time ago~\cite{Dawson:2003zu,Dawson:2002tg, Beenakker:2002nc}. Electroweak corrections were also studied~\cite{Yu:2014cka, Frixione:2014qaa, Frixione:2015zaa}, and the effect of off-shell decaying top-quarks was studied in Ref.~\cite{Denner:2015yca}. As summarized in Ref.~\cite{deFlorian:2016spz}, the best estimate for the NLO QCD (+electroweak) cross section with the Higgs mass at $m_h = 125$~GeV and the top mass $m_t = 172.5$~GeV is $498.7(507.1)^{+5.8\%}_{- 9.2\%}$~fb for $\sqrt{S} = 13$~TeV. 
Since the $t\bar{t}h$ process is already quite involved at NLO, it is unlikely that higher-order corrections will become available soon. However, the impact of soft-gluon corrections can be studied by resummation. Resummation up to next-to-leading logarithmic (NLL) accuracy of the total cross section is performed in Ref.~\cite{Kulesza:2015vda}. This is extended to next-to-next-to-leading logarithmic (NNLL) order in Ref.~\cite{Kulesza:2017ukk}, where the invariant-mass distribution of the $t\bar{t}h$ process is computed. Furthermore, electroweak corrections are included in Ref.~\cite{Kulesza:2020nfh}, where also other kinematic distributions and the effect of scale variations are studied. Using the orthogonal Soft-Collinear-Effective Theory (SCET) approach, the total cross section and several kinematic distributions have been computed to NNLL in Refs.~\cite{Broggio:2015lya,Broggio:2016lfj,Ju:2019lwp,Broggio:2019ewu}. \\
The $t\bar{t}h$ process is an interesting playground for threshold resummation, as it has three particles in the final state already at leading order (LO). This makes for a complicated phase space. Firstly, the Mellin transform of the partonic cross section cannot be performed trivially, creating a potentially severe numerical issue.  Secondly, as there are many ways to share the kinematics amongst the final state particles, multiple threshold definitions may be used. In this work we explore the impact of choosing different threshold variables on the resummation of the total cross section, the invariant-mass distribution of the final state, and the transverse-momentum distribution of the Higgs boson. We present our results at NLL accuracy, as many of the important features are already present at NLL. \\
The outline is as follows: in Section~\ref{sec:notation} we set up our notation using the $t\bar{t}h$ process at LO. In Section~\ref{sec:resum} we consider the resummation of the $t\bar{t}h$ process and introduce the threshold variables we consider in this work. Resummation in QCD is traditionally performed in Mellin space, and the resummed result needs to be transformed back to physical space to be compared against experimental results. This inverse Mellin transform needs to be performed numerically and is notoriously unstable. In Section~\ref{sec:inversemel} we introduce a new numerical method that brings better behavior to the computation of the inverse Mellin transform. This method is applied to compute the results, and the improvement in speed and the obtained numerical accuracy is shown in Section~\ref{sec:results}, along with the results for the LHC operated at $\sqrt{S} = 13$~TeV. As we will see, the threshold definitions differ in the way they include next-to-leading power (NLP) corrections, which are contributions to the resummed result that are down by one power in the threshold variable. Their role is discussed in Section~\ref{sec:nlp}. We conclude in Section~\ref{sec:discuss}.
\section{Notation and conventions}
\label{sec:notation}
\noindent To set up our notation, we first introduce the $t\bar{t}h$ process at LO. The final state may be created by two partonic processes: $g g \rightarrow t\bar{t} h$ and $q\bar{q} \rightarrow t\bar{t} h$. The initial-state momenta are indicated by $p_1$ and $p_2$, while the final-state momenta are labeled $p_t$, $p_{\bar{t}}$ and $p_h$. We define the invariants
\begin{eqnarray}
s \equiv (p_1 + p_2)^2,\qquad t_{1h} \equiv (p_1 - p_h)^2, \qquad t_{2h} \equiv (p_2 - p_h)^2, \nonumber \\
t_{1t} \equiv (p_1 - p_t)^2, \qquad t_{2t} \equiv (p_2 - p_t)^2, \nonumber \\
t_{1\bar{t}} \equiv (p_1 - p_{\bar{t}})^2, \qquad t_{2\bar{t}} \equiv (p_2 - p_{\bar{t}})^2, 
\label{eq:partonicmandel}
\end{eqnarray}
of which only $5$ are independent. The four-momentum of the Higgs boson $p_h$ is parameterized as
\begin{eqnarray}
\label{eq:higgsph}
p_h &=& (E_h, 0, |\vec{p}_{h}|\sin\theta_h, |\vec{p}_{h}|\cos\theta_h)  \\
&=& (E_h, \vec{p}_{{\rm T}}, p_{z,h}) = \left(m_{{\rm T},h}\cosh\eta, \vec{p}_{{\rm T}}, m_{{\rm T},h}\sinh\eta\right). \nonumber
\end{eqnarray}
The former parameterization is useful for the invariant-mass distribution, while the latter is more useful for the transverse-momentum one. The variable $m_{{\rm T},h}$ is the transverse mass of the Higgs boson and defined through
\begin{eqnarray}
m_{{\rm T},h} = \sqrt{p_{{\rm T}}^2+m_h^2}\,,
\end{eqnarray}
with $m_h$ the mass of the Higgs boson and $p_{{\rm T}} \equiv |\vec{p}_{{\rm T},h}|$ the transverse momentum. The pseudorapidity is indicated by $\eta$. Note that the variables $\eta$, $m_{{\rm T},h}$ and $p_{{\rm T}}$ are not mutually independent at LO. \\
One needs to parameterize the three-body phase space in a suitable way to compute the invariant-mass or transverse-momentum distribution. As is common, we separate the three-body phase space into two two-body ones
\begin{eqnarray}
\int {\rm d}\Phi_3(p_1 + p_2; p_t, p_{\bar t}, p_h) = \frac{1}{2\pi}\int {\rm d}s_{t\bar{t}} \int {\rm d}\Phi_2(p_1 + p_2; p_{t\bar{t}}, p_h)\int {\rm d}\Phi_2(p_{t\bar{t}}; p_t, p_{\bar t})\,,
\end{eqnarray}
with $p_{t\bar{t}}^2 = s_{t\bar{t}}$. The phase space of the $t\bar{t}$-system $\int {\rm d}\Phi_2(p_{t\bar{t}}; p_t, p_{\bar t})$ may be computed in the CM system of $t\bar{t}$. This system is denoted by the starred notation ($*$), and the direction of travel of the $t\bar{t}$-system in the CM system of the incoming particles is used as the $z$-axis with respect to which the angle $\theta^*$ is defined. The phase space of the $t\bar{t}$-system then results in
\begin{eqnarray}
\int {\rm d}\Phi_2(p_{t\bar{t}}; p_t, p_{\bar t}) &=& \frac{1}{\left(2\pi\right)^2} \int {\rm d} E_t^{*} \, |\vec{p}_{t}^{\,*}|^2 {\rm d}|\vec{p}_t^{\,*}| \, {\rm d}\cos\theta_t^*\, {\rm d}\phi_t^*\, \nonumber \\
&&\hspace{1.5cm}\times \,\frac{1}{2|\vec{p}_t^{\,*}|}\delta^+\left(|\vec{p}_t^{\,*}|-\sqrt{(E_t^*)^2-m_t^2}\right)\, \frac{1}{2\sqrt{s_{t\bar{t}}}}\,\delta^+\left(E_t^*-\frac{\sqrt{s_{t\bar{t}}}}{2}\right).\label{eq:ttbarframe}
\end{eqnarray}
For this to have a solution, we need that $E_t^* \geq m_t$ and therefore $\sqrt{s_{t\bar{t}}} \geq 2m_t$, which puts a lower boundary on the $s_{t\bar{t}}$ integration. The angular integrations are bounded as usual by $\theta_t^* \in [0,\pi]$ and $\phi_t^* \in [0,2\pi]$. Performing the integrations over $E_t^*$ and $|\vec{p}_t^{\,*}|$, we obtain 
\begin{eqnarray}
\int {\rm d}\Phi_2(p_{t\bar{t}}; p_t, p_{\bar t}) &=& \frac{1}{\left(2\pi\right)^2}\frac{\lambda^{1/2}\left(s_{t\bar{t}},m_t^2,m_t^2\right)}{8s_{t\bar{t}}} \int{\rm d}\cos\theta_t^* \,{\rm d}\phi_t^*\,,
\end{eqnarray}
with
\begin{eqnarray}
\lambda^{1/2}\left(x,y,z\right) = \sqrt{x^2+y^2+z^2-2xy-2xz-2yz}\,.
\end{eqnarray}
The connection between the $t\bar{t}$ CM kinematics and the partonic CM frame is detailed in Appendix~\ref{app:tth} for the convenience of the reader. \\
The phase space integration of the $(t\bar{t})h$-system ${\rm d}\Phi_2(p_1 + p_2; p_{t\bar{t}}, p_h)$ is evaluated in the partonic CM system, where the four-momenta of the initial-state particles read
\begin{eqnarray}
p_1 = \frac{\sqrt{s}}{2}\left(1,0,0,1\right),\qquad p_2 = \frac{\sqrt{s}}{2}\left(1,0,0,-1\right).
\end{eqnarray}
With this, the phase space of the $(t\bar{t})h$-system becomes
\begin{eqnarray}
\label{eq:twobodyH}
\int {\rm d}\Phi_2(p_1 + p_2; p_{t\bar{t}}, p_h) &=&\int \frac{{\rm d}^4p_{h}}{(2\pi)^2}\delta^+\left(m_{h}^2-p_{h}^2\right)\delta^+\left(s_{t\bar{t}}-\left(p_1+p_2-p_h\right)^2\right).
\end{eqnarray} 
Using these definitions, the LO fully differential partonic cross section is written as
\begin{eqnarray}
\label{eq:LOttbarH}
    {\rm d}\hat{\sigma}_{ij\rightarrow t\bar{t}h}^{\rm LO}(s) &=& \frac{1}{4\pi s}K^{ij} \sum_{\rm spin, color}\left|\mathcal{M}_{ij}\right|^2 \,{\rm d}s_{t\bar{t}}\,\, {\rm d}\Phi_2(p_1 + p_2; p_{t\bar{t}}, p_h) \,{\rm d}\Phi_2(p_{t\bar{t}}; p_t, p_{\bar t})\,. 
\end{eqnarray}
Here, the $\mathcal{M}_{ij}$ are the matrix elements for the partonic initial states $ij = q\bar{q}$ and $gg$, and $K^{ij}$ is an averaging factor for the initial state spins and colors. The matrix elements can be found in Ref.~\cite{Beenakker:2002nc}, given in terms of the partonic Mandelstam invariants of Eq.~\eqref{eq:partonicmandel}. From this, the hadronic differential cross section follows as
\begin{eqnarray}
\label{eq:hadronicttbar}
{\rm d}\sigma_{t\bar{t}h}^{\rm LO} &=&  \sum_{i,j}\int_{\left(2m_t+m_h\right)^2}^S {\rm d}s \int_0^1{\rm d}x_1 f_i(x_1,\mu_F^2)\int_0^1{\rm d}x_2 f_j(x_2,\mu_F^2) \, {\rm d}\hat{\sigma}_{ij\rightarrow t\bar{t}h}^{\rm LO}(s) \,\delta\left(s-x_1 x_2 S\right),
\end{eqnarray}
where  $S$ is the hadronic CM energy squared, $f_{i/j}(x_{1/2},\mu_F^2)$ denote the parton distribution functions at the factorization scale $\mu_F$, and $x_1$ and $x_2$ are the parton-momentum fractions.  Having set up our notation, we are now ready to turn our attention to the resummation of the $t\bar{t}h$ production process. 

\section{Resummation of the $t\bar{t}h$ process}
\label{sec:resum}
Resummation is traditionally done in Mellin space (or $N$-space). This conjugate space serves to factorize the dynamics of the soft radiation from that of the hard scattering, and to separate the PDF contribution from the partonic coefficient function. To this end, the Mellin transform must be performed with respect to a hadronic threshold variable, which is generically denoted by $\tau$ for now. The hadronic threshold variable is a dimensionless number that is weighted by the hadronic center of mass energy $S$, i.e.~$\tau = M^2/S$, where $M^2$ parameterizes a certain edge of the phase space.  Resummation for the $t\bar{t}h$ process up to NLL was performed in Ref.~\cite{Kulesza:2015vda}, using the absolute mass of the final state $M^2 = (2m_t + m_h)^2$ to parameterize the partonic threshold. In Ref.~\cite{Kulesza:2017ukk,Kulesza:2020nfh}, the formalism was extended to NNLL using a fixed invariant mass: $M^2 = (p_t + p_{\bar{t}} + p_h)^2$. Here we resum the $t\bar{t}h$ cross section up to NLL, with the aim of exploring the impact of using such different parameterizations, but also extend their work by formulating three additional threshold variables.  \\
After the transformation to $N$-space is performed, we may replace the partonic matrix element with its resummed version. By doing this, the resummed differential cross section in $N$ space reads~\cite{Kulesza:2015vda,Kulesza:2017ukk}
\begin{eqnarray}\label{eq:replace}
{\rm d}\sigma_{t\bar{t}h}^{\rm res}(N) &=& \sum_{i,j} f_i(N+1,\mu_F^2) f_j(N+1,\mu_F^2) \,{\rm Tr}\left[\mathbf{H}_{ij\rightarrow t\bar{t}h}(N)\mathbf{S}_{ij\rightarrow t\bar{t}h}(N+1)\right]\\
&& \hspace{3.5cm}\times\, \Delta_i(N+1,M^2/\mu_F^2,M^2/\mu_R^2)\Delta_j(N+1,M^2/\mu_F^2,M^2/\mu_R^2)\,. \nonumber
\end{eqnarray}
The moments of the parton distribution functions are defined in the standard way
\begin{eqnarray}
 f_k(N+1,\mu_F^2) = \int_0^1{\rm d}x\, x^N\, f_k(x,\mu_F^2)\,.
\end{eqnarray}
The trace of Eq.~\eqref{eq:replace} acts in color-tensor space, and implicitly includes the averaging over the initial-state spins and colors ($1/(4N_c^2)$ for $q\bar{q}$ channel or $1/(4(N_c^2-1)^2)$ for the $gg$ channel). The matrix $\mathbf{H}_{ij\rightarrow t\bar{t}h}$, decomposed in color-tensor space, contains the hard-scattering contributions. The soft-collinear enhancements are captured by the functions $\Delta_i(N,M^2/\mu_F^2,M^2/\mu_R^2)$, and the soft wide-angle contributions are included via $\mathbf{S}_{ij\rightarrow t\bar{t}h}$ (written in the same basis as $\mathbf{H}_{ij\rightarrow t\bar{t}h}$). The matrices $\mathbf{H}$ and $\mathbf{H}$ implicitly depend on the factorization scale $\mu_F$ and the renomalization scale $\mu_R$. Note the mismatch between the moments of the hard function, and the soft and collinear functions. This is caused by the form of the $\delta$-function in Eq.~\eqref{eq:hadronicttbar}. \\ 
To separate the PDFs from the partonic cross section, one uses a partonic threshold variable reading $\rho = M^2/Q^2 \simeq M^2/s = \tau/(x_1 x_2)$. Here, we have used that $Q^2 = z s$, where $(1-z)$ is the fraction of $s$ that is radiated away by soft gluon emissions (i.e. $z\simeq 1$). This fraction $(1-z)$ can be further decomposed into one fraction being radiated by the wide-angle emissions, and one by the initial-state jet functions (see e.g.~\cite{Laenen:1998qw}). Note that $z$ is different from $\rho$ in general, unless one sets $M^2 = Q^2$. \\
The soft-collinear enhancements are generated by the integral
\begin{eqnarray}
\label{eq:deltaiint}
\Delta_i(N,M^2/\mu_F^2,M^2/\mu_R^2) = {\rm exp}\left[\int_{0}^{1}{\rm d}z\, \frac{z^{N-1}-1}{1-z}\int_{\mu_F^2}^{M^2(1-z)^2}\frac{{\rm d}q^2}{q^2}A_i\left(\alpha_s(q^2)\right)\right].
\end{eqnarray}
The coefficient $A_i$ is a power series in the coupling $\alpha_s(q^2)$. Note that we set the upper limit of the integral to $M^2(1-z)^2$, which is an approximation. We will come back to this point in Section~\ref{sec:subsecinv}.
To NLL accuracy, the integral results in
\begin{eqnarray}
\label{eq:deltai}
\Delta_i(N,M^2/\mu_F^2,M^2/\mu_R^2) = {\rm exp}\left[\frac{1}{\alpha_s}g^{(1)}_i(\lambda) + g^{(2)}_i(\lambda,M^2/\mu_F^2,M^2/\mu_R^2)\right],
\end{eqnarray}
with $\lambda \equiv \alpha_s b_0 \ln\bar{N} \equiv \alpha_s b_0 \ln\left(N{\rm e}^{\gamma_E}\right) $, $\alpha_s \equiv \alpha_s(\mu_R^2)$ and $b_0$ the first-order coefficient of the QCD $\beta$-function (see Appendix~\ref{app:definitions}).
The $g^{(i)}$ functions are collected in Appendix~\ref{app:definitions}. In $N$-space and up to NLL, the resummed soft function $\mathbf{S}$ is given by a solution of the renormalization group equation (RGE)~\cite{Kidonakis:1998nf} and takes the form
\begin{eqnarray}
\mathbf{S}_{ij\rightarrow t\bar{t}h}(N) = \mathbf{U}^{\dagger}_{ij\rightarrow t\bar{t}h}\mathbf{S}^{(0)}_{ij\rightarrow t\bar{t}h} \mathbf{U}_{ij\rightarrow t\bar{t}h}\,. \label{eq:softres1}
\end{eqnarray}
The matrix $\mathbf{S}^{(0)}_{ij\rightarrow t\bar{t}h}$ is the boundary condition for the RGE. The logarithmic enhancements are captured by the evolution matrix $\mathbf{U}$: 
\begin{eqnarray}
\label{eq:softint}
\mathbf{U}_{ij\rightarrow t\bar{t}h} = {\rm P}\,{\rm exp}\left[\int_{\mu_R^2}^{M^2/\bar{N}^2}\frac{{\rm d}q^2}{2q^2}\mathbf{\Gamma}_{ij\rightarrow t\bar{t}h}(\alpha_s(q^2))\right].
\end{eqnarray}
Here, ${\rm P}$ denotes the path ordering in the variable $q$, and $\mathbf{\Gamma}_{ij\rightarrow t\bar{t}h}(\alpha_s(q^2))$ is the soft anomalous dimension that has the perturbative expansion
\begin{eqnarray}
\mathbf{\Gamma}_{ij\rightarrow t\bar{t}h}(\alpha_s(q^2)) = \left(\frac{\alpha_s(q^2)}{\pi}\right)\mathbf{\Gamma}^{(1)}_{ij\rightarrow t\bar{t}h} +  \left(\frac{\alpha_s(q^2)}{\pi}\right)^2\mathbf{\Gamma}^{(2)}_{ij\rightarrow t\bar{t}h} + \dots. 
\end{eqnarray}
The explicit form of the soft anomalous dimension is collected in Appendix~\ref{app:definitions}. The path ordering is not needed if the soft anomalous dimension matrix is diagonal. In such cases, we can simply solve the integral up to NLL to obtain
\begin{eqnarray}
\mathbf{U}_{ij\rightarrow t\bar{t}h} = {\rm exp}\left[\mathbf{\Gamma}^{(1)}_{ij\rightarrow t\bar{t}h} \frac{\ln(1-2\lambda)}{2\pi b_0}\right],
\end{eqnarray}
The resummed soft function at NLL can then be written as
\begin{eqnarray}
\mathbf{S}_{ij\rightarrow t\bar{t}h} = \mathbf{S}^{(0)}_{ij\rightarrow t\bar{t}h}{\rm exp}\left[ \left(\mathbf{\Gamma}^{(1)}_{ij\rightarrow t\bar{t}h}+\mathbf{\Gamma}^{(1)\dagger}_{ij\rightarrow t\bar{t}h}\right)\frac{\ln(1-2\lambda)}{2\pi b_0}\right]. \label{eq:softres}
\end{eqnarray}
However, this simple form can only be obtained if indeed the soft anomalous dimension becomes diagonal in the threshold limit, which will not be the case for every threshold variable that we consider. If it is not the case, we need to diagonalize it in order to reduce the path ordered exponential to an ordinary exponential. We make use of the method outlined in~\cite{Kidonakis:1998nf}, and introduce a matrix $\mathbf{R}$ such that
\begin{eqnarray}
\mathbf{\Gamma}_{\mathbf{R}, {ij\rightarrow t\bar{t}h}}^{(1)} = \mathbf{R}^{-1}\mathbf{\Gamma}^{(1)}_{ij\rightarrow t\bar{t}h}\mathbf{R}\,,
\end{eqnarray}
with $\mathbf{\Gamma}_{\mathbf{R}, {ij\rightarrow t\bar{t}h}, lm}^{(1)} = \lambda_l\delta_{lm}$ and $\lambda_l$ the $l^{\rm th}$ eigenvalue of $\mathbf{\Gamma}_{{ij\rightarrow t\bar{t}h}}^{(1)}$. The other two matrices $\mathbf{S}$ and $\mathbf{H}$ then also need to be written in this basis. This procedure is summarized in Appendix~\ref{app:definitions}. \\
The hard function can also be written as a perturbative series in $\alpha_s$. In order to perform NLL resummation one only needs to know the lowest-order contribution $\mathbf{H}^{(0)}_{ij\rightarrow t\bar{t}h}$, which can be obtained after decomposing the LO matrix element for $t\bar{t}h$ production into $s$-channel color bases, one for each channel~\cite{Kidonakis:1997gm}. This is a standard technique, and the resulting expressions for $\mathbf{H}^{(0)}_{ij\rightarrow t\bar{t}h}$ and $\mathbf{S}^{(0)}_{ij\rightarrow t\bar{t}h}$ are collected in Appendix~\ref{app:definitions}. The resulting resummed hadronic distribution is then
\begin{eqnarray}\label{eq:NLLresumhad}
{\rm d}\sigma_{t\bar{t}h}^{\rm NLL}(N) &=& \sum_{i,j} f_i(N+1,\mu_F^2) f_j(N+1,\mu_F^2) \\
&&\times\, {\rm Tr}\left[\mathbf{H}^{(0)}_{ij\rightarrow t\bar{t}h}(N)\mathbf{S}^{(0)}_{ij\rightarrow t\bar{t}h}(N+1)\right]\, {\rm exp}\left[\frac{2}{\alpha_s}g_{i}^{(1)}(\lambda)+2g_i^{(2)}\left(\lambda,M^2/\mu_F^2,M^2/\mu_R^2\right)\right]\, \nonumber \\
&\equiv&  \sum_{i,j} f_i(N+1,\mu_F^2) f_j(N+1,\mu_F^2) {\rm d}\hat{\sigma}^{\rm NLL}_{ij\rightarrow t\bar{t}h}(N)\,,
\end{eqnarray}
where we leave the dependence of ${\rm d}\hat{\sigma}^{\rm NLL}_{ij\rightarrow t\bar{t}h}$ on $M^2$, $\mu_R^2$ and $\mu_F^2$ implicit. 
The resummation-improved distributions can be obtained by matching the resummed result to the fixed-order result as follows
\begin{eqnarray}
{\rm d}\sigma^{\rm NLL + NLO}_{t\bar{t}h+X} = \left({\rm d}\sigma^{\rm NLL}_{t\bar{t}h+X} - {\rm d}\sigma^{\rm NLL}_{t\bar{t}h+X}\Bigg|_{\rm NLO}\right) + {\rm d}\sigma^{\rm NLO}_{t\bar{t}h+X}\,,
\end{eqnarray}
where ${\rm d}\sigma^{\rm NLL}_{t\bar{t}h+X}\Big|_{\rm NLO}$ indicates the resummed result truncated to NLO. This involves expanding the resummation exponents to $\mathcal{O}(\alpha_s)$ and performing an inverse Mellin transform of the resummed result to transform it back to physical space, i.e.
\begin{eqnarray}
\label{eq:inverse}
{\rm d}\sigma^{\rm NLL}_{t\bar{t}h+X} - {\rm d}\sigma^{\rm NLL}_{t\bar{t}h+X}\Bigg|_{\rm NLO} &=& \sum_{i,j}\frac{1}{2\pi i}\int_{c-i\infty}^{c+i\infty}{\rm d}N \tau^{-N}\, f_i(N+1,\mu_F^2)\, f_j(N+1,\mu_F^2) \\
&&\hspace{4cm}\times\,\left[{\rm d}\hat{\sigma}_{ij\rightarrow t\bar{t}h}^{\rm NLL}(N) -{\rm d}\hat{\sigma}_{ij\rightarrow t\bar{t}h}^{\rm NLL}(N)\Big|_{\rm NLO}\right], \nonumber
\end{eqnarray}
with $c$ a real number. \\
In what follows, we demonstrate how different parameterizations of the threshold variable may be chosen, depending on the distribution one is interested in. We start by examining the threshold behavior of the invariant-mass distribution, followed by that of the transverse-momentum distribution. Of course, one may integrate the resulting resummed distributions to obtain the full cross section. 
\subsection{Threshold definitions for the invariant-mass distribution}
\label{sec:subsecinv}
For the invariant-mass distribution, we parameterize the two-body phase space of Eq.~\eqref{eq:twobodyH} using the first expression of Eq.~\eqref{eq:higgsph} for the momentum of the Higgs boson. Using the azimuthal symmetry of the matrix element, we can write 
\begin{eqnarray}
\label{eq:invmassPS}
\int {\rm d}\Phi_2(p_1 + p_2; p_{t\bar{t}}, p_h) &=&\frac{1}{2\pi}\frac{|\vec{p}_h|}{4\sqrt{s}} \int {\rm d}\cos\theta_h\,,
\end{eqnarray} 
with 
\begin{eqnarray}
 E_h = \frac{s+m_h^2-s_{t\bar{t}}}{2\sqrt{s}} \geq m_h\,, \quad |\vec{p}_h| = \frac{\lambda^{1/2}(s,s_{t\bar{t}},m_h^2)}{2\sqrt{s}}\,.
\end{eqnarray}
We aim to achieve factorization of the partonic coefficient function and the PDFs after Mellin transforming  Eq.~\eqref{eq:hadronicttbar} by choosing a suitable threshold variable. The first option is to define $M^2 = (2m_t+m_h)^2$ such that
\begin{eqnarray}
\label{eq:absthres}
 \rho_{\rm abs} \equiv \frac{\left(2m_t+m_h\right)^2}{s} \simeq \frac{\left(2m_t+m_h\right)^2}{Q^2},\quad \tau_{\rm abs} \equiv \frac{\left(2m_t+m_h\right)^2}{S}\,,
\end{eqnarray}
with $Q^2$ the invariant mass of the final state, i.e.~$Q^2 = (p_t+p_{\bar{t}}+p_h)^2$ ($=s$ at LO). We are interested in the invariant-mass distribution. To this end, we Mellin transform Eq.~\eqref{eq:hadronicttbar} with respect to $\tau_{\rm abs}$, and rewrite the argument of the $\delta$-function as
\begin{eqnarray}
\label{eq:solvedelta}
 \delta\left(s-x_1x_2S\right) = \frac{1}{s}\frac{(2m_t+m_h)^2}{S}\delta\left(\tau_{\rm abs} - x_1x_2\rho_{\rm abs}\right) = \frac{1}{s}\,\tau_{\rm abs}\,\delta\left(\tau_{\rm abs} - x_1x_2\rho_{\rm abs}\right).
\end{eqnarray}
We also make a variable transform from $s$ to $\rho_{\rm abs}$,
\begin{eqnarray}
\int_{(2m_t+m_h)^2}^S\frac{{\rm d}s}{s} = 
\int_{\tau_{\rm abs}}^1\frac{{\rm d}\rho_{\rm abs}}{\rho_{\rm abs}}\,,
\end{eqnarray}
where the factor of $1/s$ stems from Eq.~\eqref{eq:solvedelta}. After these two steps, we may interchange the order of the $\tau_{\rm abs}$ and $\rho_{\rm abs}$ integrations, and obtain
\begin{eqnarray}
{\rm d}\sigma_{t\bar{t}h}^{{\rm LO, abs}}(N) = \sum_{i,j}f_i(N+1,\mu_F^2)f_j(N+1,\mu_F^2)\int_0^1\frac{{\rm d}\rho_{\rm abs}}{\rho_{\rm abs}}\rho_{\rm abs}^{N}\,{\rm d}\hat{\sigma}^{\rm LO}_{ij\rightarrow t\bar{t}h}\left((2m_t+m_h)^2/\rho_{\rm abs}\right).
\end{eqnarray}
By now introducing an integral over $Q^2$ using $Q^2 \simeq s = (2m_t+m_h)^2/\rho_{\rm abs}$ (valid at LO and in the limit of $z\rightarrow 1$)
\begin{eqnarray}
 \int {\rm d}Q^2\, \delta\left(Q^2 - \frac{(2m_t+m_h)^2}{\rho_{\rm abs}}\right),
\end{eqnarray}
we rewrite
\begin{eqnarray}
\label{eq:hadronicttbarABS}
{\rm d}\sigma_{t\bar{t}h}^{\rm LO, abs}(N) &=&  \sum_{i,j} f_i(N+1,\mu_F^2) f_j(N+1,\mu_F^2)\\
&& \hspace{0.2cm} \times \int{\rm d}Q^2 \int_{0}^1\frac{{\rm d}\rho_{\rm abs}}{\rho_{\rm abs}}\, \delta\left(Q^2-\frac{(2m_t+m_h)^2}{\rho_{\rm abs}}\right) \,\rho^{N}_{\rm abs}\,{\rm d}\hat{\sigma}_{ij\rightarrow t\bar{t}h}^{\rm LO}((2m_t+m_h)^2/\rho_{\rm abs})\,. \nonumber
\end{eqnarray}
Solving the $\delta$-function for $\rho_{\rm abs}$ leads us to the invariant mass distribution
\begin{eqnarray}
\label{eq:hadronicINVABS}
\frac{{\rm d}\sigma_{t\bar{t}h}^{\rm LO,abs}}{{\rm d}Q} (N) &=&  \frac{2}{Q}\,\sum_{i,j} f_i(N+1,\mu_F^2) f_j(N+1,\mu_F^2) \, \rho_{{\rm abs}}^N\,{\rm d}\hat{\sigma}_{ij\rightarrow t\bar{t}h}^{\rm LO}(Q^2)\Big|_{ Q^2 = (2m_t+m_h)^2/\rho_{\rm abs}}\,.\,\,\,\,
\end{eqnarray}
The integration bounds on $s_{t\bar{t}}$ become $\rho_{\rm abs}$-dependent and read
\begin{equation}
\label{eq:rhodepstt}
    4m_t^2\, <\, s_{t\bar{t}}\, <\, \left(\frac{2m_t+m_h}{\sqrt{\rho_{\rm abs}}}-m_h\right)^2\,.
\end{equation}
As noted before, $\rho_{\rm abs}$ was used in Ref.~\cite{Kulesza:2015vda} to compute the resummed total cross section to NLL accuracy.  \\
Now we come back to a point that was raised before, namely the approximation applied to the upper limit in the integral of Eq.~\eqref{eq:deltaiint}. One can show using phase-space arguments~\cite{Forte:2002ni} that the upper limit of this integral actually reads $s(1-z)^2/z$. Two approximations are made to end up with Eq.~\eqref{eq:deltaiint} that are valid up to leading power in the threshold variable: $1/z \simeq 1$ and $s \simeq M^2$. It is trivial to see that the first replacement may be made for the $z\rightarrow 1$ limit valid in the limit where all gluons are soft, which is the only region where one can guarantee the validity of Eq.~\eqref{eq:deltaiint}. The $z\rightarrow 1$ limit is isolated by the $N\rightarrow \infty$ limit. By taking $N\rightarrow \infty$, the largest contribution of Eq.~\eqref{eq:hadronicttbarABS} comes from $\rho_{\rm abs} \simeq 1$. Therefore, we replace $s \rightarrow M^2$. Corrections of this replacement are of $\mathcal{O}(1/N)$, i.e.~of next-to-leading power. Therefore, by choosing different threshold variables, one is actually probing the importance of partial sub-leading power corrections.  \\
Since we are after the invariant-mass distribution, we have to integrate over the invariant mass of the $t\bar{t}$ pair. Upon taking a closer look at its integration limits, we may find our second threshold variable as $M^2 = \left(\sqrt{s_{t\bar{t}}}+m_h\right)^2$ such that
\begin{eqnarray}
\label{eq:sttthres}
 \rho_{s_{t\bar{t}}} \equiv \frac{\left(\sqrt{s_{t\bar{t}}}+m_h\right)^2}{s}\,,\quad \tau_{s_{t\bar{t}}} \equiv \frac{\left(\sqrt{s_{t\bar{t}}}+m_h\right)^2}{S}\,.
\end{eqnarray}
It is easy to see how this threshold definition arises if one switches around the order of the $s_{t\bar{t}}$ and $s$ integrations. That is, we may write
\begin{eqnarray}
\label{eq:sttswitch}
\int^S_{(2m_t+m_h)^2}{\rm d}s \int^{\left(\sqrt{s}-m_h\right)^2}_{4m_t^2}{\rm d}s_{t\bar{t}} =  \int^{\left(\sqrt{S}-m_h\right)^2}_{4m_t^2}{\rm d}s_{t\bar{t}} \int_{\left(\sqrt{s_{t\bar{t}}}+m_h\right)^2}^{S}{\rm d}s\,,
\end{eqnarray}
where indeed the lower limit of the $s$-integration indicates an edge of phase space that depends on the  value of $s_{t\bar{t}}$.
Secondly, we make a variable transform from $s$ to $\rho_{s_{t\bar{t}}}$ to obtain
\begin{eqnarray}
\label{eq:otherthreshold}
{\rm d}\sigma_{t\bar{t}h}^{{\rm LO,}s_{t\bar{t}}} &=&  \tau_{s_{t\bar{t}}} \sum_{i,j} \frac{1}{4\pi s}K^{ij}  \int_{4m_t^2}^{\left(\sqrt{S}-m_h\right)^2} {\rm d}s_{t\bar{t}} \int_0^1{\rm d}x_1 f_i(x_1,\mu_F^2)\int_0^1{\rm d}x_2 f_j(x_2,\mu_F^2)\\
&&  \times \, \int_{\tau_{s_{t\bar{t}}}}^1 \frac{{\rm d}\rho_{s_{t\bar{t}}}}{\rho_{s_{t\bar{t}}}} \, \sum_{\rm spin, color}\left|\mathcal{M}_{ij}\right|^2 \, {\rm d}\Phi_2(p_1 + p_2; p_{t\bar{t}}, p_h) \,{\rm d}\Phi_2(p_{t\bar{t}}; p_t, p_{\bar t}) \,\delta\left(\tau_{s_{t\bar{t}}}-x_1 x_2 \rho_{s_{t\bar{t}}}\right). \nonumber 
\end{eqnarray}
We then perform the Mellin transform of Eq.~\eqref{eq:otherthreshold} with respect to $\tau_{s_{t\bar{t}}}$, which uses the $\delta$-function in Eq.~\eqref{eq:otherthreshold}. The invariant-mass distribution is obtained by again introducing the integral
\begin{eqnarray}
\label{eq:invdelta}
 \int {\rm d}Q^2\, \delta\left(Q^2-\frac{\left(\sqrt{s_{t\bar{t}}}+m_h\right)^2}{\rho_{s_{t\bar{t}}}}\right)\,.
\end{eqnarray}
This $\delta$-function may be used to solve the integral over $\rho_{s_{t\bar{t}}}$ such that the invariant-mass distribution becomes
\begin{eqnarray}
\label{eq:hadronicINVstt}
\frac{{\rm d}\sigma_{t\bar{t}h}^{{\rm LO,}s_{t\bar{t}}}}{{\rm d}Q} (N) &=&  \frac{2}{Q}\sum_{i,j} \frac{1}{4\pi s}K^{ij} f_i(N+1,\mu_F^2)\,f_j(N+1,\mu_F^2)  \int_{4m_t^2}^{\left(Q-m_h\right)^2} {\rm d}s_{t\bar{t}} \, \rho_{s_{t\bar{t}}}^N\,\\
&& \times \, \, \sum_{\rm spin, color}\left|\mathcal{M}_{ij}\right|^2 \, {\rm d}\Phi_2(p_1 + p_2; p_{t\bar{t}}, p_h) \,{\rm d}\Phi_2(p_{t\bar{t}}; p_t, p_{\bar t}) \Big|_{\rho_{s_{t\bar{t}}} = (\sqrt{s_{t\bar{t}}}+m_h)^2/Q^2}\,.\nonumber
\end{eqnarray}
As one can see, the integration limits on $s_{t\bar{t}}$ are now threshold-variable independent, and directly depend on $Q$ instead (which guarantees that with $\rho_{s_{t\bar{t}}} \leq 1$, Eq.~\eqref{eq:invdelta} has a solution).  \\
The third and final option is obtained by setting $M^2 = Q^2$, with
\begin{eqnarray}
\label{eq:rhoQ2}
 \rho_{Q^2} \equiv \frac{Q^2}{s} \,,\quad \tau_{Q^2} = \frac{Q^2}{S}\,,
\end{eqnarray}
which was used in Ref.~\cite{Kulesza:2017ukk} to compute the resummed invariant-mass distribution to NNLL accuracy.
With this definition and using the same steps as before, the invariant-mass distribution is easily obtained as 
\begin{eqnarray}
\label{eq:hadronicINVQ2}
\frac{{\rm d}\sigma_{t\bar{t}h}^{{\rm LO,}Q^2}}{{\rm d}Q} (N) &=&  \frac{2}{Q}\,\sum_{i,j} f_i(N+1,\mu_F^2) f_j(N+1,\mu_F^2) \, {\rm d}\hat{\sigma}_{ij\rightarrow t\bar{t}h}^{\rm LO}(Q^2)\,.
\end{eqnarray}
Obviously, the resulting LO invariant mass distributions are independent of the parameterization of the threshold variable. However, the resummed result is impacted by this choice because of the upper limit in Eq.~\eqref{eq:deltaiint} and Eq.~\eqref{eq:softint}. Using $\rho=\rho_{\rm abs}$, the soft-anomalous-dimension matrices (Eq.~\eqref{eq:gammaqq} and \eqref{eq:gammagg}) become diagonal in the threshold limit where $\rho_{\rm abs}\rightarrow 1$. The off-diagonal components contribute at $\mathcal{O}(1/N)$. We will explicitly include these in our numerical analysis in Section~\ref{sec:results} to analyze the numerical impact of this approximation. The soft-anomalous-dimension matrices are not diagonalized by the threshold limit for  $\rho_{s_{t\bar{t}}}$ and $\rho_{Q^2}$. 
\subsection{Threshold definitions for the Higgs transverse-momentum distribution}
\label{sec:subsectrans}
To compute the transverse-momentum spectrum, it is more convenient to use $p_h$ as given in the second expression of Eq.~\eqref{eq:higgsph} to write down the phase space of the $(t\bar{t})h$-system. We obtain
\begin{eqnarray}
\int {\rm d}\Phi_2(p_1 + p_2; p_{t\bar{t}}, p_h) &=& \frac{1}{(2\pi)^2}\int m_{{\rm T},h}{\rm d}m_{{\rm T},h}\, {\rm d}^2\vec{p}_{{\rm T}}\,{\rm d}\eta\, \nonumber \\
&&\hspace{2cm}\times\,\delta^+\left(m_{h}^2-m_{{\rm T},h}^2+p_{T,h}^2\right)\delta^+\left(s_{t\bar{t}}-\left(p_1+p_2-p_h\right)^2\right)\nonumber \\
&=& \frac{1}{\pi}\int \,{\rm d}p_{{\rm T}}\,  \frac{1}{4\sqrt{s}}\frac{p_{{\rm T}}}{p_{z,h}}\,. 
\end{eqnarray}
To find solutions to the $\delta^+$-distributions, we need that $m_{{\rm T},h} \geq m_h$ and
\begin{equation}
s_{t\bar{t}} \leq s + m_h^2 - 2\sqrt{s}\,m_{{\rm T},h}\,.
\end{equation}
Since $p_{z,h}$ is obtained via $\sinh\eta = \sqrt{\cosh\eta^2 -1}$ with $\cosh\eta = (s+m_h^2-s_{t\bar{t}})/(2\sqrt{s}m_{{\rm T},h})$, we need to multiply the final result by $2$ to also take into account the $p_{z,h} < 0$ configuration (for which $\theta_h > \pi/2$). We have included this factor in the equation above. We may then turn around the integration order of the $s_{t\bar{t}}$ and $s$ integrations as in Eq.~\eqref{eq:sttswitch}, resulting in
\begin{eqnarray}
&&\int_{(m_{{\rm T},h}+m_{{\rm T},4m_t^2})^2}^{S}{\rm d}s \,\delta(s-x_1 x_2 S)\int_{4m_t^2}^{s+m_h^2-2\sqrt{s}m_{{\rm T},h}} {\rm d}s_{t\bar{t}}\nonumber  \\
&&\hspace{3cm}=\int_{4m_t^2}^{S+m_h^2-2\sqrt{S}m_{{\rm T},h}}{\rm d}s_{t\bar{t}} \int_{\left(m_{{\rm T},h}+ \sqrt{p_{{\rm T}}^2+s_{t\bar{t}}}\right)^2}^{S} {\rm d}s\, \delta(s-x_1 x_2 S)\,,
\end{eqnarray}
with $m_{{\rm T},4m_t^2} = \sqrt{p_{{\rm T}}^2+4m_t^2}$, in analogy to the definition of $m_{{\rm T},h}$ (Eq.~\eqref{eq:higgsph}). 
There now arise two natural partonic threshold variables for a resummed transverse-momentum distribution:
\begin{eqnarray}
\label{eq:xstt}
    &1.&\quad x_{{\rm T},s_{t\bar{t}}}^2 = \frac{\left(m_{{\rm T},h}+m_{{\rm T},s_{t\bar{t}}}\right)^2}{s}\,, \text{ with } m_{{\rm T},s_{t\bar{t}}} = \sqrt{s_{t\bar{t}}+p_{t\bar{t},T}^2} = \sqrt{s_{t\bar{t}}+p_{{\rm T}}^2} \,, \\
\label{eq:x4mt2}
    &2.&\quad x_{{\rm T},4m_t^2}^2 = \frac{\left(m_{{\rm T},h}+m_{{\rm T}, 4m_t^2}\right)^2}{s}\,, \text{ with } m_{{\rm T},4m_t^2} = \sqrt{4m_t^2+p_{{\rm T}}^2}\,. 
\end{eqnarray}
The first option sets $M^2 = \left(m_{{\rm T},h}+m_{{\rm T},s_{t\bar{t}}}\right)^2$, while for the second option we have the phase-space-boundary parameterization $M^2 = \left(m_{{\rm T},h}+m_{{\rm T},4m_t^2}\right)^2$. Near $p_{{\rm T}} = 0$, $x_{{\rm T},4m_t^2}^2$ reduces to the absolute mass definition of the threshold with $\rho_{\rm abs} = \frac{(2m_t + m_h)^2}{s}$, while $x_{{\rm T},s_{t\bar{t}}}^2$ reduces to the the invariant mass definition with $\rho_{s_{t\bar{t}}} = \frac{(\sqrt{s_{t\bar{t}}}+ m_h)^2}{s}$. The first option ensures that the $s_{t\bar{t}}$ integration boundary is independent of the threshold variable. Using the first option, we may write $s = \frac{\left(m_{{\rm T},h}+m_{{\rm T},s_{t\bar{t}}}\right)^2}{x_{{\rm T},s_{t\bar{t}}}^2}$, such that
\begin{eqnarray}
\label{eq:mellin}
\int_{\left(m_{{\rm T},h}+ m_{{\rm T},s_{t\bar{t}}}\right)^2}^{S} {\rm d}s\, \delta(s-x_1 x_2 S) = \int_{X_{{\rm T},s_{t\bar{t}}}^2}^1\frac{{\rm d}x_{{\rm T},s_{t\bar{t}}}^2}{x_{{\rm T},s_{t\bar{t}}}^2}\, \delta\left(1 - \frac{x_1 x_2 x_{{\rm T},s_{t\bar{t}}}^2}{X_{{\rm T},s_{t\bar{t}}}^2}\right), 
\end{eqnarray}
with $X_{{\rm T},s_{t\bar{t}}}^2 =  \frac{\left(m_{{\rm T},h}+m_{{\rm T},s_{t\bar{t}}}\right)^2}{S}$ the hadronic threshold variable. For the second option, we have to use an upper boundary of the $s_{t\bar{t}}$ integration that depends on the threshold variable. As shown in Appendix~\ref{sec:softanomalousdim}, the soft-anomalous-dimension matrices are diagonal in the threshold limit for  $x^2_{{\rm T},4mt^2}$ but not for $x^2_{{\rm T},s_{t\bar{t}}}$. \\

\noindent We have now set up our notation for the resummation of the $t\bar{t}h$ process in $N$-space, and formulated different threshold definitions for the invariant-mass and transverse-momentum distribution. In what follows, we will present our numerical results obtained using the various threshold definitions for the invariant-mass distribution, transverse-momentum distribution, and the total cross section. An overview of these definitions (and scale choices) will be given in Section~\ref{sec:results} (Table~\ref{tab:summary} on page \pageref{tab:summary}). First, however, we discuss a technical issue: performing the inverse Mellin transform. 
\section{The inverse Mellin transform}
\label{sec:inversemel}
Stabilizing the numerical evaluation of the inverse Mellin transform is a notorious problem in direct-QCD resummation. Most commonly, the so-called Minimal Prescription (MP) method is used to handle the integral in Eq.~\eqref{eq:inverse}, which consists of bending the contour towards the negative real axis with a large angle. This method works well if one has access to both the analytical Mellin transforms of the PDFs and that of the partonic cross section. For a complicated cross section already at LO like that of $t\bar{t}h$, the latter is not achievable. This presents numerical stability issues. The method that we introduce here reduces the oscillations of the integrand in the complex plane, and thereby helps to stabilize the numerical evaluation of the inverse Mellin transform. \\ 
\noindent Despite it often being mentioned in this context, the \emph{original} MP as introduced in Ref.~\cite{Catani:1996yz} is \emph{not} a numerical prescription that tells one how to numerically compute the inverse Mellin transform. Instead, the MP is a definition for the existence of the inverse Mellin transform of the resummed formula. Before we discuss the issues of performing the inverse Mellin transform numerically, we briefly recall why the MP is needed to obtain an analytically viable result in the first place, as it is often confused in the literature with the numerical prescription. 
\subsection{Analytical considerations of the inverse Mellin transform}
Let us first consider the definition of a Mellin transform~\cite{doi:10.1002/9781118032770.ch9,mellin}. Given a function $g(t)$,  assume that its integral up to a finite (real) $a$ is bounded
\begin{eqnarray}
\int_0^a {\rm d}t\, |g(t)| < \infty\,,
\end{eqnarray}
and that $|g(t)| \leq K\, {\rm e}^{c_1 t}$ for $t\rightarrow \infty$ with $K$ and $c_1$ real constants and $K > 0$. For such functions, a one-sided Laplace transform exists and is defined by
\begin{eqnarray}
\label{eq:laplaceoneside}
g(N) \equiv \mathcal{L}^+\left[g(t)\right] \equiv \int_0^{\infty}{\rm d}t\, {\rm e}^{-Nt}\,g(t)\,.
\end{eqnarray}
The resulting function $g(N)$ is analytic for ${\rm Re}[N] > c_1$. From this, we may derive the Mellin transform of a function $f(x)$ for $x\in[0,1]$. To this end, we set $t = -\ln(x)$ and $f(x) \equiv g(-\ln(x))$, such that
\begin{eqnarray}
\label{eq:ourmellin}
f(N) \equiv \mathcal{M}\left[f(x)\right] \equiv \int_0^{1}{\rm d}x\, x^{N-1}\,f(x)\,.
\end{eqnarray}
This is the form of the Mellin transform that we have encountered above. \\
However, we may also get the Mellin transform from a two-sided Laplace transform by combining it with the one-sided Laplace transform of $g(t)$ for $t\in(-\infty,0]$. The latter object reads
\begin{eqnarray}
\label{eq:laplaceotherside}
\mathcal{L}^-\left[g(t)\right] &\equiv& \int_{-\infty}^{0}{\rm d}t\, {\rm e}^{-Nt}g(t) =  \int_{0}^{\infty}{\rm d}t\, {\rm e}^{Nt}g(-t)\,,
\end{eqnarray}
with the requirement
\begin{eqnarray}
\int_0^a {\rm d}t\, |g(-t)| < \infty\,.
\end{eqnarray}
The Laplace transform of $g(t)$ with $t\in(-\infty,0]$ exists and is analytic for ${\rm Re}[N] < c_2$ if $|g(t)| \leq K'\,{\rm e}^{c_2 t}$ for $t\rightarrow -\infty$ with $K'$ and $c_2$ real constants with $K' > 0$. Setting $t = -\ln(x)$ and $f(x) \equiv g(-\ln(x))$ in Eq.~\eqref{eq:laplaceotherside}, and combining it with Eq.~\eqref{eq:laplaceoneside}, we obtain
\begin{eqnarray}
\mathcal{M}_{\infty}\left[f(x)\right] \equiv \int_0^{\infty}{\rm d}x\, x^{N-1}\,f(x)\,,
\end{eqnarray}
which holds for $c_1 < {\rm Re}[N] < c_2$. We have used the subscript `${\infty}$' to distinguish this Mellin transform from the one in Eq.~\eqref{eq:ourmellin}. \\
In order for $g(N)$ to represent a Laplace transform of the function $g(t)$ with $t \in [0,\infty)$, we need to require that
\begin{eqnarray}
\label{eq:reqMellin1}
\left|g(N)\right| &\equiv& \left|\int_0^{\infty} {\rm d}t\, {\rm e}^{-Nt}\, g(t)\right|
\leq \int_0^{\infty} {\rm d}t\, {\rm e}^{-{\rm Re}[N]t}\, \left|g(t)\right| \leq  \frac{K}{{\rm Re}[N]-c_1}\,.
\end{eqnarray}
\noindent This needs to hold for all ${\rm Re}[N] > c_1$. It straightforwardly follows that the same requirement is imposed on $f(N)$ if $f(N)$ represents a Mellin transform of a function $f(x)$ with $x \in [0,1]$. For a two-sided Laplace transform of $g(t)$ with $t \in (-\infty,\infty)$, we obtain an additional requirement on $g(N)$ (or on $f(N)$ for a Mellin transform of a function $f(x)$ with $x\in[0,\infty]$) that reads
\begin{eqnarray}
\label{eq:reqMellin2}
\left|g(N)\right|  \leq  \frac{K'}{c_2 - {\rm Re}[N]}\,,
\end{eqnarray}
which needs to hold for all ${\rm Re}[N] < c_2$. \\
The ${\rm Re}[N]$ domain for which the Mellin transform $f(N)$ exists and is analytic, and the domain in which its inverse describes the original function $f(x)$, is called the strip of definition. The strip of definition is either $c_1 < {\rm Re}[N] < c_2$ for $f(x)$ with $x\in [0,\infty)$, or ${\rm Re}[N] > c_1$ if $x$ is restricted within the interval $[0,1]$. The inverse Mellin transform must be taken over a straight line that runs from $c-i\infty$ to $c+i\infty$, as already shown in Eq.~\eqref{eq:inverse}. The value $c$ can be chosen arbitrarily, as long as it lies within the strip of definition, which is simply a consequence of Cauchy's theorem. If we pick a value of $c$ that lies outside the strip of definition, there is no guarantee that the inverse will return the original function. \\
We now turn our attention to the resummed cross section. The large-$N$-approximated resummed exponent introduces two branch cuts. These may already be observed in $g^{(1)}_i(\lambda)$ of Eq.~\eqref{eq:deltai}, as this function cannot be evaluated below ${\rm Re}[N] < 0$ and beyond ${\rm Re}[N]>N_{L} = \exp\left(\frac{1}{2\alpha_s b_0}-\gamma_E\right)$. Therefore, the strip of definition is given by $0 < {\rm Re}[N] < N_{L}$. Note that, besides the resummed exponent, the $N$-space PDFs and partonic cross section could also introduce additional poles in the ${\rm Re}[N] > 0$ domain, and place additional constraints on the values that $c$ is allowed to take. \\
However, now a problem appears, as such a bounded strip of definition \emph{only} holds for a Mellin transform that is performed on a function $f(x)$ with $x \in [0,\infty)$, rather then $x \in [0,1]$. This is caused by the fact that we cannot find a finite value for $K$ for which Eq.~\eqref{eq:reqMellin1} holds for any resummed partonic cross section $\hat{\sigma}(N)$. Therefore, $\hat{\sigma}(N)$ \emph{cannot represent the Mellin transform of any resummed partonic cross section $\hat{\sigma}(\rho)$ with $0 \leq \rho \leq 1$}. Instead, the resummed partonic cross section receives contributions from the non-physical domain where $\rho > 1$, which might be problematic if these terms grow large. The original MP~\cite{Catani:1996yz} then consists of \emph{defining} the hadronic resummed cross section in physical space through Eq.~\eqref{eq:inverse}. This is validated in Ref.~\cite{Catani:1996yz} by proving that the contribution of the domain where $\rho > 1$ is suppressed by a factor ${\rm e}^{-HQ(1-\tau)/\Lambda_{\rm QCD}}$, with $H\sim \ln(Q/\Lambda_{\rm QCD})$ a slowly varying positive function. Since $\Lambda_{\rm QCD} = \mathcal{O}(1)$~GeV, $\tau \neq 1$, and $Q= \mathcal{O}(100)$~GeV represents the hard scale of the process, the contribution of the domain with $\rho > 1$ is indeed negligible. The original MP tackles the problem of the formally non-existent inverse Mellin transform. However, there is a second problem to worry about: that of the numerical convergence of the inverse Mellin transform. 
\subsection{Numerical stability issues of the inverse Mellin transform}
To implement threshold resummation numerically, one needs to numerically perform the integration on the domain ${\rm Im}[N] \in (-\infty,\infty)$. This leads to large oscillations (as will be shown later), which are difficult to stabilize.  One option, often referred to as the MP parameterization in literature, is to double the contribution of the integral in Eq.~\eqref{eq:inverse} in the ${\rm Im}[N] > 0$ plane, with $N$ being parameterized as
\begin{eqnarray}
\label{eq:cmp}
N = C_{\rm MP} + y\exp\left(i\phi_{\rm MP}\right)\,\, \text{with}\,\, 0<y<\infty\,,\,\, 0< C_{\rm MP} < N_L\,,\,\, \text{and}\,\,\, \frac{\pi}{2} \leq \phi_{\rm MP} < \pi\,.\,\,\,\,\,
\end{eqnarray}
Note that this parameterization in principle has no relation to the original MP, which is the method that proves that the contributions of Eq.~\eqref{eq:inverse} outside the physical domain with $\rho > 1$ are exponentially suppressed, as summarized above. However, the reason why it is known in the literature as the MP parameterization is that it uses the same principle: it consists of bending the integration contour towards the negative real axis. However, for the orginal MP, the bending towards the negative axis is done infinitesimally ($\phi_{\rm MP} = \mathcal{O}(\epsilon)$), while $\phi_{\rm MP} =  \mathcal{O}(1)$ in the MP parameterization above. The integral can formally not depend on the value of $C_{\rm MP}$ and $\phi_{\rm MP}$. However, choosing these parameters with care can be helpful for an efficient numerical evaluation. By choosing $\phi_{\rm MP} >\frac{\pi}{2}$ (for $y > 0$), one introduces an exponential damping to the factor $\tau^{-N}$ of Eq.~\eqref{eq:inverse}. This means that the integrand is exponentially suppressed for $\tau < 1$ and ${\rm Re}[N] \rightarrow -\infty$. The value of $C_{\rm MP}$ has to be chosen inside the strip of definition, and in practice, it is usually chosen to be around $2$.  \\
If the Mellin transforms of either the PDFs or the partonic cross section needs to be evaluated numerically, the MP parameterization presents a problem. Firstly, as we have no handle on the analytical continuation of the Mellin transform, we do not know the strip of definition and therefore we do not know a good value for $C_{\rm MP}$, nor can we be certain that points outside the strip of definition will not be sampled by bending the contour towards the negative real axis of $N$.  \\
The more practical issue however is that the numerical Mellin transform of the PDFs (or partonic cross section) does not converge along the path of integration for ${\rm Re}[N] < 0$, as then $x^{N}$ with $x<1$ will become increasingly large for $x \rightarrow 0$. Secondly, $x^{N}$ oscillates heavily for ${\rm Im}[N] \rightarrow \infty$ if no exponential damping is introduced. It follows that in cases where we do not have access to the analytic forms of $f_i(N,\mu_F^2)$ and $\hat{\sigma}(N)$, there is an overall factor that reads
\begin{eqnarray}
\label{eq:expgrowth}
\left(\frac{x_1 x_2 \rho}{\tau}\right)^N  &=& {\rm exp}\left[N\ln\left(\frac{x_1 x_2\rho}{\tau}\right)\right].
\end{eqnarray}
The argument of the logarithm can be larger or smaller than $1$ for fixed $\tau$. When the argument becomes larger than $1$, we would like to have a contour where ${\rm Re}[N]<0$, as then the integral can converge numerically. On the other hand, if the argument is smaller than $1$, we need ${\rm Re}[N]>0$. These two requirements cannot hold simultaneously, although both domains, for constant $\tau$, are probed in Eq.~\eqref{eq:inverse}. \\
\noindent One way in which the numerical stability issue may be circumvented is by setting $(x_1 x_2 \rho)/\tau > 1 $ to define the integration domain for $\rho$, $x_1$ and $x_2$, and confirm that the contribution that originates from $(x_1 x_2 \rho)/\tau \leq 1 $ is negligible by checking this explicitly. For the PDFs, one may employ the \emph{derivative method}~\cite{Kulesza:2002rh} to introduce an $\mathcal{O}(1/N)$ (or higher, depending on how many derivatives are taken) suppression to the numerical oscillations. For resummation it is assumed that ${\rm Re}[N]\rightarrow \infty$, hence additional factors of $\mathcal{O}(1/N)$ will reduce the absolute size of the oscillations. However, there are worries whether one may trust the result of this method for general processes, especially when employing higher derivatives (see Appendix~\ref{app:deriv} for a justification of this statement). \\
In this work, we will make an analytical fit to the PDFs, using the functional form
\begin{eqnarray}
\label{eq:fit}
&&y = 1-2\sqrt{x}\,, \\
&& xf(x) = A(1-x)^{a_1}x^{a_2}(1+b y + c (2 y^2-1))+B(1-x)^{a_3}x^{a_4}(1 + C x^{a_5})\,.  \nonumber 
\end{eqnarray}
We demand that the fitted function lies within the $1\sigma$ error as given by the LHAPDF~\cite{Buckley:2014ana} grid implementation of the PDFs in the entire domain. By fitting the PDFs, we remove the numerical factor of $x_{1,2}^N$ that could result in exponential growth. We now also know the strip of definition for the Mellin transform of the PDFs. More specifically, we require
\begin{eqnarray}
{\rm Re}[a_{1,3}] > -1\,, \quad {\rm Re}[a_2+N] > 0\,, \quad {\rm Re}[a_4+N] > 0\,, \quad {\rm Re}[a_4+a_5+N] > 0\,,
\end{eqnarray}
\noindent where the first set of conditions is respected in the fitting procedure. The last three conditions define the strip of definition for the Mellin transform of the PDFs. \\
To reduce the numerical stability issues of the Mellin transform of the partonic cross section, we invented a novel deformation method, which is explored in the next sub-section.
\subsection{The deformation method}
\label{sec:deformation}
\noindent This sub-section covers a novel method to perform the Mellin transform numerically, which results in an additional suppression factor of $1/N$, but also removes part of the numerical oscillations present in Eq.~\eqref{eq:expgrowth}. Consider the Mellin transform of the partonic cross section
\begin{eqnarray}
\hat{\sigma}_{ij}\left(N\right)  &=& \int_{0}^1 {\rm d}\rho\, \rho^{N-1} \hat{\sigma}_{ij}\left(\rho\right). \label{eq:xsecmellin}
\end{eqnarray}
We make the following change of variables
\begin{eqnarray}
\label{eq:cov}
\rho = {\rm exp}\left[\frac{w}{N}\right]\text{, where } w \in \left(-N\infty, 0\right] \text{ for } \rho\in \left[0,1\right] \text{ and } N \in \mathbb{C}\,.
\end{eqnarray}
The new variable is $w$, which is complex-valued and $N$-dependent. The Jacobian leads to a $1/N$ suppression
\begin{eqnarray}
\frac{{\rm d}\rho}{{\rm d}w} = \frac{1}{N}\,{\rm exp}\left[\frac{w}{N}\right].
\end{eqnarray}
\noindent Then, for each value for $N = C_{\rm MP} + iy$, we integrate over a straight line in the lower left part of the complex $w$ plane. This means that Eq.~\eqref{eq:xsecmellin} now becomes
\begin{eqnarray}
\label{eq:transform1}
\hat{\sigma}_{ij}\left(N\right)  &=& \int_{-N\infty}^0 {\rm d}w\, \frac{{\rm e}^w}{N}\, \hat{\sigma}_{ij}\left({\rm exp}\left[\frac{w}{N}\right]\right).
\end{eqnarray}
\begin{figure}[t]
  \centering
    \includegraphics[width=0.7\textwidth]{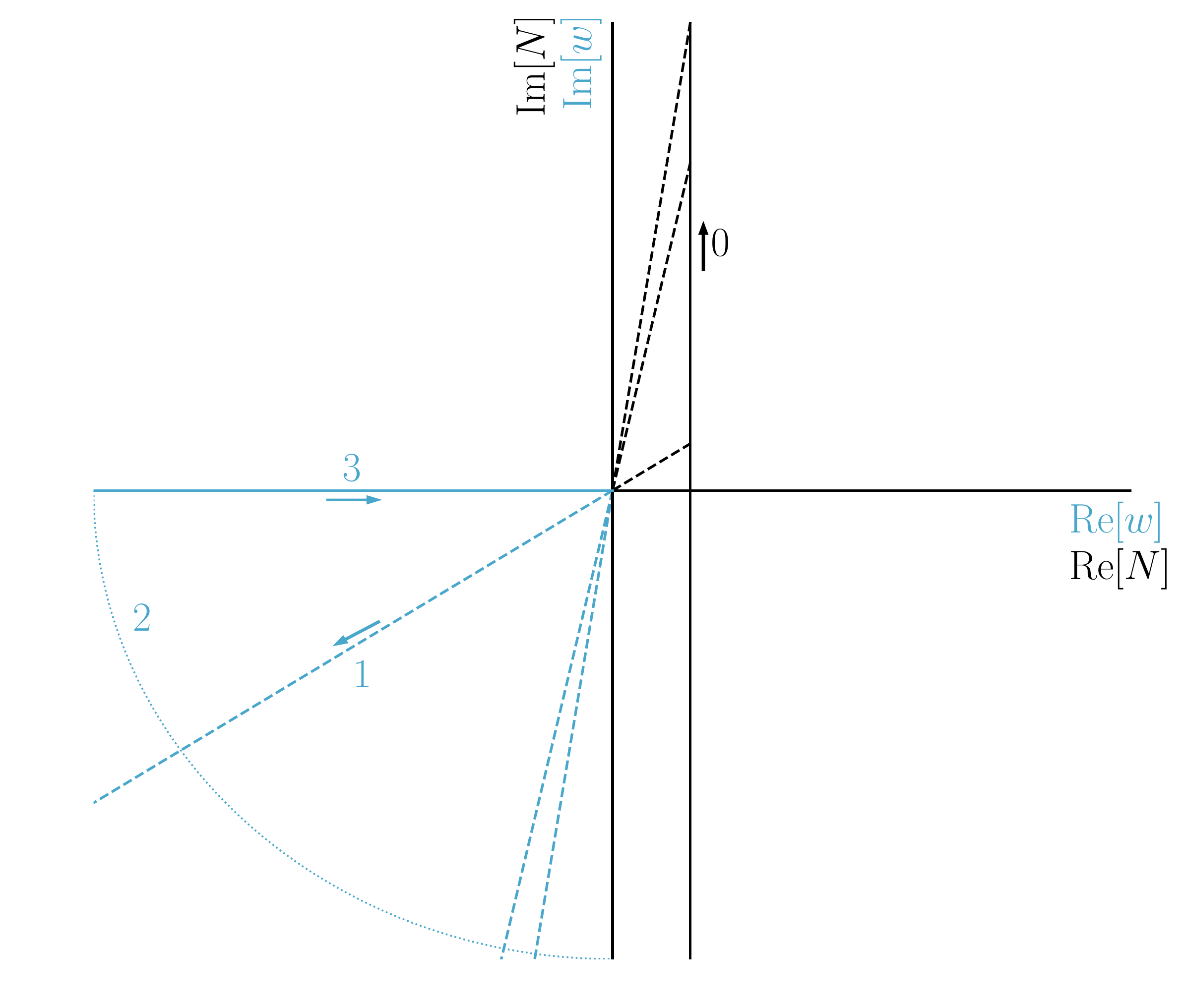}
  \caption{Schematic representation of the change of variables of Eq.~\eqref{eq:cov}. The solid black line (indicated by $0$) corresponds to the standard integration path for the inverse Mellin transform, while the dashed blue line ($1$) corresponds to the integration path of $w$ after the change of variables. As $w$ is $N$ dependent, we obtain a different integration path for each value of $N$, as indicated by the other dashed blue lines. We then deform this path to the negative real axis (solid blue line, indicated by $3$) for all values of $N$. }
\label{fig:newmethod}
\end{figure}

\noindent In Fig.~\ref{fig:newmethod}, this integration path is indicated by the dashed blue lines for different values of $N$. These lines extend to ${\rm Re}[w] = -\infty$. This integration path does not converge numerically due to large oscillations that are induced by the imaginary part of $w$. However, by Cauchy's theorem, we may deform the integral to the real negative axis, where we can compute the integral. The total contour integral equals $0$, as there are no poles enclosed, thus
\begin{eqnarray}
\oint {\rm d}w f(w) = 0\, ,
\end{eqnarray}
where $f(w) = \frac{{\rm e}^w}{N}\hat{\sigma}\left({\rm exp}\left[\frac{w}{N}\right]\right)$ (see Eq.~\eqref{eq:transform1}).
The contour integral consists of three segments, represented in Fig.~\ref{fig:newmethod} by the labels $1$, $2$, and $3$. For $N = C_{\rm MP} + iy$, these paths can respectively be parameterized by
\begin{align}
\text{Segment $1$:}\qquad w &= t\,{\rm e}^{i\theta_{\max}}&;&\, -R < t < 0 \nonumber \\
\text{Segment $2$:}\qquad w &= R\,{\rm e}^{i\theta}&;&\, \pi < \theta < \theta_{\rm max} = \pi+\tan^{-1}\left(\frac{y}{C_{\rm MP}}\right) \label{eq:paramcov}\\
\text{Segment $3$:}\qquad w &= w &;&\, 0 > w > -R\,, \nonumber 
\end{align}
when $y > 0$ with $\theta_{\rm max} < \frac{3\pi}{2}$, and 
\begin{align}
\text{Segment $1$:}\qquad w &= t\,{\rm e}^{i\theta_{\min}}&;&\, -R < t < 0 \hspace{4.4cm} \nonumber \\
\text{Segment $2$:}\qquad w &= R\,{\rm e}^{i\theta}&;&\, \pi > \theta > \theta_{\rm min} = \pi+\tan^{-1}\left(\frac{y}{C_{\rm MP}}\right)\\
\text{Segment $3$:}\qquad w &= w &;&\, 0 > w > -R\,,  \nonumber  
\end{align}
when $y < 0$ with $\theta_{\rm min} > \frac{\pi}{2}$ (not shown in Fig.~\ref{fig:newmethod}). The radius of the arc of integration path $2$ is parameterized by $R$, which has to be taken to $\infty$ to obtain the original path. For $y>0$, the contribution of segment $2$ is
\begin{eqnarray}
\label{eq:arccontri}
iR\int_{\pi}^{\theta_{\rm max}}{\rm d} \theta\, {\rm e}^{i\theta}f\left(w = R {\rm e}^{i\theta}\right) = iR\int_{\pi}^{\theta_{\rm max}}{\rm d} \theta\, {\rm e}^{i\theta}\frac{1}{N}\,{\rm exp}\left[R{\rm e}^{i\theta}\right]\hat{\sigma}\left({\rm exp}\left[\frac{R {\rm e}^{i\theta}}{N}\right]\right).\,\,\,\,\,\,\,\,\,\,
\end{eqnarray}
The absolute upper bound of this integral may be estimated. By setting $h(R,\theta) \equiv \hat{\sigma}\left({\rm exp}\left[\frac{R {\rm e}^{i\theta}}{N}\right]\right)/N$ we write
\begin{eqnarray}
\left|iR\int_{\pi}^{\theta_{\rm max}}{\rm d} \theta\, {\rm e}^{i\theta}f\left(w = R {\rm e}^{i\theta}\right)\right| &\leq& R\int_{\pi}^{\theta_{\rm max}}{\rm d} \theta\, \left|{\rm e}^{i\theta}{\rm exp}\left[R{\rm e}^{i\theta}\right]h(R,\theta)\right|  \\
&\leq& R\int_{\pi}^{\theta_{\rm max}}{\rm d} \theta\, \left|{\rm e}^{R\cos\theta}\right|\,\left|h(R,\theta)\right|.  \nonumber 
\end{eqnarray}
\noindent Since $\theta \in \{\pi, \theta_{\rm max}\}$, where $\theta_{\rm max} < 3\pi/2$, the bounds on $\cos\theta$ are $-1\leq\cos\theta < 0$. Therefore, we arrive at 
\begin{eqnarray}
\left|iR\int_{\pi}^{\theta_{\rm max}}{\rm d} \theta\, {\rm e}^{i\theta}f\left(w = R {\rm e}^{i\theta}\right)\right| &\leq& R\,{\rm e}^{-\alpha R}\int_{\pi}^{\theta_{\rm max}}{\rm d} \theta \,\left|h(R,\theta)\right|,
\end{eqnarray}
with $0 < \alpha < 1$.
We assume that the function $h(R,\theta)$ is bounded, as $\hat{\sigma}$ represents a cross section. Hence, we observe that the integral of Eq.~\eqref{eq:arccontri} is exponentially suppressed. We consequently find that the contribution of segment $2$ vanishes for $R\rightarrow \infty$ for $\theta_{\max} < 3\pi/2$. Similar arguments hold for $y < 0$. \\
It follows that the contribution from the real axis is precisely opposite to that of the original integration path. Therefore, we can deform the contour away from segment $1$, and remove the oscillatory behavior that is present in the factor ${\rm e}^w$ of Eq.~\eqref{eq:cov} where $w$ is parameterized by $t\,{\rm e}^{i\theta_{\rm max}}$ (see Eq.~\eqref{eq:paramcov}). The new integration path is indicated by segment $3$ in Fig.~\ref{fig:newmethod}, where $w$ is now integrated from $-\infty$ to $0$. The resulting Mellin transform of the partonic cross section reads
\begin{eqnarray}
\hat{\sigma}_{ij}(N) = \int_{-\infty}^0 {\rm d}w\,  \frac{\exp[w]}{N}\, \hat{\sigma}_{ij}\left({\rm exp}\left[\frac{w}{N}\right]\right).
\end{eqnarray}
In this procedure, we have removed the increasingly large oscillatory behavior from the Mellin transform for ${\rm Re}[N]\rightarrow -\infty$ and ${\rm Im}[N]\rightarrow \pm\infty$, as now all $N$-dependence is captured in the variable $\rho = {\rm exp}\left[\frac{w}{N}\right]$, whose absolute value cannot grow bigger than $1$. Moreover, we have introduced an exponential damping via the factor ${\rm e}^w$ with $w \leq 0$, and introduced an extra suppression factor of $\frac{1}{N}$. \\
One important improvement over the traditional methods is that we can parameterize $N = C_{\rm MP} \pm iy$ with $y \geq 0$ to perform the inverse Mellin transform, as the tilted contour is not needed to obtain numerical convergence. Therefore, we do not have to worry about the validity of the Mellin transform outside the strip of definition, as this domain is never reached in the numerical integration. Secondly, we do not need to artificially introduce a cut-off on $x_1 x_2 \rho$ to stabilize the numerical integration. However, we should note that by deforming the contour, $w$ becomes real, but $\rho$ becomes complex. Therefore, this method can only be used if we have access to the analytical $\rho$-behavior of the matrix element. In particular, this method is not suited if the matrix element is obtained numerically. 
\section{Numerical results}
\label{sec:results}
\begin{table}[t]
\centering
\begin{tabular}{ |c| c| c| }
\hline
Distribution & Threshold definition & Scale choices ($\mu_F = \mu_R$)\\
\hline
$Q$ & $\rho_{\rm abs}$ (Eq.~\eqref{eq:absthres}), $\rho_{s_{t\bar{t}}}$ (Eq.~\eqref{eq:sttthres}), $\rho_{Q^2}$ (Eq.~\eqref{eq:rhoQ2}) &  $\mu_{\rm low}$, $\mu_{\rm high}$, $\mu_M$\\
$p_T$ & $x^2_{{\rm T},s_{t\bar{t}}}$ (Eq.~\eqref{eq:xstt}), $x_{{\rm T},4m_t^2}^2$ (Eq.~\eqref{eq:x4mt2}) &  $\mu_{\rm low}$, $\mu_{\rm high}$, $\mu_M$, $\mu_{H_T}$\\
\hline
\end{tabular}
\caption{Summary of threshold definitions and scale choices.}
\label{tab:summary}
\end{table}

In this section we explore the numerical results one obtains by setting different threshold variables as defined in Section~\ref{sec:subsecinv} and~\ref{sec:subsectrans}. As shown above, these variables result in different values for $M$ that parameterize the upper limit of Eq.~\eqref{eq:deltaiint} and Eq.~\eqref{eq:softint}, and their impact on the results via terms proportional to $\ln(M^2/\mu^2)$ in the resummed functions can be partially undone by choosing suitable values for the renormalization and factorization scales. The recommendations of Ref.~\cite{Denner:2047636} specify to use $\mu_R = \mu_F =\mu \equiv m_t + m_h/2$ for the computation of the total cross section. In Ref.~\cite{deFlorian:2016spz}, where a comparison of the $t\bar{t}h$ cross section is performed using different parton showers, a central scale choice is used of either $\mu = \left(m_{{\rm T},t}m_{{\rm T},\bar{t}}m_{{\rm T},h}\right)^{1/3}$ or $\mu = H_T/2$ with $H_T = m_{{\rm T},t}+ m_{{\rm T},\bar{t}}+ m_{{\rm T},h}$. Following these recommendations, and considering that we integrate over the phase space of the $t\bar{t}$-system, we examine the results for the following scale choices:
\begin{itemize}
    \item Fixed scale choice of $\mu_{\rm low} = (2m_t+m_h)/2$,
    \item Fixed scale choice of $\mu_{\rm high} = 2m_t+m_h$,
    \item Dynamical scale choice of $\mu_M = Q$ for the invariant-mass distribution, and $\mu_M = m_{{\rm T},4m_t^2}+m_{{\rm T},h}$ for the transverse-momentum distribution, 
    \item Dynamical scale choice of $\mu_{H_T} = \sqrt{m_{{\rm T},4m_t^2}m_{{\rm T},h}}$ (only used for the transverse-momentum distribution). 
\end{itemize}
The explicit scale logarithms in the NLL resummation function $g^{(2)}$ cancel if we use $\mu_M$ with $\rho_{Q^2}$ or $x_{{\rm T},4m_t^2}^2$, and if we use $\mu_{\rm high}$ with $\rho_{\rm abs}$. Since we probe the effect of different central scale choices, we refrain from varying $\mu_R$ and $\mu_F$ independently. \\
\noindent In our presentation of the numerical LHC results, we use the fitted form (Eq.~\eqref{eq:fit}) of the central member of the NNLO PDF4LHC15 PDF set~\cite{Butterworth:2015oua}, and take the CM energy equal to 13~TeV. We use \texttt{MadGraph5\_aMC@NLO}~\cite{Alwall:2014hca} to obtain the NLO fixed-order result. Before we discuss the matched result, we first compare the numerical methods introduced in the previous section for the unmatched total NLL resummed cross section with $\rho = \rho_{s_{t\bar{t}}}$. 

\subsection{Comparison of numerical methods}

\begin{figure}[t]
\centering
\mbox{
\begin{subfigure}{0.5\textwidth}\includegraphics[width=\textwidth]{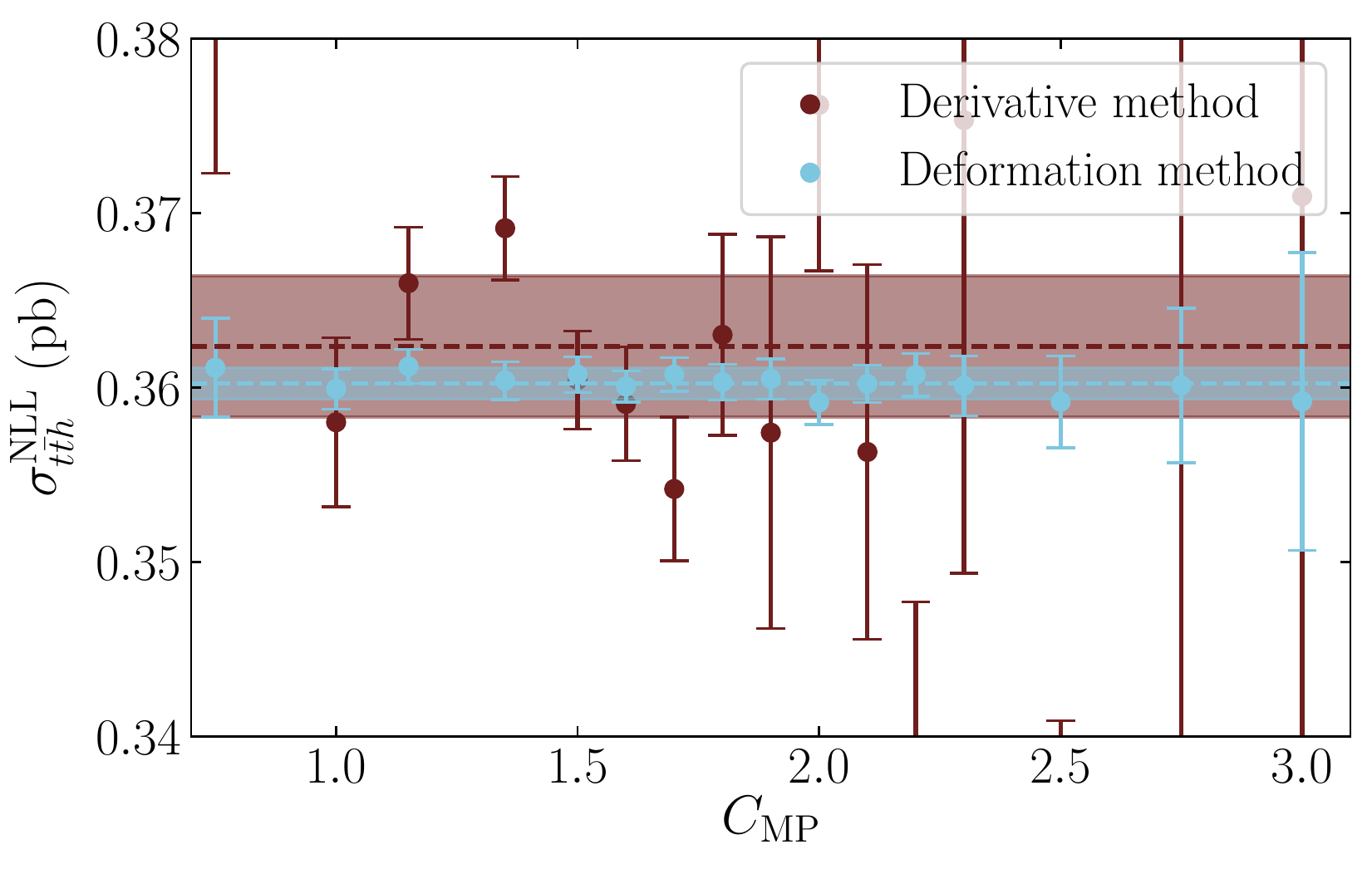}
\end{subfigure}
\begin{subfigure}{0.51\textwidth}\includegraphics[width=\textwidth]{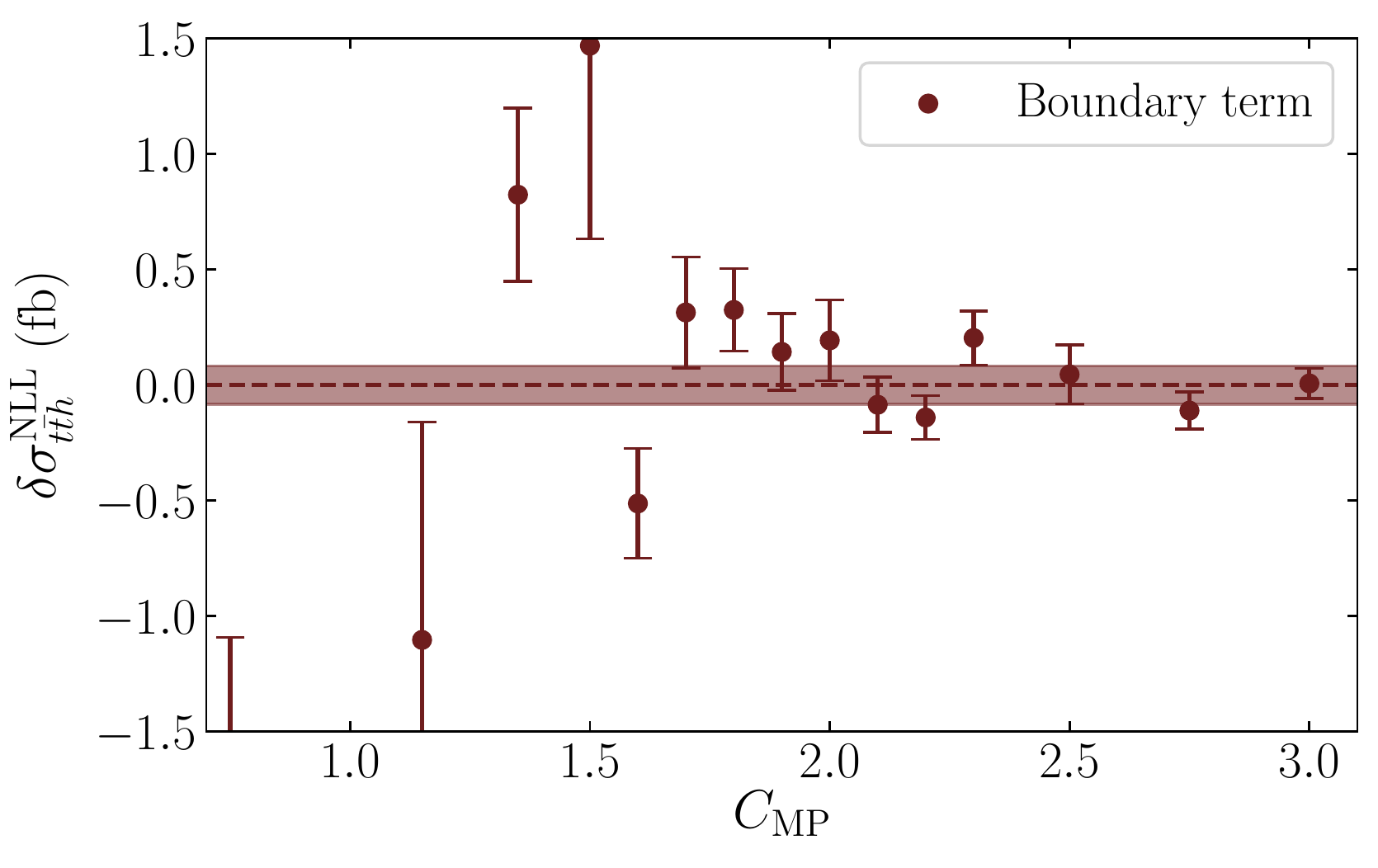}
\end{subfigure}}
  \caption{Left: The NLL cross section with $\rho = \rho_{s_{t\bar{t}}}$ and $\mu = \mu_{\rm low}$ for different values of $C_{\rm MP}$ and the numerical error bars, computed using the derivative (red) and the deformation (blue) methods. The dashed line indicates the average cross section obtained via Eq.~\eqref{eq:weightedav}, and the band the weighted numerical error. Right: Contribution of the boundary terms of the derivative method for different values of $C_{\rm MP}$. The dashed red line indicates the average boundary contribution, and the band the error computed via Eq.~\eqref{eq:weightedav}.}
\label{fig:ttHboundary}
\end{figure}

\label{app:comparison}
\noindent Here we compare the deformation method (Section~\ref{sec:deformation}) and the derivative method (Appendix~\ref{app:deriv}). We use the unmatched resummed NLL cross section for this comparison, set $\rho = \rho_{s_{t\bar{t}}}$ and $\mu = \mu_{\rm low}$. As covered in Appendix~\ref{app:deriv}, the derivative method may be used to obtain a suppression factor of $1/N^4$, which helps the inverse-Mellin transform to converge numerically. Three values of $\phi_{\rm MP}$ are probed ($\pi/2$, $5\pi/8$ and $3\pi/4$) for a range of $C_{\rm MP}$ values between $0.75$ and $3.0$. For the deformation method we fix $\phi_{\rm MP} = \pi/2$, and have the same range of $C_{\rm MP}$ values. The resulting cross sections are shown in Fig.~\ref{fig:ttHboundary}, where for the derivative method we only show the result for each $C_{\rm MP}$ value with the minimal error obtained after varying over $\phi_{\rm MP}$. One may immediately observe that the deformation method leads to a more stable result over a wider range of $C_{\rm MP}$ values. \\
We also compute a weighted average and weighted error via the standard method
\begin{eqnarray}
\label{eq:weightedav}
    \sigma_{pp\rightarrow t\bar{t}h}^{\text{average}} \,=\, \frac{\mathlarger{\sum}_{i}\,\, {\sigma_{pp\rightarrow t\bar{t}h, i}/\left(\sigma_{pp\rightarrow t\bar{t}h,i}^{{\text{error}}}\right)^2}}{\mathlarger{\sum}_{i}\,\, 1/\left(\sigma_{pp\rightarrow t\bar{t}h,i}^{{\text{error}}}\right)^2}\,,  \quad\,\,\,\, \sigma_{pp\rightarrow t\bar{t}h}^{\text{average error}} \,=\, \frac{\mathlarger{\sum}_{i}\,\, 1/\sigma_{pp\rightarrow t\bar{t}h,i}^{{\text{error}}}}{\mathlarger{\sum}_{i}\,\, 1/\left(\sigma_{pp\rightarrow t\bar{t}h,i}^{{\text{error}}}\right)^2}\,,\,\,\,\,\,
\end{eqnarray}
where the sum runs over $i$ values of $C_{\rm MP}$. The average cross sections in Eq.~\eqref{eq:weightedav} are constructed in such a way that values with lower numerical errors are favored. The resulting average cross section using the derivative method is $\sigma_{pp\rightarrow t\bar{t}h}^{\rm deriv.} =  0.36089 \pm 0.00372$~pb. That for the deformation method is $\sigma_{pp\rightarrow t\bar{t}h}^{\rm deform.} =  0.36015 \pm 0.00071$~pb, which has a similar average value as the former method, but the numerical error is smaller by roughly a factor of $5$. \\
For the derivative method, the lower bounds on $x_1$, $x_2$ and $\rho$ need to be chosen such that $x_1 x_2 \rho / \tau > 1$, otherwise the numerical integration is not stable. In Appendix~\ref{app:deriv} we expressed a worry that such a lower bound may create a boundary contribution. This contribution can be computed explicitly, as we know it must match the contribution in the domain where $x_1 x_2 \rho / \tau < 1$. To compute it, we use values of $\phi_{\rm MP} \leq \pi/2$, as for $\phi_{\rm MP} > \pi/2$ the integral does not converge numerically due to the $x_1 x_2 \rho / \tau < 1$ requirement. The result can be observed on the right-hand side of Fig.~\ref{fig:ttHboundary}, where one sees that the contribution of the boundary term is negligible, as the average result is consistent with $0$, hence we can safely discard it.

\begin{table}[t]
\centering
\begin{tabular}{|c|c|c|} 
 \hline
 Method & Result (pb) & Function evaluations\\
 \hline
 Derivative method (average) & $0.36237 \pm 0.00405$ & $1.5 \cdot 10^6$ \\
 Deformation method (average) & $0.36024\pm 0.00090$ & $2.0 \cdot 10^5$ \\
 \hline
 Derivative method (best) & $0.36044 \pm 0.00281$ & $1.2 \cdot 10^6$ \\
 Deformation method (best) & $0.36075\pm 0.00071$ & $6.2 \cdot 10^4$ \\
 \hline
\end{tabular}
\caption{Results and number of function evaluations for the NLL $t\bar{t}h$ production cross section using the deformation or derivative numerical integration method. The top two rows indicate the average results, and the bottom two the best results.}
\label{table:xsecs}
\end{table}

\noindent We summarize the results for the computation of the NLL $t\bar{t}h$ cross section in Table~\ref{table:xsecs}, where we also indicate the number of function evaluations needed to obtain the indicated result and precision. For the average results, we quote the average number of function evaluations. We see that the deformation method leads to an error that is a factor of $4-5$ smaller when compared to the derivative method, while it needs at least a factor of $10$ less function evaluations. \\
We conclude that the deformation method leads to a substantially more stable result than the derivative method, and does so by needing less function evaluations. Therefore, in the presentation of our results below, we employ the deformation method to evaluate all threshold variables, except for $\rho_{Q^2}$, as this choice directly eliminates the Mellin-space integral over the partonic threshold variable by virtue of the $\delta$-function. Note that not only the threshold variable becomes complex for $\rho_{\rm abs}$ and $x^2_{{\rm T},4m_t^2}$ when using the deformation method: as these variables feature in the upper limit of the $s_{t\bar{t}}$ integration, the $s_{t\bar{t}}$ variable becomes complex too. We compute each matched result for the different values of $C_{\rm MP}$ as indicated in Fig.~\ref{fig:ttHboundary}, and use the average as obtained via Eq.~\eqref{eq:weightedav} as our final result. 

\subsection{The invariant-mass distribution}
\begin{figure}[th!]
\centering
\vspace{-1.5cm}

\begin{subfigure}{0.5\textwidth}\includegraphics[width=\textwidth]{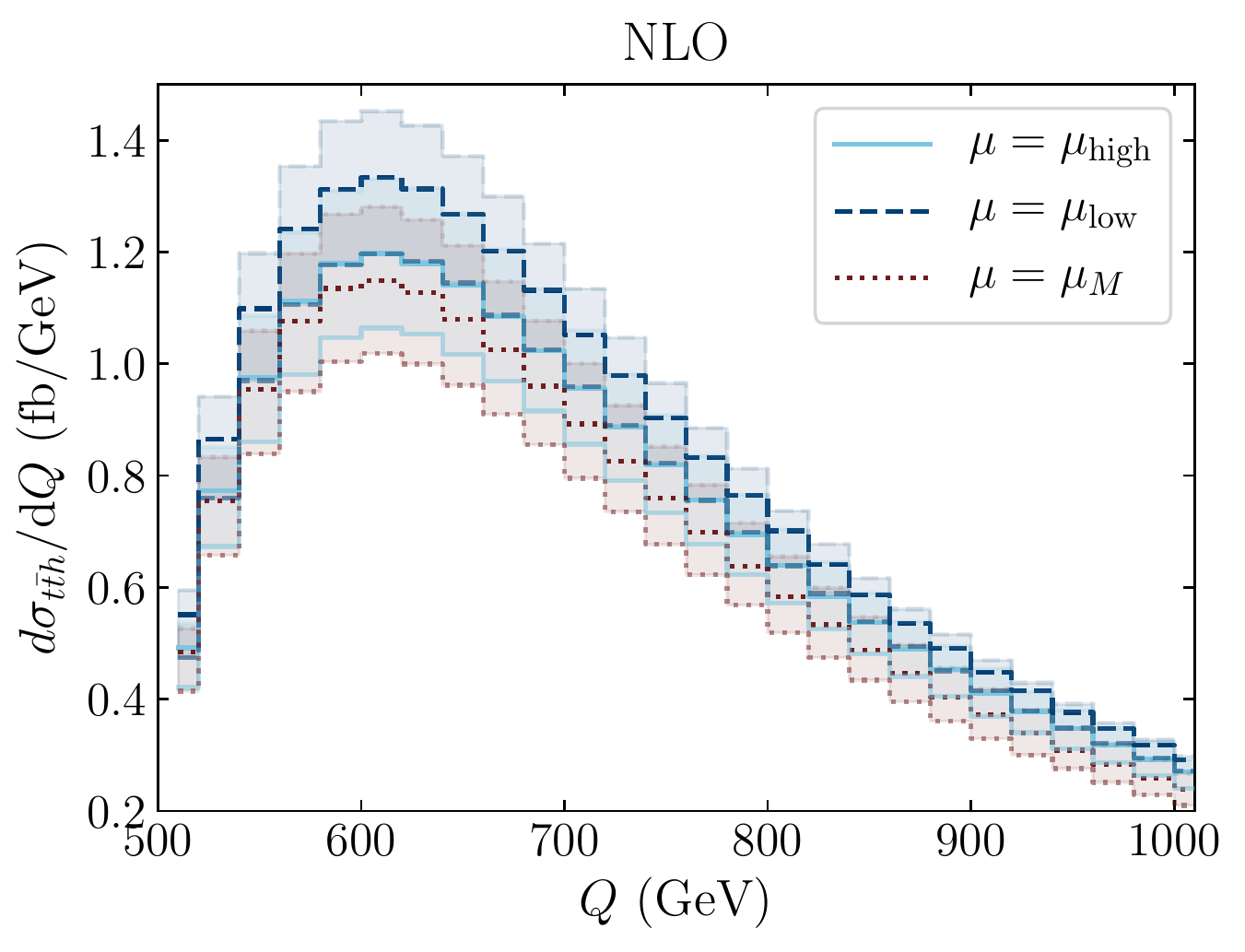}\end{subfigure}
\mbox{\begin{subfigure}{0.5\textwidth}\includegraphics[width=\textwidth]{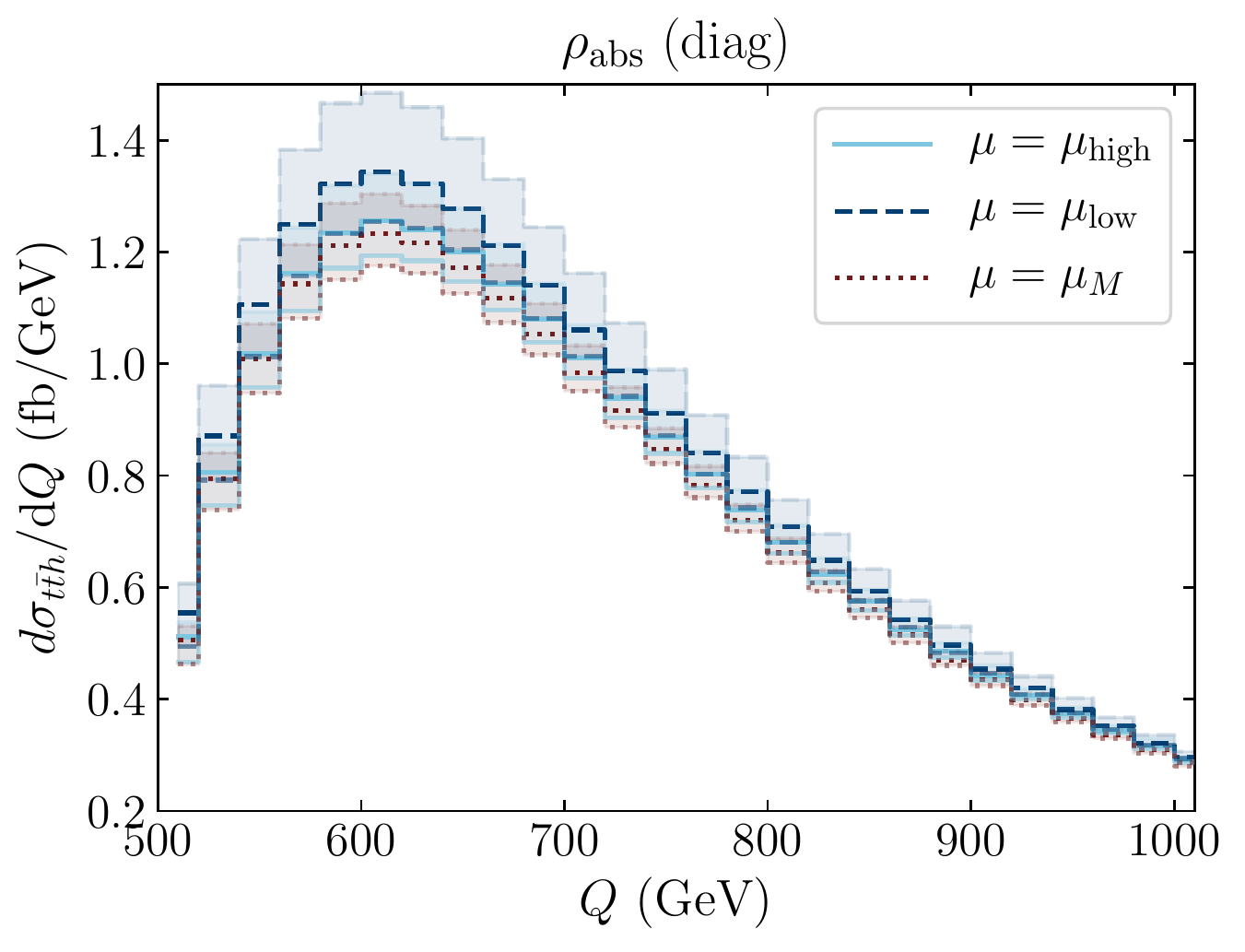}\end{subfigure}
\begin{subfigure}{0.5\textwidth}\includegraphics[width=\textwidth]{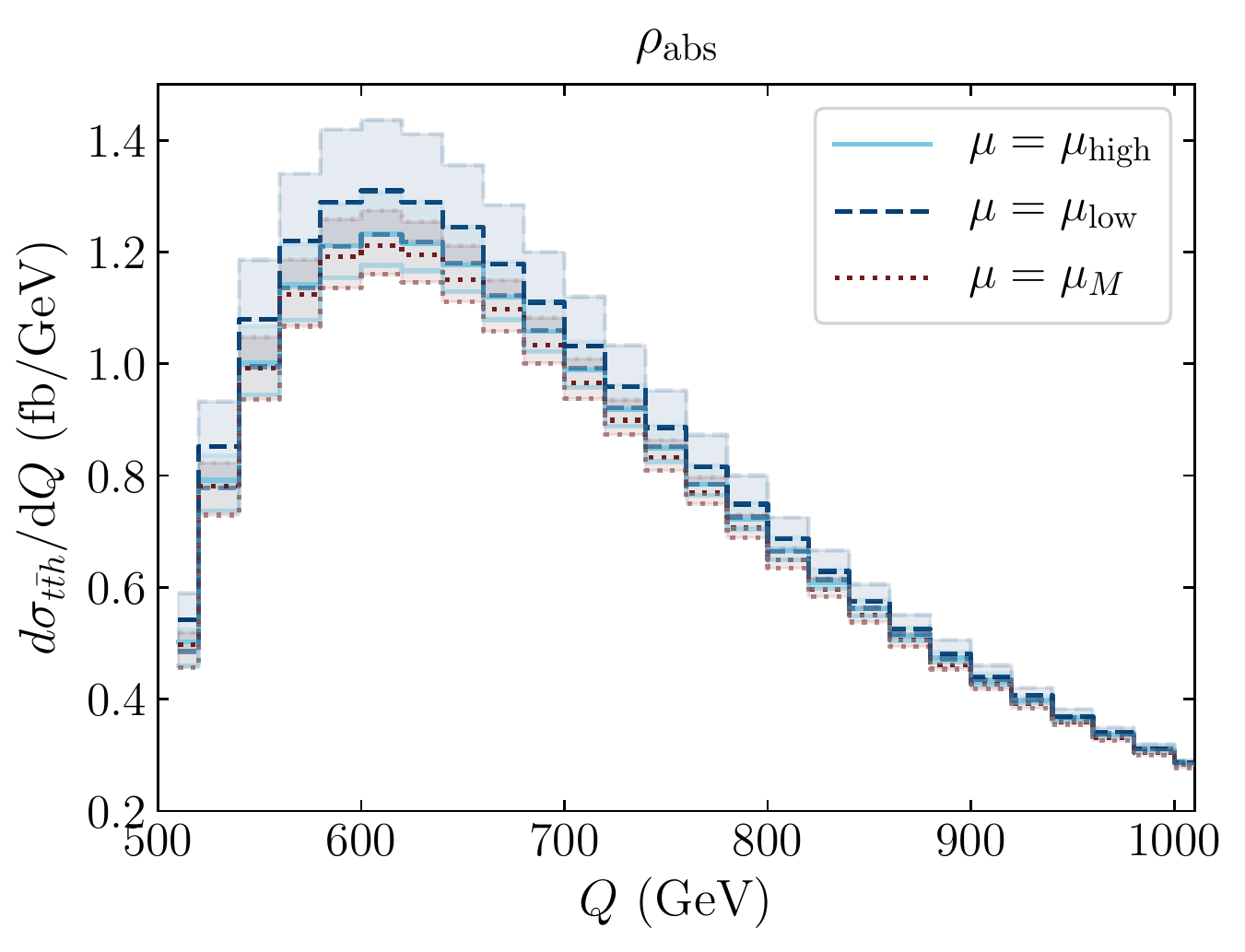}\end{subfigure}
}
\mbox{\begin{subfigure}{0.5\textwidth}\includegraphics[width=\textwidth]{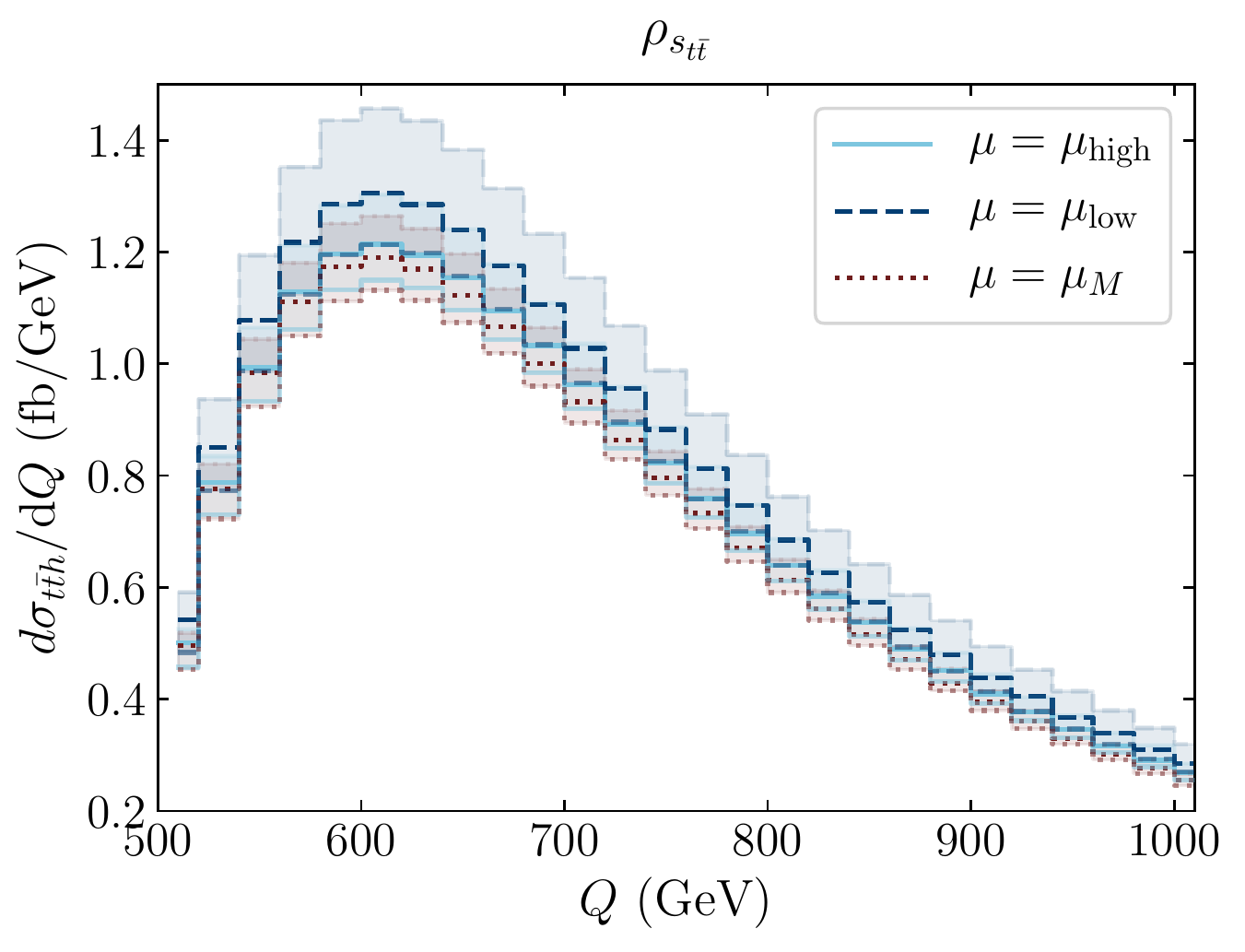}\end{subfigure}
\begin{subfigure}{0.5\textwidth}\includegraphics[width=\textwidth]{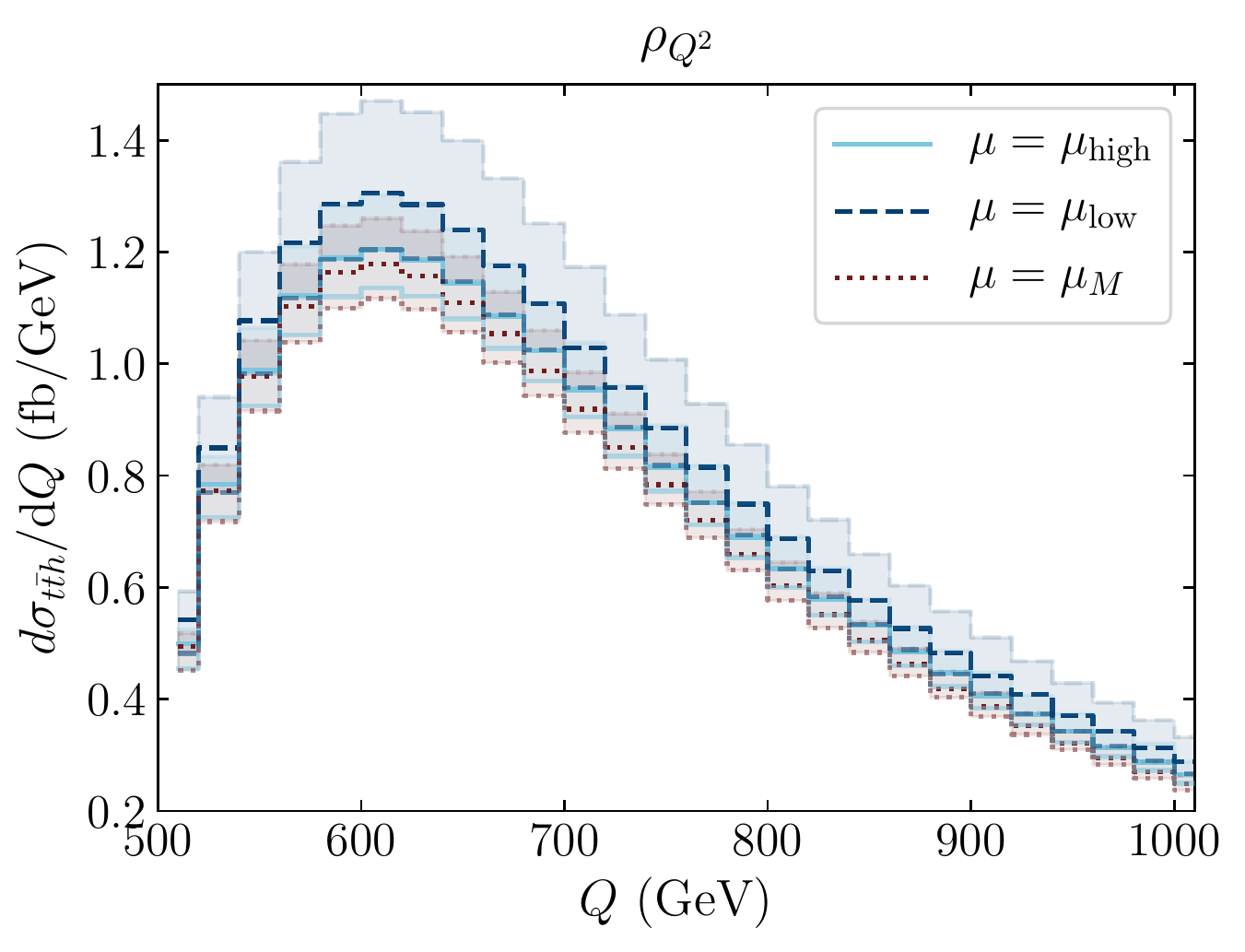}\end{subfigure}
}
\caption{Invariant-mass distributions for the scale choices $\mu = \mu_{\rm high}$ (light-blue solid), $\mu = \mu_{\rm low}$ (dark-blue dashed), and $\mu = \mu_{M} = Q$ (red dotted). Shown are the NLO result (upper panel), and the matched resummed result using $\rho_{\rm abs}$ with/without approximating the soft-anomalous dimension matrix (middle left/right), $\rho_{s_{t\bar{t}}}$ (lower left) and $\rho_{Q^2}$ (lower right). The colored bands indicate the scale uncertainty within each scale choice as obtained by varying $\mu$ between $\mu_{\rm ref}/2$ and $2\mu_{\rm ref}$, with $\mu_{\rm ref}$ the reference factorization/renormalization scale.}
\label{fig:Qinv}
\end{figure}
\begin{figure}[t]
\centering
\mbox{\begin{subfigure}{0.5\textwidth}\includegraphics[width=\textwidth]{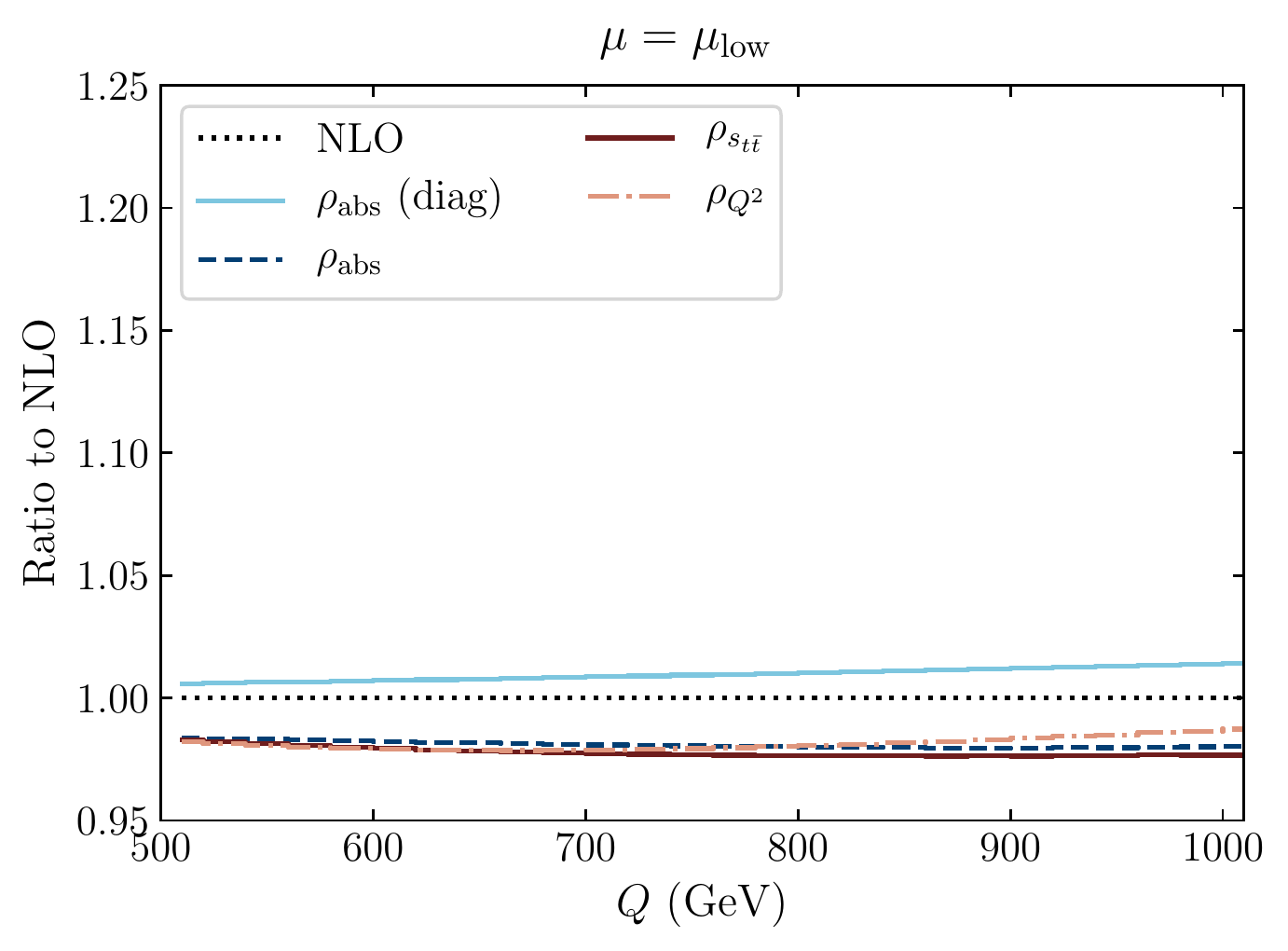}\end{subfigure}
\begin{subfigure}{0.5\textwidth}\includegraphics[width=\textwidth]{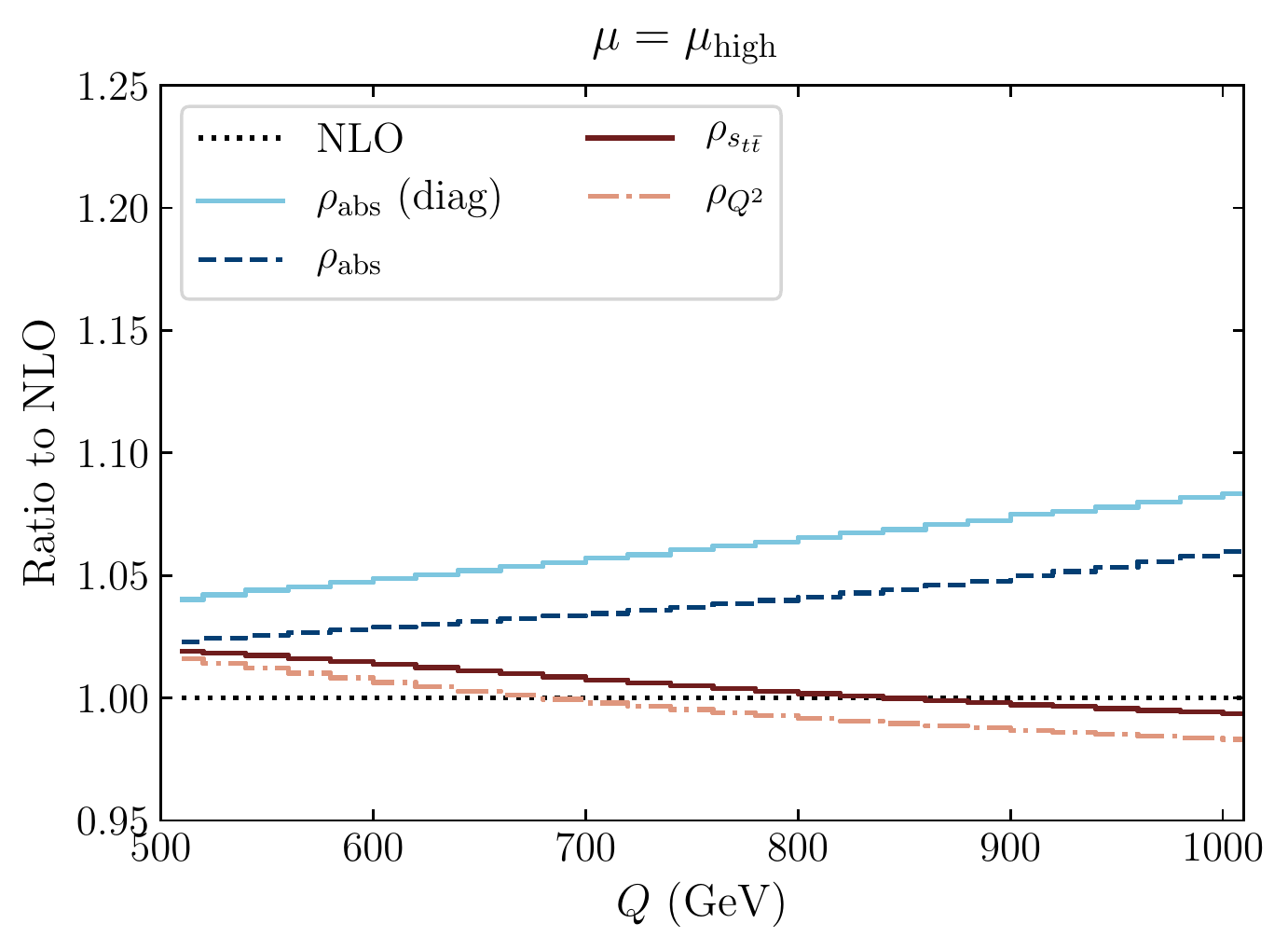}\end{subfigure}
}
\begin{subfigure}{0.5\textwidth}\includegraphics[width=\textwidth]{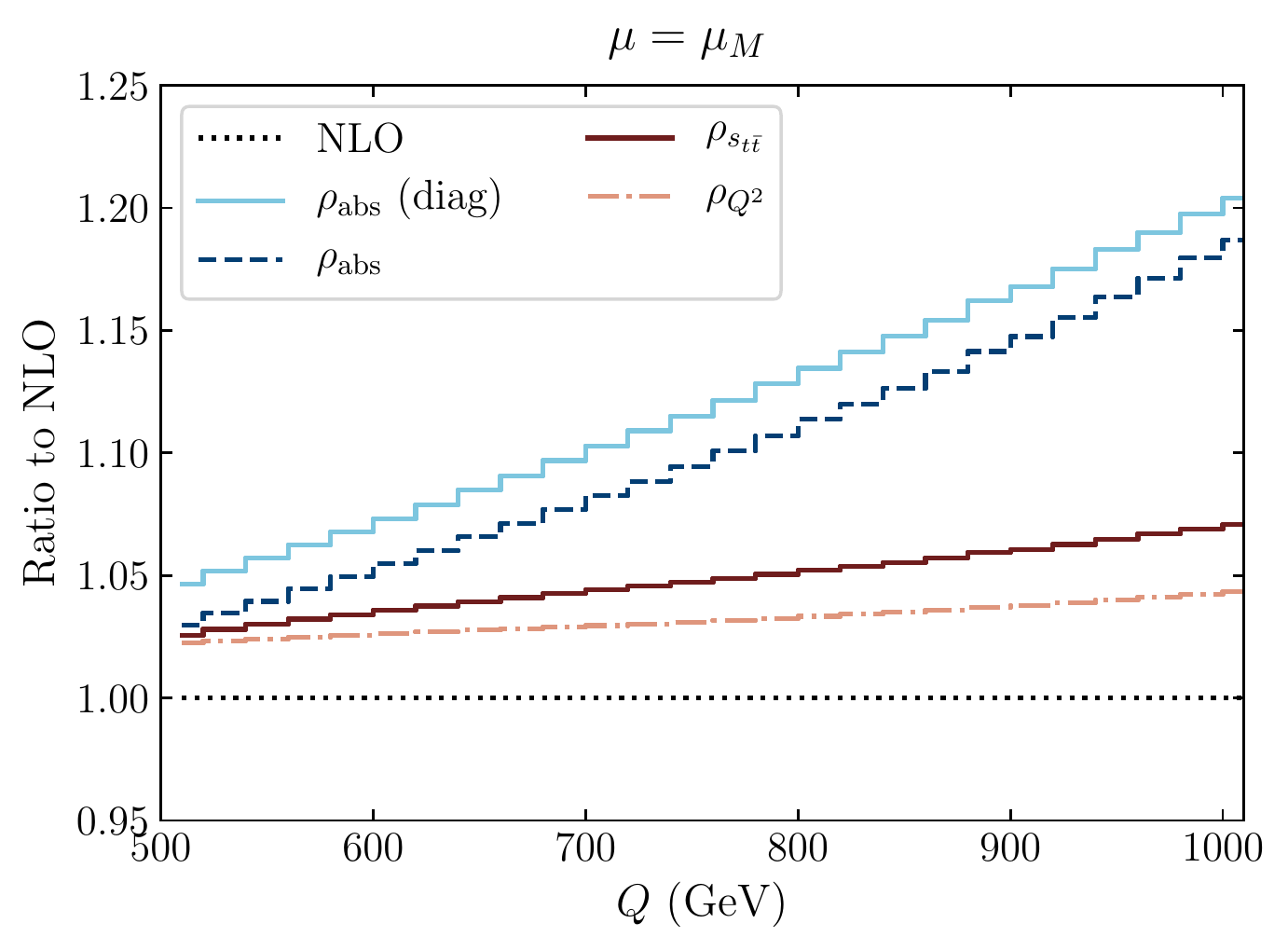}\end{subfigure}
\caption{Ratio of the various resummed invariant-mass distributions to the NLO one for the scale choices $\mu = \mu_{\rm low}$ (upper left), $\mu = \mu_{\rm high}$ (upper right), and $\mu = \mu_{M} = Q$ (below). The ratios obtained using the threshold variable $\rho_{\rm abs}$, $\rho_{s_{t\bar{t}}}$ and $\rho_{Q^2}$ are indicated by the dashed dark-blue, solid red, and dash-dotted orange lines respectively. The result that is obtained using the approximated soft-anomalous dimension matrix is indicated by the solid light-blue line.}
\label{fig:Qinvscales}
\end{figure}

We begin by discussing the invariant-mass distributions  (Fig.~\ref{fig:Qinv}). One may observe that the scale uncertainty of the NLO cross section is large. At the peak of the distribution around $Q = 610$~GeV, values for the invariant mass distribution between $1.02-1.45$~fb/GeV are found. The smallest scale variations of the NLO distribution are obtained for the choice $\mu = \mu_{\rm low}$, with ${\rm d}\sigma_{t\bar{t}h}/{\rm d}Q = 1.33^{+8.9\%}_{-10.2\%}$~fb/GeV at $Q = 610~$GeV. The resummed distributions show better behavior under scale variations. The smallest scale variations at $Q=610$~GeV are obtained by setting $\rho = \rho_{\rm abs}$ and using $\mu = \mu_M$, where we find ${\rm d}\sigma_{t\bar{t}h}/{\rm d}Q = 1.21^{+5.2\%}_{-4.2\%}$~fb/GeV. The central values of the other threshold and scale choices (excluding the one where we approximate the soft-anomalous dimension matrices) lie between ${\rm d}\sigma_{t\bar{t}h}/{\rm d}Q = 1.18-1.31$~fb/GeV, where higher values are typically found for $\mu = \mu_{\rm low}$.\\
\noindent It is remarkable that different threshold choices affect the resulting scale uncertainty. The matched distribution obtained with $\rho = \rho_{\rm abs}$ is most robust under scale variations, while that obtained with $\rho = \rho_{Q^2}$ shows the largest dependence on the scale. Especially in the tail of the distribution this effect is visible: for $\rho = \rho_{\rm abs}$, the scale uncertainty nearly vanishes at $Q = 1$~TeV. \\
\noindent By comparing the two plots in the middle panel of Fig.~\ref{fig:Qinv}, one may directly observe a difference between the distribution obtained using an approximated diagonal form of the soft-anomalous dimension matrices (labeled with $\rho_{\rm abs}$ (diag)) and that without approximating these matrices (labeled with $\rho_{\rm abs}$). Although the difference lies within the scale uncertainty of the results, we feel that it is worth to investigate this further, which we do in Section~\ref{sec:nlp}. \\
In Fig.~\ref{fig:Qinvscales} we show the ratio plots with respect to the NLO results for the various threshold definitions, where we see that the $\mathcal{O}(1/N)$ terms of the soft anomalous dimension matrices have a noticeable numerical impact, and this is visible across all three scale choices. At the scale choice of $\mu = \mu_{\rm low}$, which possesses the smallest scale variations for the NLO distribution, we see that the numerical difference across the different threshold variables is very small (although increasing slightly for large values of $Q$). For all non-approximated definitions we find a decrease of the NLO central contribution for $\mu = \mu_{\rm low}$ between $-1.5\%$ and  $-2.5\%$. \\
Setting $\mu = \mu_{\rm high}$ (upper-right panel of Fig.~\ref{fig:Qinvscales}), we find a positive correction at low values of $Q$ of roughly $+2.0\%$ for all threshold definitions. However, the correction varies widely at higher $Q$ values. The distributions obtained by setting $\rho_{s_{t\bar{t}}}$ and $\rho_{Q^2}$ give a negative correction of $-0.5\%$ and $-1.6\%$ respectively to the NLO distribution at $Q= 1$~TeV, whereas the one with $\rho_{\rm abs}$ gives a positive correction of $+5.8\%$. This may be understood as follows: by setting $\mu_{\rm high}$ and $\rho_{\rm abs}$, one cancels the scale logarithm $\ln\left(\frac{M^2}{\mu^2}\right)$ in the resummed expression (see Eq.~\eqref{eq:g2} in Appendix~\ref{app:definitions}), whose contribution is negative for $M^2 > \mu^2$ because of the negative prefactor $\ln(1-2\lambda)$ for $0 < 2\lambda < 1$. For the other two threshold definitions and with $\mu = \mu_{\rm high}$, this scale logarithm grows in size with higher values of $Q$, as then both $M = Q$ and $M = \sqrt{s_{t\bar{t}}+m_h^2}$ get larger. The numerical value of the scale logarithm obtained with $\rho = \rho_{Q^2}$ is higher than that with $\rho = \rho_{s_{t\bar{t}}}$, therefore the deviation from the $\rho=\rho_{\rm abs}$ result is largest in the former case. A similar effect is observed for $\mu= \mu_M$: there, the distribution with $\rho_{Q^2}$ shows a $+4.3\%$ correction at $Q=1$~TeV. The distribution with $\rho_{s_{t\bar{t}}}$ lies above that with a $+6.9\%$ correction, and that obtained with $\rho_{\rm abs}$ gives a very large $+17.9\%$ correction to the NLO distribution.  \\

\vspace{-1em}

\begin{figure}[t]
\centering
\mbox{
\begin{subfigure}{0.5\textwidth}\includegraphics[width=\textwidth]{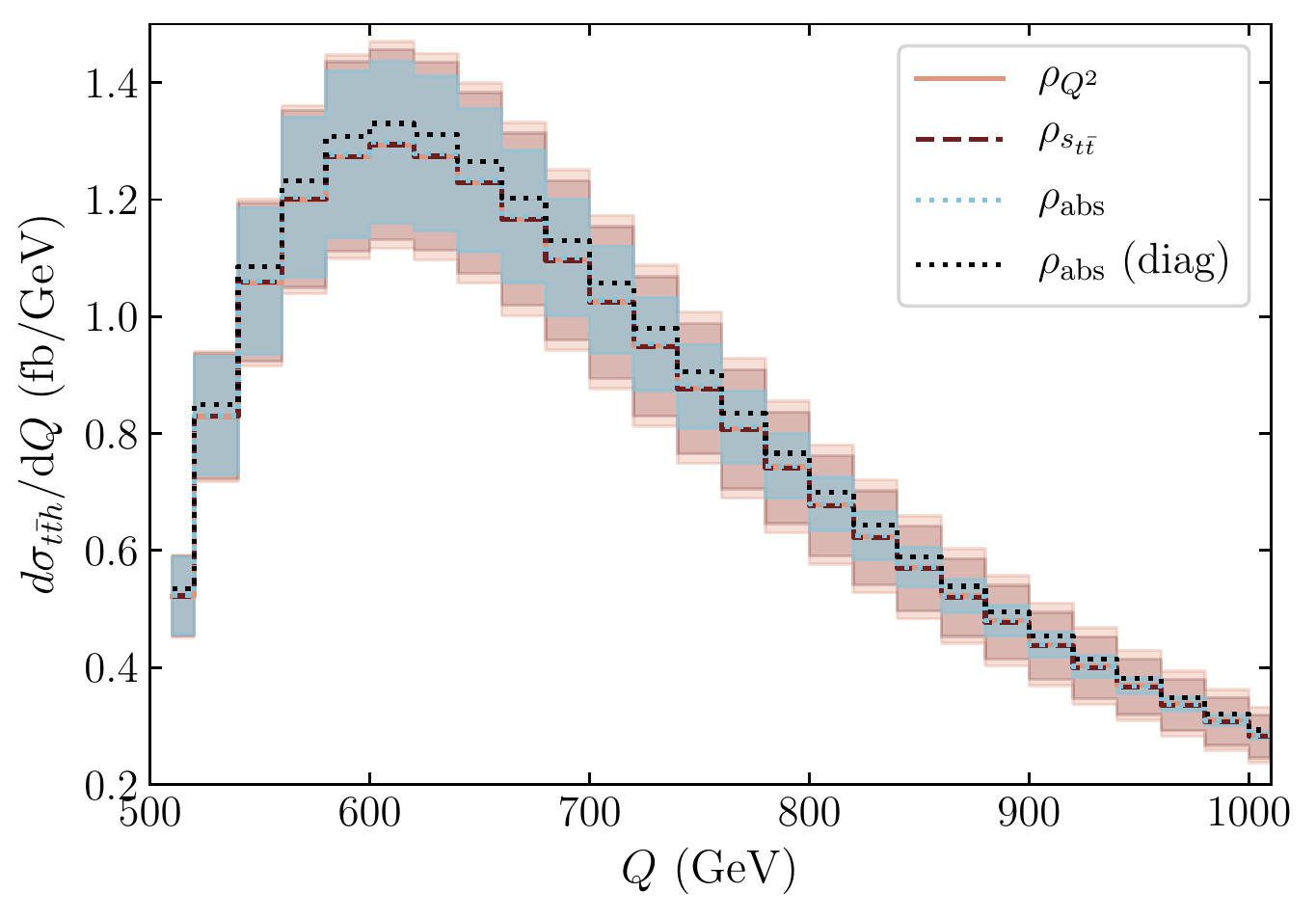}\end{subfigure}
\begin{subfigure}{0.5\textwidth}\includegraphics[width=\textwidth]{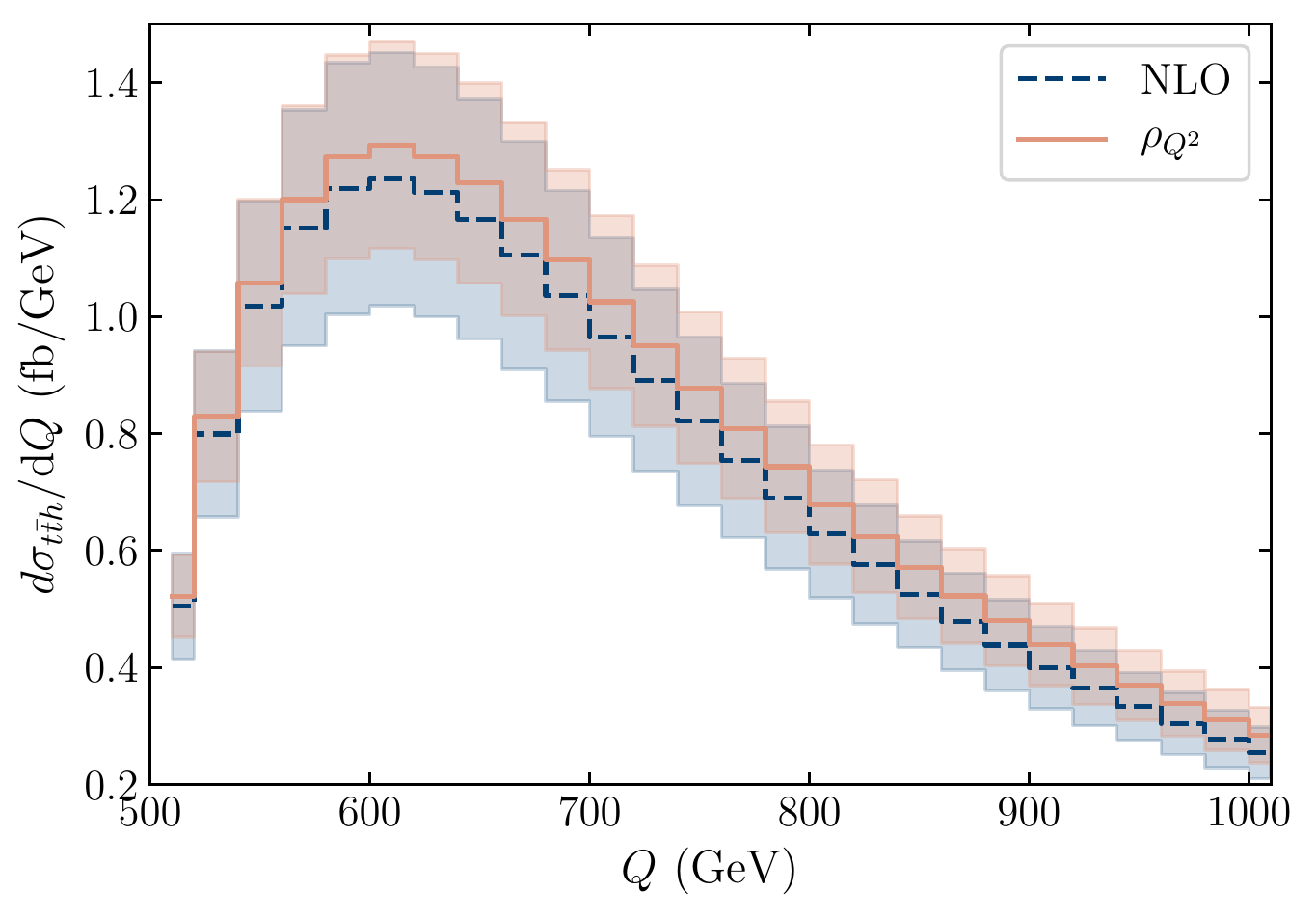}\end{subfigure}}
\caption{Right: Invariant-mass distribution of the matched resummed results using $\rho_{\rm abs}$ with/without the approximated soft-anomalous dimension matrix (black and light-blue dotted), $\rho_{s_{t\bar{t}}}$ (red dashed) and $\rho_{Q^2}$ (orange solid). The colored bands indicate the total scale uncertainty band, obtained by setting $\mu = \mu_Q$, $\mu_{\rm high}$ and $\mu_{\rm low}$, varying these between $\mu_{\rm ref}/2$ and $2\mu_{\rm ref}$, with $\mu_{\rm ref}$ the reference factorization/renormalization scale, and picking the minimum and maximum value for each value of $Q$ (Eq.~\eqref{eq:thisishowwedoit}). The central value is obtained by averaging these minimum and maximum values. The scale uncertainty band for $\rho_{\rm abs}$ (diag) is not shown. Right: Invariant-mass distribution of the NLO result (dark-blue dashed) and the matched resummed result obtained using $\rho_{Q^2}$. }
\label{fig:Qinv_complete}
\end{figure}

\noindent In contrast to what the plots in Fig.~\ref{fig:Qinvscales} may suggest, the numerical difference of using different threshold variables is actually small at the peak of the distribution: we obtain a central value of ${\rm d}\sigma_{t\bar{t}h}/{\rm d}Q = 1.18-1.23$~fb/GeV for all three definitions at $\mu = \mu_M$ and  $\mu = \mu_{\rm high}$, and around ${\rm d}\sigma_{t\bar{t}h}/{\rm d}Q = 1.31$~fb/GeV for $\mu = \mu_{\rm low}$. We therefore see that the scale choice is of bigger importance than the threshold choice. Slightly larger discrepancies between using different threshold variables are found in the tail of the distributions. Especially the difference between using $\rho_{\rm abs}$ and $\rho_{s_{t\bar{t}}}$ or $\rho_{Q^2}$ grows for the scale choices $\mu = \mu_{\rm high}$ and $\mu_{\mu_M}$: at $Q = 910$~GeV we find that ${\rm d}\sigma_{t\bar{t}h}/{\rm d}Q \simeq 0.39$~fb/GeV for $\rho_{s_{t\bar{t}}}$ and $\rho_{Q^2}$, while ${\rm d}\sigma_{t\bar{t}h}/{\rm d}Q \simeq 0.44$~fb/GeV for $\rho_{\rm abs}$. Interestingly, we do obtain a consistent result between all three threshold definitions at $\mu = \mu_{\rm low}$ with ${\rm d}\sigma_{t\bar{t}h}/{\rm d}Q \simeq 0.44$~fb/GeV.  \\
In Fig.~\ref{fig:Qinv_complete}, we show the invariant mass distributions for all threshold parameterizations with their complete scale uncertainty. That is, for each $Q$ value we have the minimum (min) and maximum (max) values
\begin{eqnarray}
\label{eq:thisishowwedoit}
\frac{{\rm d} \sigma_{t\bar{t}h}}{{\rm d}Q}\Bigg|_{\rm min/max} = {\rm min/max}\left[\sum_{k = 1/2,1,2}\,\,\,\sum_{\mu' = \mu_Q,\mu_{\rm high},\mu_{\rm low}}\frac{{\rm d} \sigma_{t\bar{t}h}}{{\rm d}Q}\Bigg|_{\mu = k\mu'}\right],
\end{eqnarray}
while the central value is obtained by simply averaging the maximum and minimum values. On the left-hand side of Fig.~\ref{fig:Qinv_complete} we observe that a near-perfect agreement is found on the central value between choosing different threshold parameterizations. The scale uncertainties do differ, and the smallest scale uncertainty is found for $\rho = \rho_{\rm abs}$. By approximating the soft-anomalous dimension matrices we do find a noticeably different result of around $+3\%$ for all $Q$ values. However, this difference lies well within the scale uncertainty of the results obtained without this approximation. On the right-hand side of Fig.~\ref{fig:Qinv_complete}, one sees that the correction from NLL resummation obtained with respect to the averaged NLO distribution is around $+4.7\%$ at the peak of the distribution, which grows to about $+12.1\%$ in the tail of the distribution. In Section~\ref{sec:nlp} we will further discuss the role of $\mathcal{O}(1/N)$ corrections, but first we turn our attention to the transverse-momentum distributions. \\

\subsection{The transverse-momentum distribution}

\begin{figure}[th!]
\centering
\mbox{\begin{subfigure}{0.5\textwidth}\includegraphics[width=\textwidth]{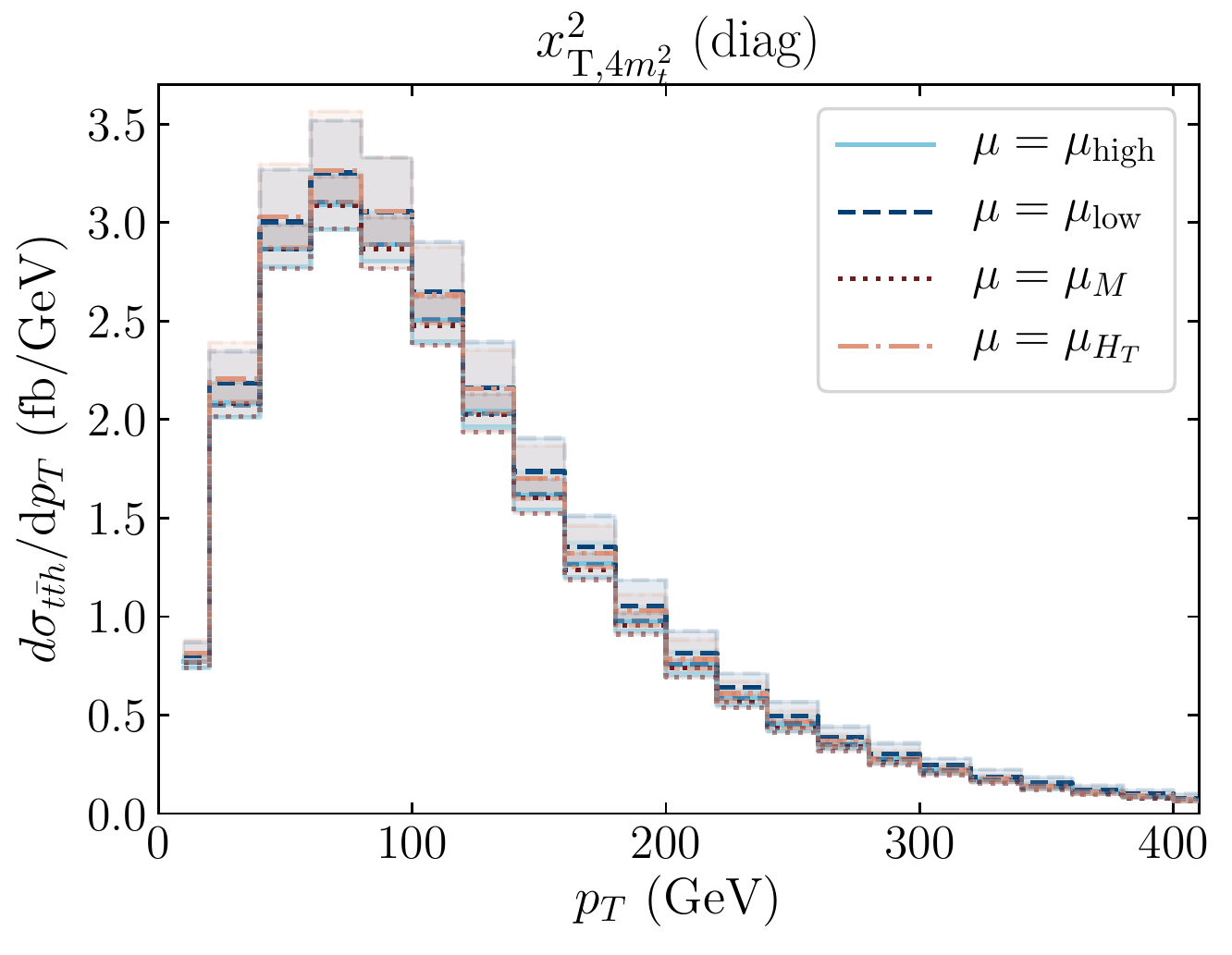}\end{subfigure}
\begin{subfigure}{0.5\textwidth}\includegraphics[width=\textwidth]{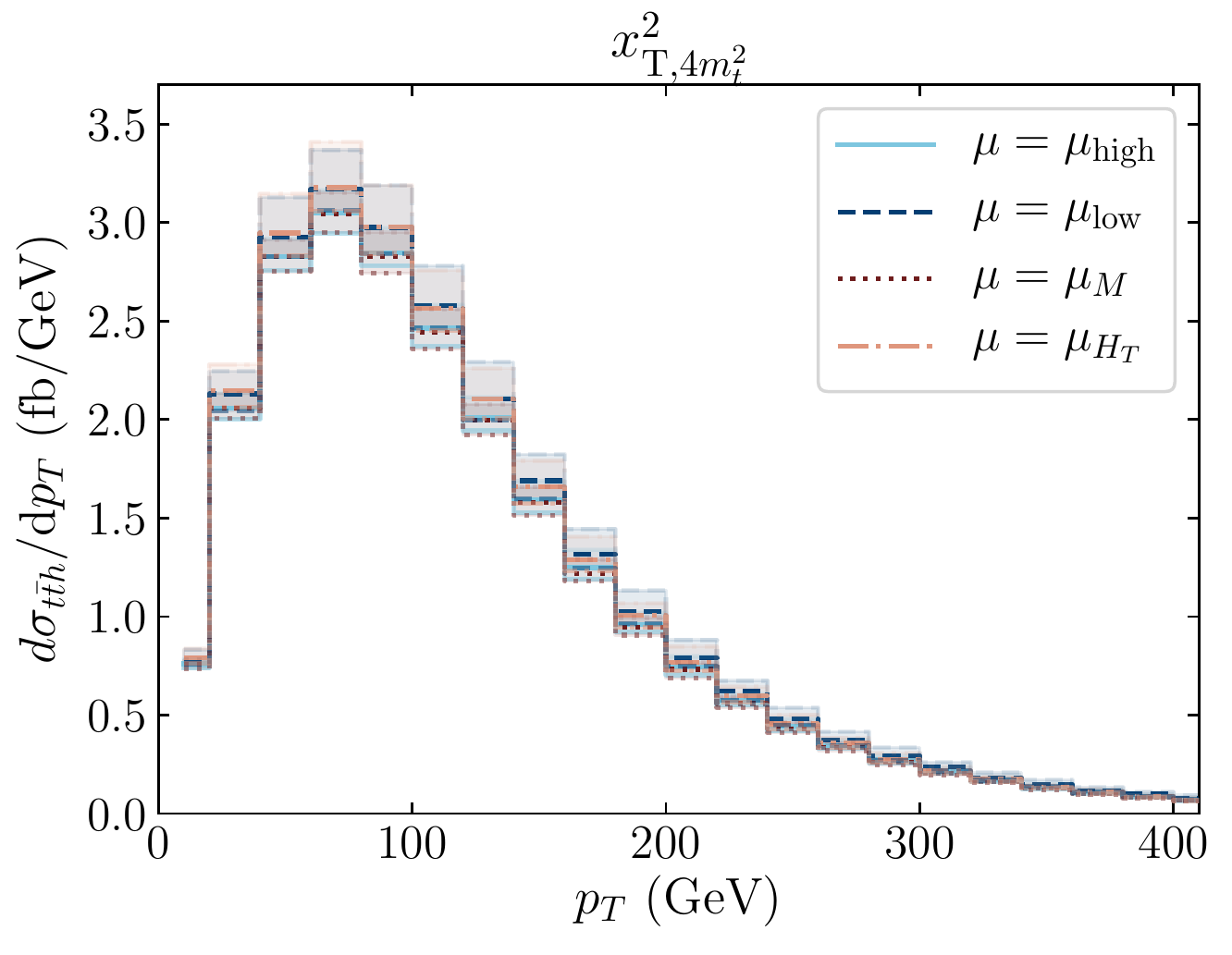}\end{subfigure}}
\mbox{\begin{subfigure}{0.5\textwidth}\includegraphics[width=\textwidth]{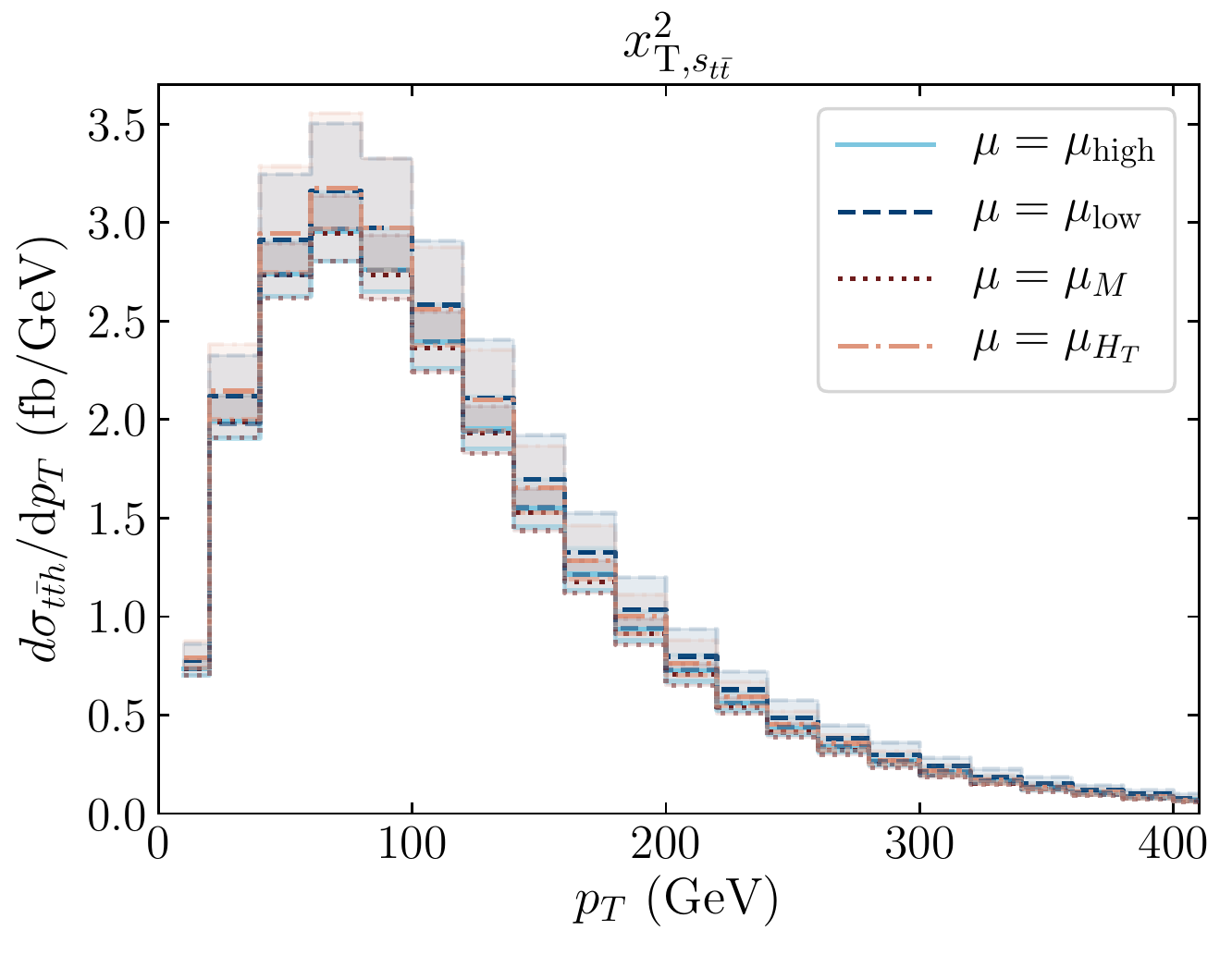}\end{subfigure}
\begin{subfigure}{0.5\textwidth}\includegraphics[width=\textwidth]{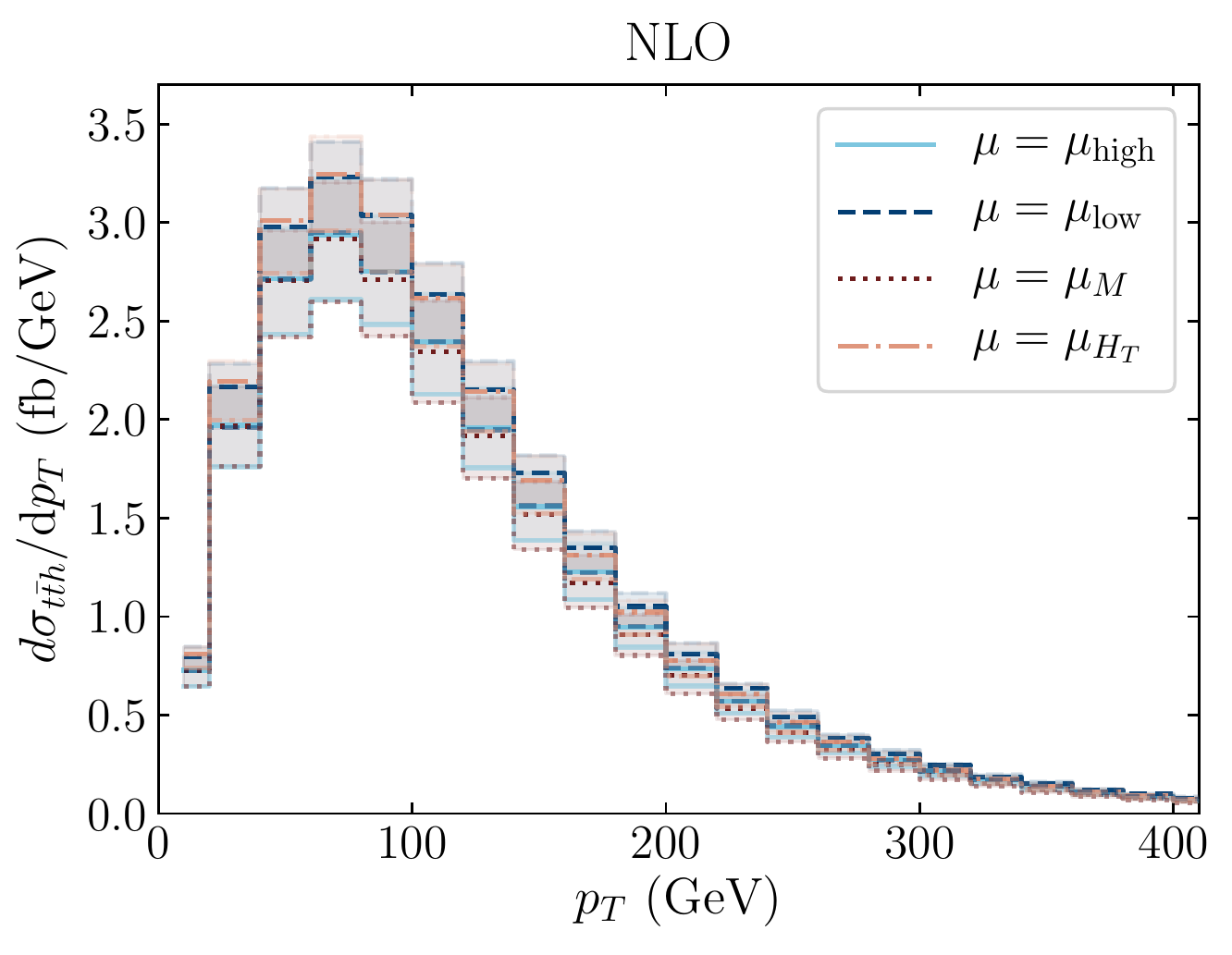}\end{subfigure}
}
\caption{Transverse-momentum distributions for the scale choices $\mu = \mu_{\rm high}$ (light-blue solid), $\mu = \mu_{\rm low}$ (dark-blue dashed), $\mu = \mu_{M}$ (red dotted), and $\mu = \mu_{H_T}$ (orange dash-dotted). Shown are the matched resummed results using $x^2_{{\rm T},4m_t^2}$ with/without approximating the soft-anomalous dimension matrix (upper left/right), $x^2_{{\rm T},s_{t\bar{t}}}$ (lower left) and the NLO result without resummation (lower right). The colored bands indicate the scale uncertainty within each scale choice as obtained by varying $\mu$ between $\mu_{\rm ref}/2$ and $2\mu_{\rm ref}$, with $\mu_{\rm ref}$ the reference factorization/renormalization scale.}
\label{fig:pTdist}
\end{figure}

\begin{figure}[t!]
\centering
\mbox{\begin{subfigure}{0.5\textwidth}\includegraphics[width=\textwidth]{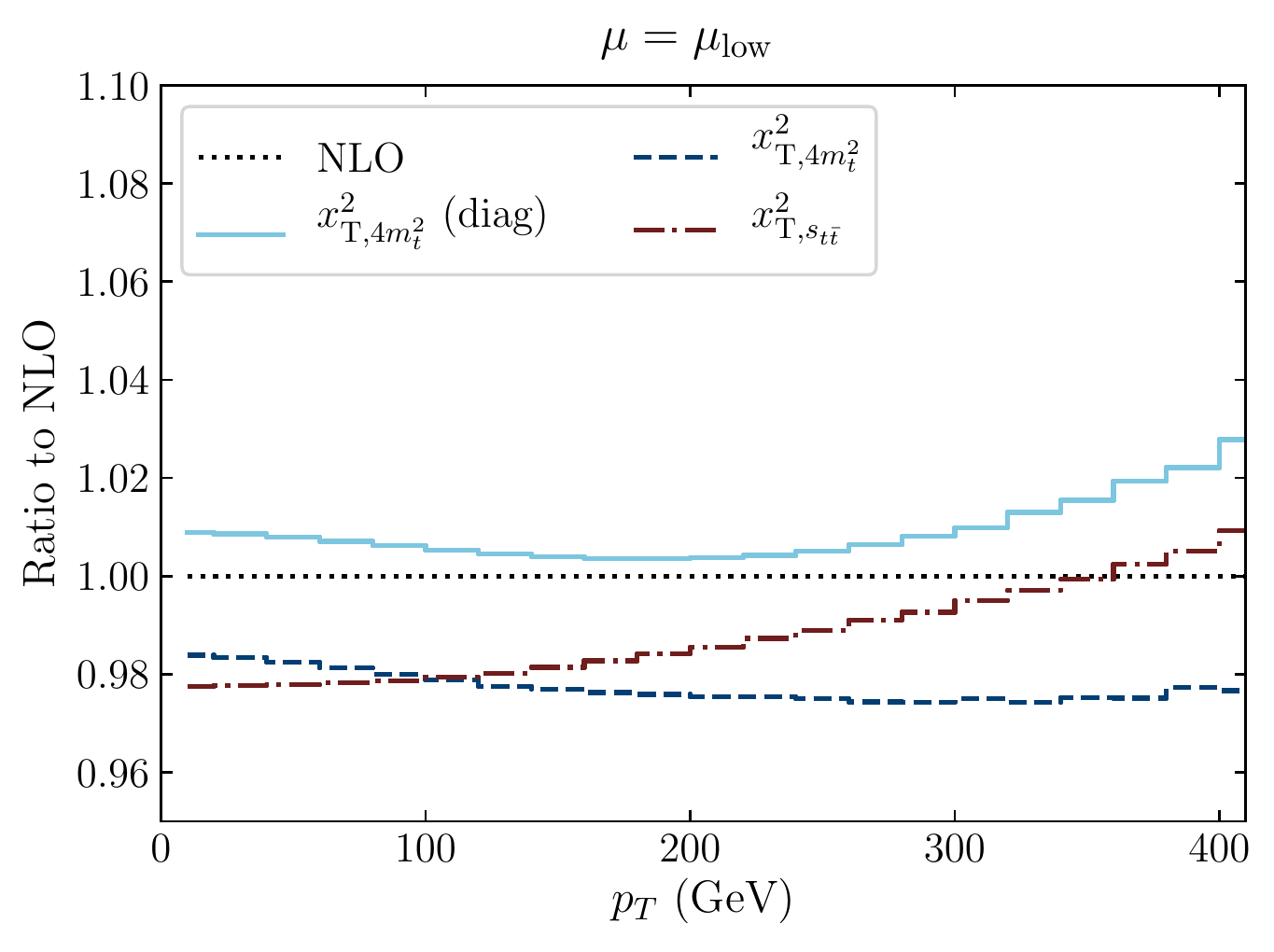}\end{subfigure}
\begin{subfigure}{0.5\textwidth}\includegraphics[width=\textwidth]{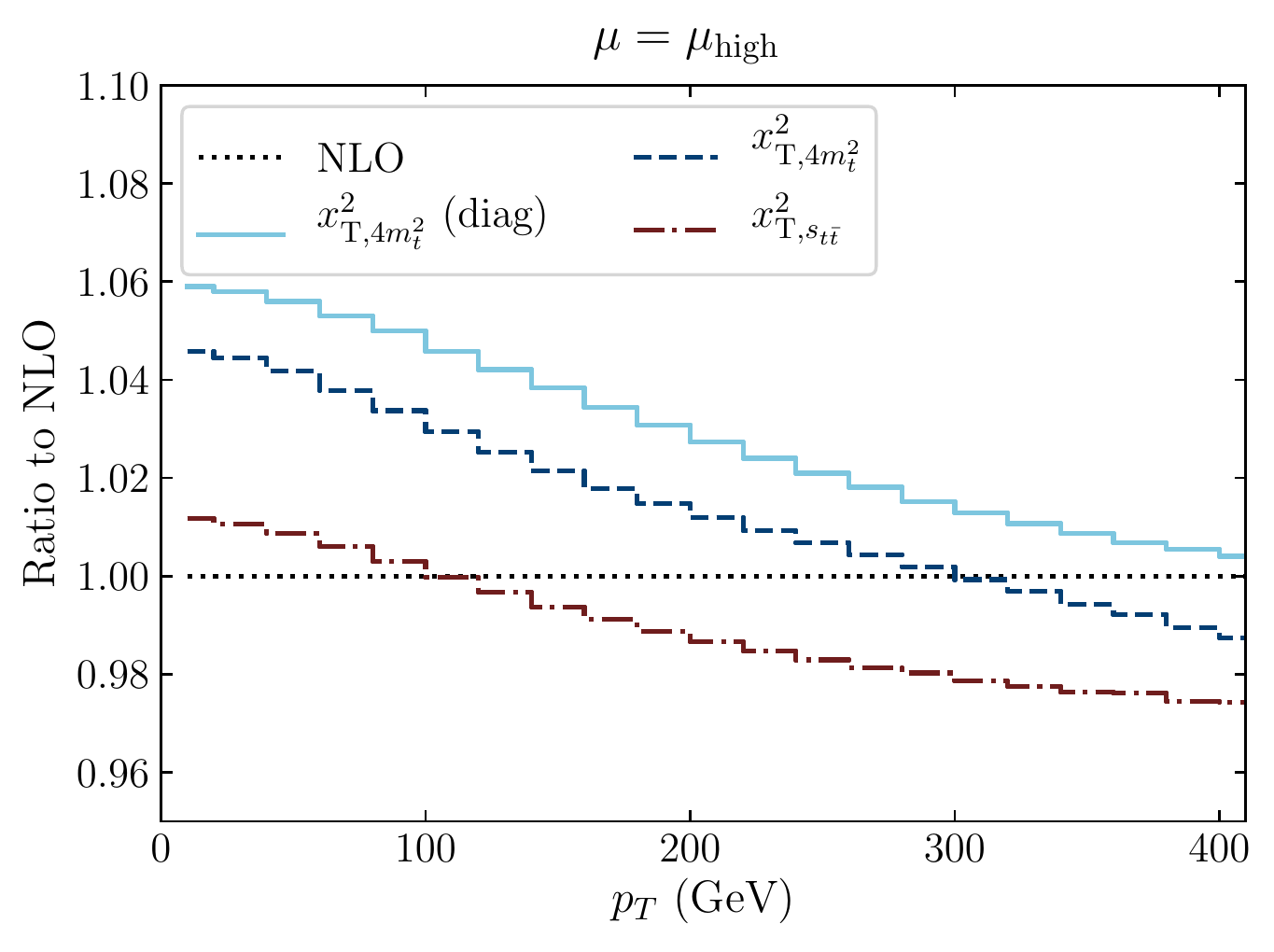}\end{subfigure}
}
\mbox{\begin{subfigure}{0.5\textwidth}\includegraphics[width=\textwidth]{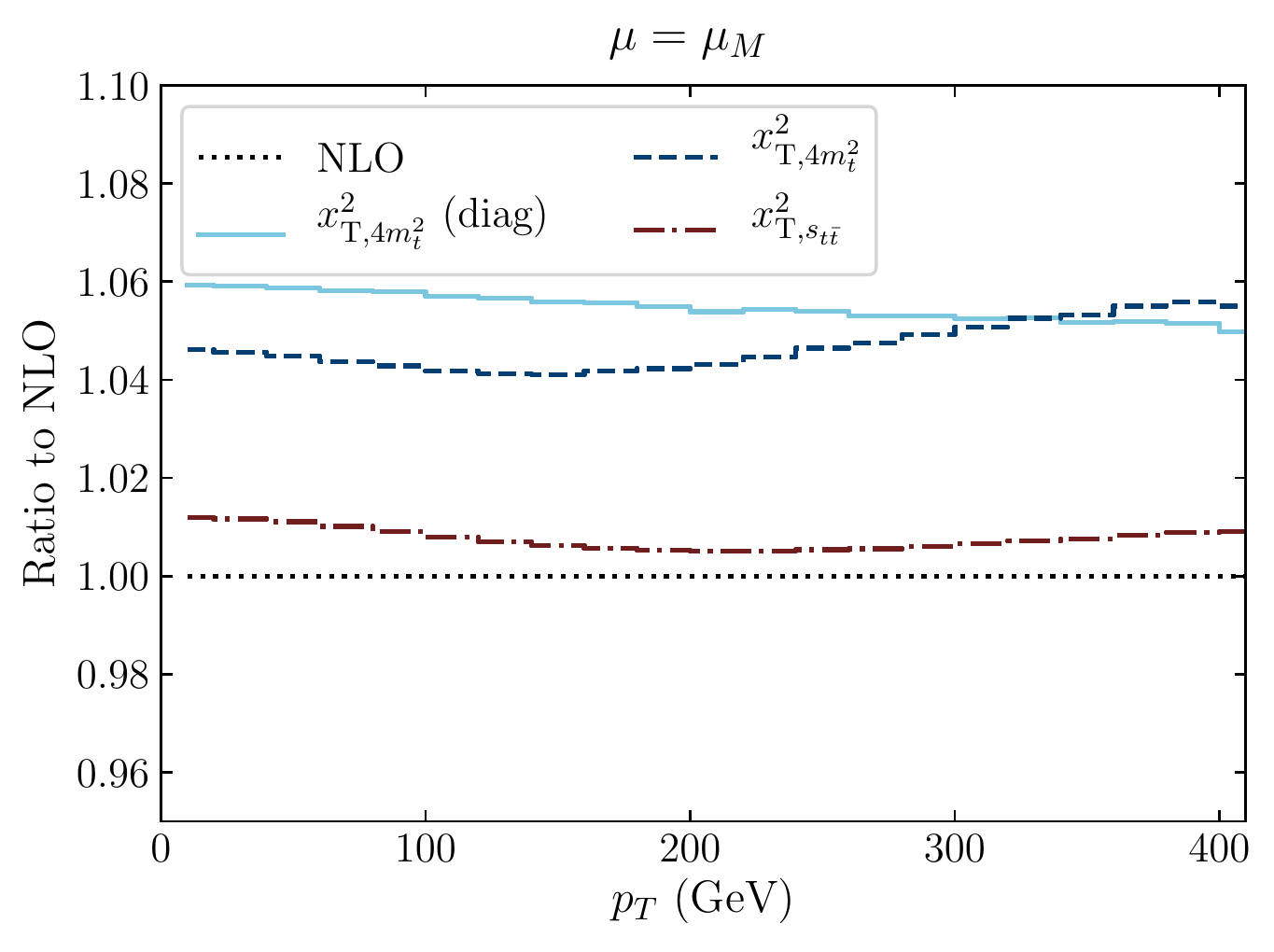}\end{subfigure}
\begin{subfigure}{0.5\textwidth}\includegraphics[width=\textwidth]{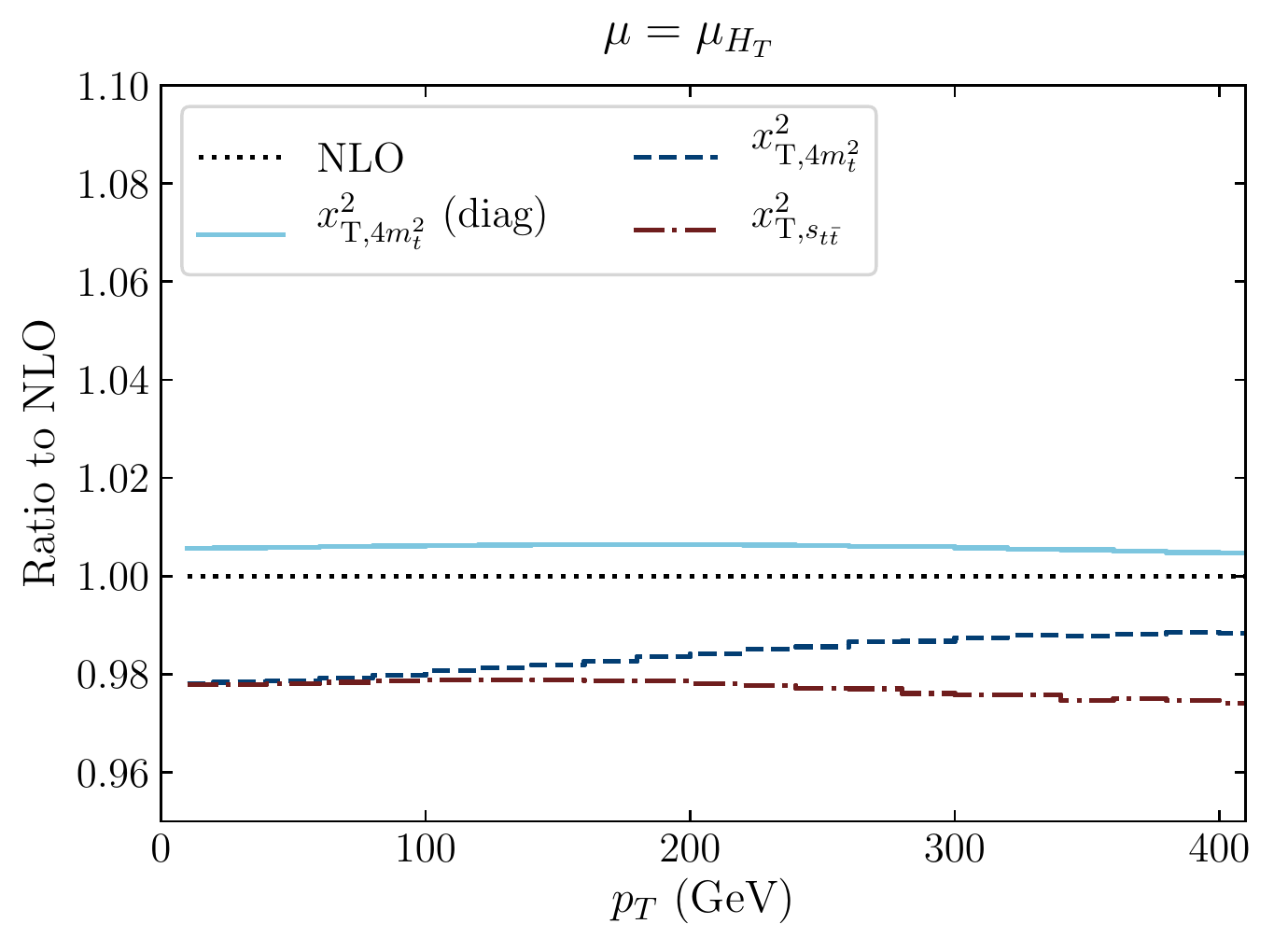}\end{subfigure}}
\caption{Ratios of the various resummed transverse momentum distributions to the NLO one for the scale choices $\mu = \mu_{\rm low}$ (upper left), $\mu = \mu_{\rm high}$ (upper right), $\mu = \mu_{M}$ (lower left), and $\mu = \mu_{H_T}$ (lower right). The ratios obtained using the threshold variable $x^2_{{\rm T},4m_t^2}$ and $x^2_{{\rm T},s_{t\bar{t}}}$  are indicated by the dashed dark-blue and the dash-dotted red lines respectively. The result that is obtained using the approximated form of the soft-anomalous dimension matrices is indicated by the solid light-blue line.}
\label{fig:pTscales}
\end{figure}

\noindent We now consider the impact of resummation on the transverse-momentum of the Higgs boson. The results are shown in Fig.~\ref{fig:pTdist}. As before, let us first take a look at the NLO distributions. At the peak of the distribution around $p_T = 70$~GeV we observe (lower-right panel of Fig.~\ref{fig:pTdist}) that the results are similar for $\mu = \mu_{\rm high}$ and $\mu = \mu_{\rm M}$, and those setting $\mu = \mu_{\rm low}$ and $\mu = \mu_{\rm H_T}$ are similar too. The reason for this is clear: when $p_{\rm T}$ is small, $m_{{\rm T},4m_t^2}\simeq 2m_t$ and $m_{{\rm T},h}\simeq m_{h}$. Therefore, $\mu_M \simeq  \mu_{\rm high}$ for small $p_T$, and the NLO distributions at small $p_T$ for these scale choices lie close in value. The discrepancy steadily grows with higher $p_T$ values. Similarly, $\mu_{H_T} \simeq \mu_{\rm low}$ at low $p_T$ values, where the best agreement between the two scales is found for $p_T \simeq 100$~GeV. This is directly reflected in the results. The NLO distribution shows a substantial scale uncertainty at the peak of the distribution, and at $p_T = 70$~GeV the central results vary between ${\rm d}\sigma_{t\bar{t}h}/{\rm d}p_{\rm T} = 2.91$~fb/GeV and ${\rm d}\sigma_{t\bar{t}h}/{\rm d}p_{\rm T} = 3.24$~fb/GeV. The smallest scale uncertainty on the interval $\mu \in [\mu_{\rm ref}/2, 2\mu_{\rm ref}]$ at $p_T = 70$~GeV is found by setting $\mu_{\rm ref} = \mu_{\rm low}$ where we obtain  ${\rm d}\sigma_{t\bar{t}h}/{\rm d}p_T = 3.23^{+5.6\%}_{-8.7\%}$~fb/GeV.  \\
\noindent The scale uncertainty of the distribution is again reduced by matching the NLL resummation to the NLO fixed-order result. Using $x_{{\rm T},4m_t^2}^2$ (upper-right panel) we obtain the least dependence on the scale choice: the distribution looks very similar across the four different scale choices and its variations on the interval $\mu \in [\mu_{\rm ref}/2, 2\mu_{\rm ref}]$. At $p_T = 70$~GeV we find central values between ${\rm d}\sigma_{t\bar{t}h}/{\rm d}p_T = 3.04-3.17$~fb/GeV. The smallest dependence on varying $\mu_{\rm ref}$ is found by setting $\mu_{\rm ref} = \mu_{M}$, in which case we obtain ${\rm d}\sigma_{t\bar{t}h}/{\rm d}p_T = 3.04^{+3.6\%}_{-3.1\%}$~fb/GeV at $p_T = 70$~GeV. The scale uncertainties grow when using $x_{{\rm T},s_{t\bar{t}}}^2$ as our threshold variable (lower-left panel). There, the smallest scale uncertainty at $p_T = 70$~GeV is again found for $\mu = \mu_{M}$, resulting in ${\rm d}\sigma_{t\bar{t}h}/{\rm d}p_T = 2.94^{+6.5\%}_{-4.7\%}$~fb/GeV. The upper scale uncertainty therefore has increased a bit with respect to the NLO distribution, while the lower scale uncertainty has roughly halved.\\
Again we find a noticeable impact by approximating the kinematics of the soft-anomalous dimension matrices (upper-left panel), although the difference with the result obtained by using the unapproximated form of these matrices again lies within the scale uncertainty. In Fig.~\ref{fig:pTscales} we show the obtained resummed correction for each of the central-scale choices and threshold variables. The scale logarithm of the NLL resummation vanishes by using $\mu_M$ and $x_{{\rm T},4m_t^2}^2$ (lower-left panel, blue dashed line). The correction with respect to the NLO distribution is then between $4.5-5.5\%$, while upon approximating the soft-anomalous dimension matrices the correction is around $6.0-5.0\%$ from low to high $p_T$. Interestingly, and in contrast to the invariant-mass case, we now find that this approximation does not only affect the overall size of the resummed correction, but also the shape of the distribution, as it does not lead to a constant difference between the curves labeled by $x_{{\rm T},4m_t^2}^2$ and $x_{{\rm T},4m_t^2}^2$ (diag). When instead $x_{{\rm T},s_{t\bar{t}}}^2$ is used with the same scale choice, the correction with respect to the NLO distribution is reduced to about $0.5-1.2\%$. The difference between the matched resummed result obtained with $x_{{\rm T},4m_t^2}^2$ or $x_{{\rm T},s_{t\bar{t}}}^2$ is fairly constant over all $p_T$ values, from which we infer that the scale logarithm $\ln((m_{{\rm T},s_{t\bar{t}}}+m_{{\rm T},h})^2/\mu_M^2)$ changes minimally when $p_T$ varies. This suggests that the dominant contribution to the transverse-momentum distribution is picked up by a value of $s_{t\bar{t}}$ that is roughly constant between different $p_T$ values. We will come back to this point in Section~\ref{sec:nlp}. \\
The results obtained with $\mu_{\rm high}$ (upper-right panel) are similar to the $\mu_M$ results at small $p_T$, but quickly deviate from those results when $p_T$ is increased. This is due to the ratio $(m_{{\rm T},t}+m_{{\rm T},h})/\mu_{\rm high}$ (with $m_{{\rm T},t}$ either $m_{{\rm T},4m_t^2}$ or $m_{{\rm T},s_{t\bar{t}}}$), which increases as $p_T$ grows. The most stable results are found for $\mu = \mu_{H_T}$ (lower-right panel), where we find a correction with respect to the NLO result at small $p_T$ values of $-2.1\%$, while for larger $p_T$ values we find a correction between $-1.5\%$ and $-2.5\%$. The correction obtained by approximating the soft-anomalous dimension matrices is constant at about $+0.5\%$.  We observe little dependence on the choice of threshold variable for the scale choices $\mu = \mu_{\rm low}$ and $\mu_{H_T}$. Especially at low values of $p_T$, we find that the two threshold choices return a similar correction of around $-2\%$. For larger values of the transverse momentum, we find that the correction changes sign for the threshold choice $x_{{\rm T},s_{t\bar{t}}}^2$ for $\mu = \mu_{\rm low}$ (upper-left panel, red dash-dotted line). Interestingly, a similar upward trend is also observed for the result obtained after approximating the kinematics of the soft-anomalous dimension matrices in the threshold limit (for the same $\mu$ choice), but does not happen for the other threshold choices. \\

\begin{figure}[t!]
\centering
\mbox{
\begin{subfigure}{0.5\textwidth}\includegraphics[width=\textwidth]{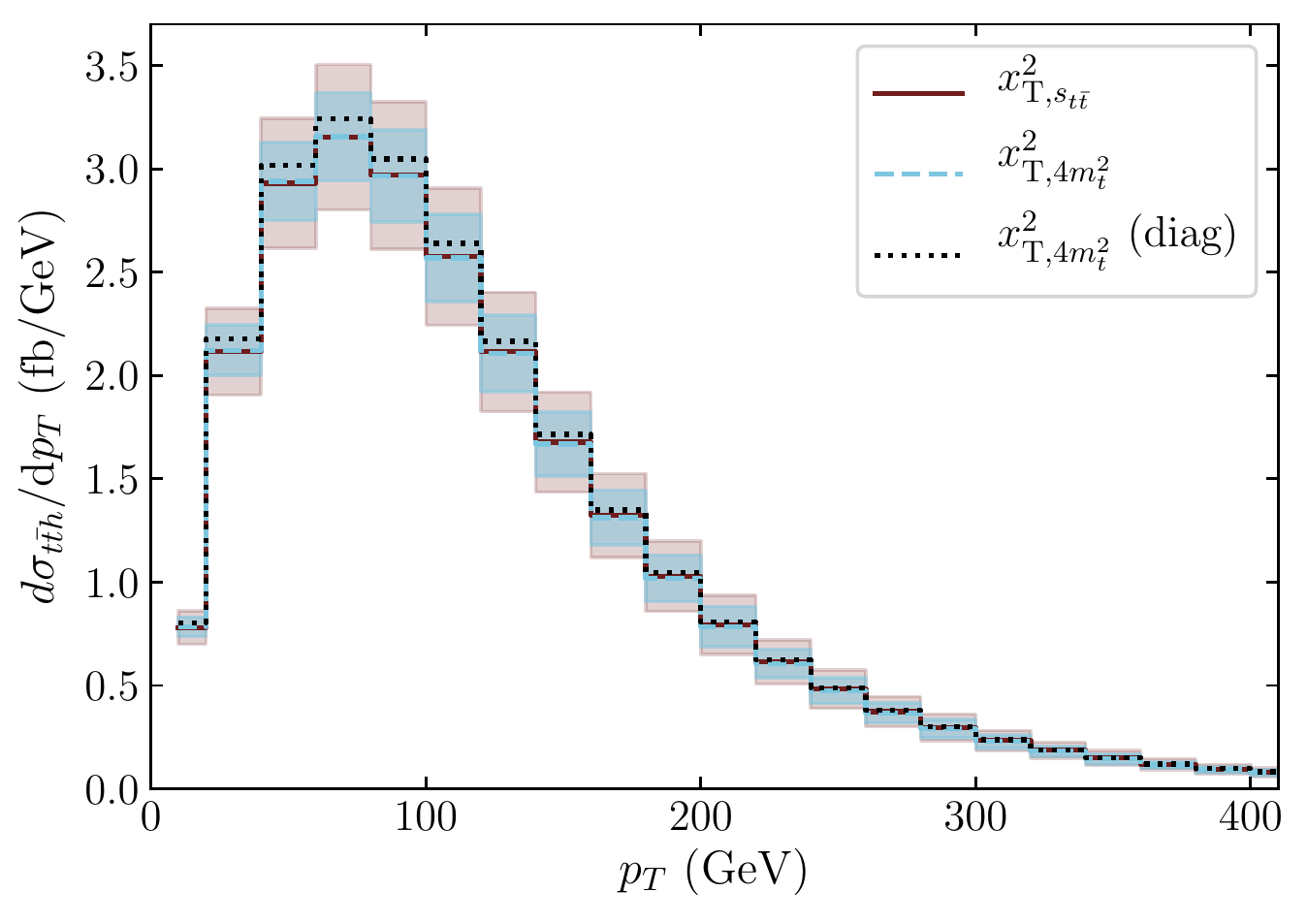}\end{subfigure}
\begin{subfigure}{0.5\textwidth}\includegraphics[width=\textwidth]{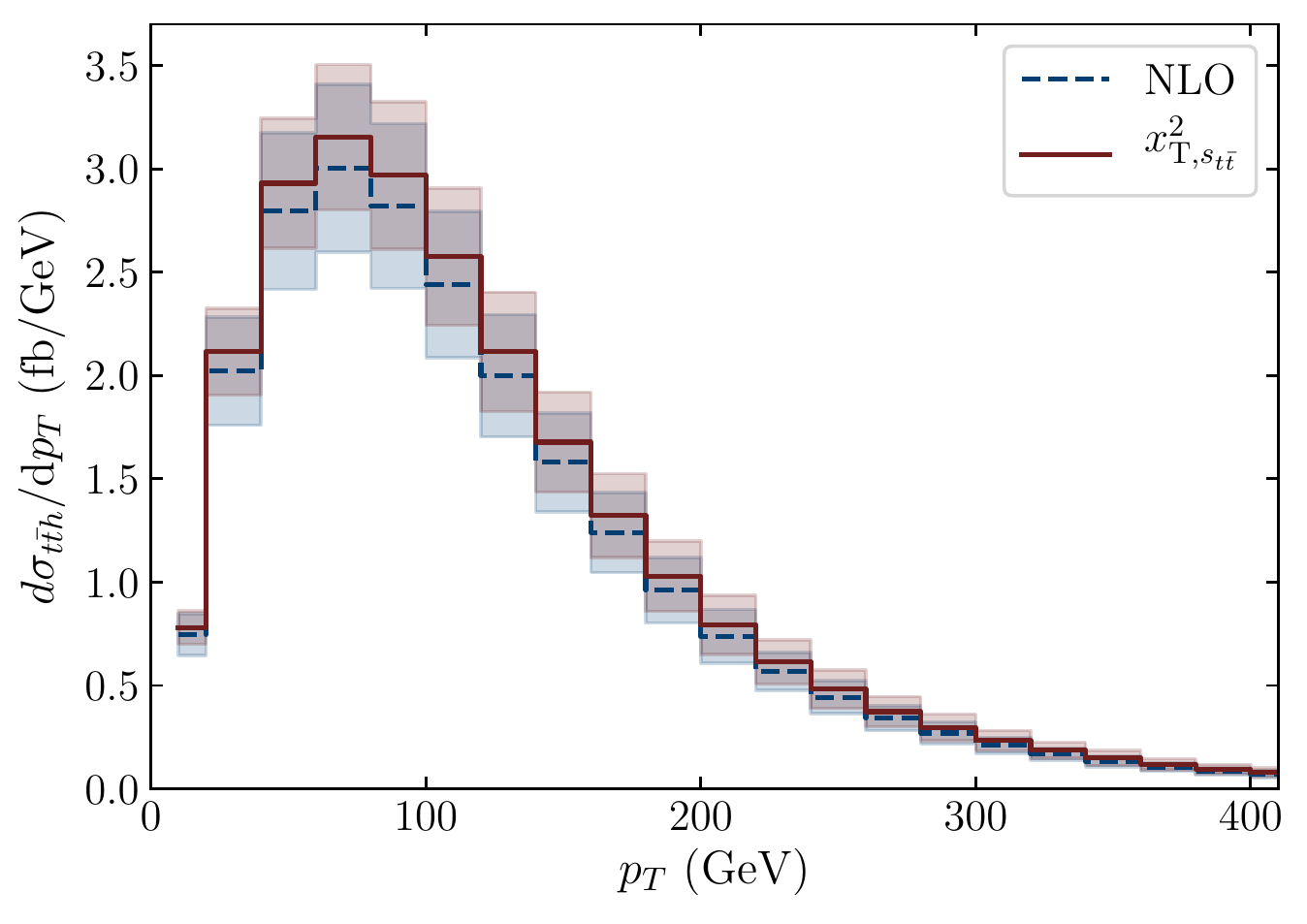}\end{subfigure}}
\caption{Left: Transverse-momentum distribution of the matched resummed results using $x^2_{{\rm T},4m_t^2}$ with/without the approximated soft-anomalous dimension matrix (black dotted and light-blue dashed) and $x^2_{{\rm T},s_{t\bar{t}}}$ (red solid). The colored bands indicate the total scale uncertainty band, obtained as before (Eq.~\eqref{eq:thisishowwedoit} with $Q\rightarrow p_T$ and $\mu' \in \mu_M$, $\mu_{H_T}$, $\mu_{\rm high}$, $\mu_{\rm low}$). The scale uncertainty band for $x^2_{{\rm T},4m_t^2}$ (diag) is not shown. Right: Transverse-momentum distribution of the NLO result (dark-blue dashed) and the matched result for $x_{{\rm T},s_{t\bar{t}}}^2$. }
\label{fig:pT_complete}
\end{figure}
\vspace{-1.0em}

\noindent As for the invariant-mass distribution, the ratio plots in Fig.~\ref{fig:pTscales} are again slightly misleading: the resummed results show a consistent picture when averaged over all scale choices. This may be observed in Fig.~\ref{fig:pT_complete}, where we see that the use of either $x_{{\rm T},4m_t^2}$ or  $x_{{\rm T},s_{t\bar{t}}}^2$ results in the same average value of the transverse-momentum distribution when considering all scale choices, although the former choice leads to smaller scale uncertainties. As for the invariant-mass distribution, the effect of approximating the soft anomalous dimension matrices is noticeable, but lies within the scale uncertainty. On the right-hand side of Fig.~\ref{fig:pT_complete}, one can see that the scale uncertainty of the resummed result is slightly smaller than that of the NLO result. The central result at $p_T=70$~GeV has increased with respect to the NLO one with $+5.2\%$, whereas the correction in the tails of the distribution grows to $+10.5\%$. The correction obtained with the approximated soft-anomalous dimension matrices is $+8.1\%$ at the peak and around $15-18\%$ in the tails. These large corrections are not observed in Fig.~\ref{fig:pTscales}, as there the ratio plots are weighted by the NLO distribution with the same scale choice. 

\subsection{Total cross section}
\noindent To conclude these results, and before moving on to the discussion of the $\mathcal{O}(1/N)$ effects, we briefly comment on the total cross section obtained using various scale choices and threshold variables. These values are obtained by numerically integrating the resummed expressions for either $Q$ or $p_T$ with the various threshold-variable definitions. At NLO, the smallest scale uncertainties are indeed found by setting $\mu = \mu_{\rm low}$, for which we obtain $\sigma_{t\bar{t}h}^{\rm NLO} = 0.499^{+5.8\%}_{-9.2\%}$ pb (in accordance with the NLO cross section reported in Table~229 of Ref.~\cite{deFlorian:2016spz}). If instead we again vary the scales around all values of $\mu$ that are shown in Table~\ref{tab:summary}, we find $\sigma_{t\bar{t}h}^{\rm NLO, average} = 0.449^{+17.6\%}_{-17.6\%}$ pb. The NLL resummed and matched result, averaged over all $\mu$ vales and threshold choices is $\sigma_{t\bar{t}h}^{\rm NLL, average} = 0.492^{+12.9\%}_{-12.9\%}$ pb. This lies higher than the averaged NLO result ($+9.6\%$), but is close to the NLO result obtained for $\mu = \mu_{\rm low}$. The scale uncertainty is reduced slightly. The option that shows the smallest scale uncertainties is obtained by setting $\rho_{\rm abs}$ with $\mu_{\rm high}$, which results in $\sigma_{t\bar{t}h,\mu = \mu_{\rm high}}^{{\rm NLL}, \rho_{\rm abs}} = 0.476^{+3.4\%}_{-2.7\%}$ pb, followed by that obtained setting $\mu=\mu_{\rm low}$:  $\sigma_{t\bar{t}h,\mu = \mu_{\rm low}}^{\rm NLL, \rho_{\rm abs}} = 0.491^{+5.6\%}_{-3.3\%}$ pb~\footnote{Note that these values differ from those obtained in Ref.~\cite{Kulesza:2015vda}, as there another PDF set is used and they used the approximated form of the soft anomalous dimension matrices. Comparisons with Ref.~\cite{Kulesza:2017ukk,Kulesza:2020nfh} are complicated since a different PDF accuracy (NLO instead of NNLO) is used in those references. On the other hand, the comparison with the soft-collinear effective theory results presented in Ref.~\cite{Broggio:2015lya,Broggio:2016lfj,Broggio:2019ewu} is complicated due to the presence of multiple scales (the soft and hard scale), which are not free scale choices in the direct QCD framework.}. The average value obtained by approximating the kinematics of the soft anomalous dimension matrices is 
$\sigma_{t\bar{t}h}^{\rm NLL, diag} = 0.506^{+7.3\%}_{-7.3\%}$ pb. Therefore, as for the two distributions that we have considered, the effect of approximating the soft anomalous dimension matrices is noticeable, but the resulting value for the total cross section lies well within the scale uncertainty band of the resummed result when considering all scales and threshold definitions. 

\section{Role of NLP corrections}
\label{sec:nlp}
In the previous section, it became clear that the parameterically subleading $\mathcal{O}(1/N)$ contributions of the soft-anomalous dimension matrices show a noticeable impact on the resulting distributions. In this section we further study the role of these $\mathcal{O}(1/N)$ contributions. The partonic cross section is evaluated as a function of the partonic center of mass energy squared $s$, with $s = x_1 x_2 S$. The partonic threshold region is the region in which the partonic threshold parameter $\rho$ is close to $1$. All values of $x_1$ and $x_2$ between $\tau$ and $1$ are however accessible, so whether or not resummation is actually relevant depends on which values of $x_1$ and $x_2$ give the \emph{dominant contribution} to the total hadronic cross section. The region where this dominant contribution originates from in $N$-space may be estimated with a saddle-point argument (as was done in Refs.~\cite{Bonvini:2014qga,Bonvini:2012an,Bonvini:2010tp} for the single Higgs and DY production processes), which we review here and apply to the production of $t\bar{t}h$. \\
We first clarify that the large-$N$ limit \emph{only} does not apply when arguing the validity of approximations made for the threshold variable $\rho$. The validity of the large-$N$ approximation that one employs to compute the effects of soft-gluon contributions, leading for example to the resummation functions $g^{(i)}$, is unchanged. By using the large-$N$ limit to compute the integral in e.g.~Eq.~\eqref{eq:deltaiint}, one essentially isolates the $z\rightarrow 1$ limit, which is the only region where the results of this integral can be trusted. The integral needs to be adjusted with subleading-power contributions away from the $z\rightarrow 1$ limit if one also wants to assess the $\mathcal{O}(1/N)$ contribution of this integral. However, since we have no control over all-order $\mathcal{O}(1/N)$ effects, it is safest to truncate the result of this integral at $\mathcal{O}(1)$. This line of reasoning \emph{does not apply} to the inverse Mellin transform, as we have no reason to assume that the final resummed result is dominated by the \emph{partonic} threshold limit $\rho \rightarrow 1$ for all variables that we formulated above. Indeed: we will find that small ($\mathcal{O}(1)$) values of $N$ dominate the final result, so in general we expect that effects originating from regions away from the $\rho\rightarrow 1$ (and similarly $x_i \rightarrow 1$) limit do affect the final result. \\
Denoting the resummed hadronic distribution of interest as ${\rm d}\sigma^{\rm NLL}_{pp\rightarrow t\bar{t}h}$ we have
\begin{eqnarray}
\label{eq:hadronic1}
{\rm d}\sigma^{\rm NLL}_{pp\rightarrow t\bar{t}h} = \sum_{i,j} \int_{\mathcal{C}}\frac{{\rm d}N}{2\pi i}\,\tau^{-N}\,\mathcal{L}_{ij}(N)\,\int_0^1{\rm d}\rho\, \rho^{N-1}\,{\rm d}\hat{\sigma}^{\rm NLL}_{ij\rightarrow t\bar{t}h}\,,
\end{eqnarray}
with $\rho$ and $\tau$ the generic partonic and hadronic threshold variables, and we have defined $\mathcal{L}_{ij}(N) = f_i(N+1,\mu_F^2)f_j(N+1,\mu_F^2)$. We now consider the resummation up to LL, which means that our expression for the resummed partonic distribution becomes simply
\begin{eqnarray}
{\rm d}\hat{\sigma}^{\rm LL}_{ij\rightarrow t\bar{t}h}  \equiv {\rm d}\hat{\sigma}^{\rm LO}_{ij\rightarrow t\bar{t}h}\,{\rm exp}\left[\frac{2}{\alpha_s}g_i^{(1)}(\lambda)\right],
\end{eqnarray}
with $i=j=g$ or $i=\bar{j}=q$. We may then rewrite Eq.~\eqref{eq:hadronic1} to 
\begin{eqnarray}
\label{eq:hadronic}
{\rm d}\sigma^{\rm LL}_{pp\rightarrow t\bar{t}h} &=& \sum_{i,j}\int_{\mathcal{C}}\frac{{\rm d}N}{2\pi i}\,\int_0^1 \frac{{\rm d}\rho}{\rho}\,{\rm d}\hat{\sigma}^{\rm LO}_{ij\rightarrow t\bar{t}h} \nonumber \\
&& \hspace{4cm}\times{\rm exp}\left[-N\ln\left(\frac{\tau}{\rho}\right)+\ln(\mathcal{L}_{ij}(N))+ \frac{2}{\alpha_s}g_i^{(1)}(\lambda)\right]\, \nonumber \\
&\equiv& \sum_{i,j}\int_{\mathcal{C}}\frac{{\rm d}N}{2\pi i}\,\int_0^1\frac{{\rm d}\rho}{\rho}\,{\rm d}\hat{\sigma}^{\rm LO}_{ij\rightarrow t\bar{t}h}\,{\rm exp}\left[E_{ij\rightarrow t\bar{t}h}(N)\right].
\label{eq:hihi}
\end{eqnarray}

\begin{figure}[t]
\centering
\mbox{\begin{subfigure}{0.5\textwidth}\includegraphics[width=\textwidth]{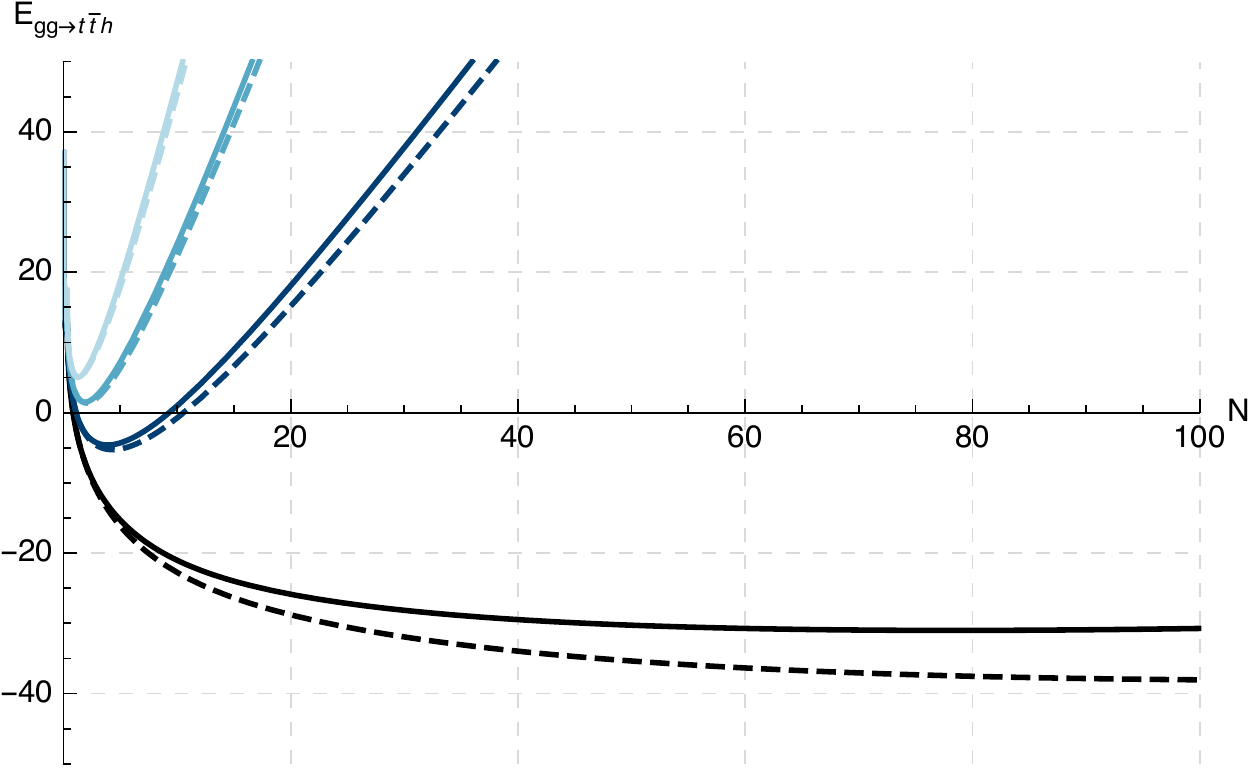}\end{subfigure}
\begin{subfigure}{0.5\textwidth}
\includegraphics[width=\textwidth]{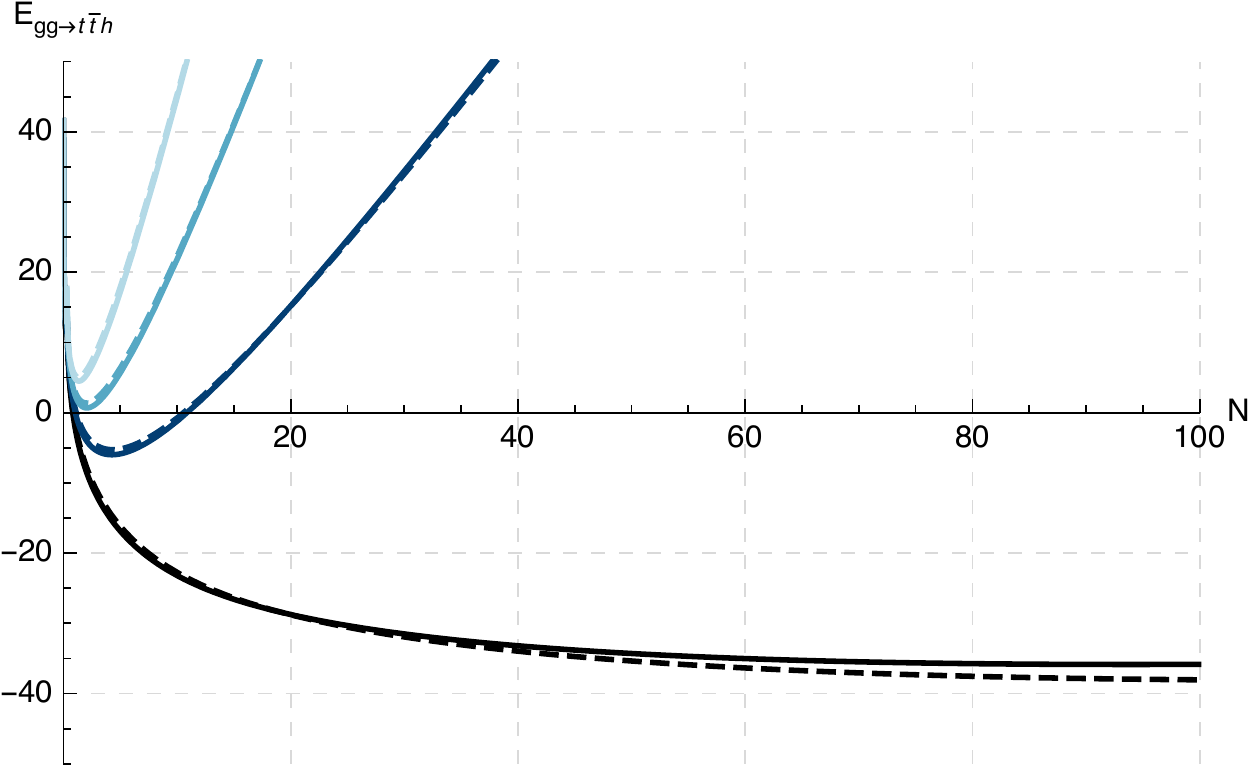}\end{subfigure}}
\caption{$E_{ij\rightarrow t\bar{t}h}$ with (solid) and without (dashed) the $g^{(n)}$ contributions with $n = 1,2$ for the $gg$-initiated channel. On the left-hand side, we set $\ln(M^2/\mu^2) = 0$, while on the right-hand side we use $\ln(M^2/\mu^2) = 3.1$. We show  $\tau/\rho = 0.9$ (black), $0.1$ (dark blue), $0.01$ (blue), $0.001$ (lightest blue). }
\label{fig:diffEtth}
\end{figure}
\begin{figure}[t]
\centering
\mbox{\begin{subfigure}{0.5\textwidth}\includegraphics[width=\textwidth]{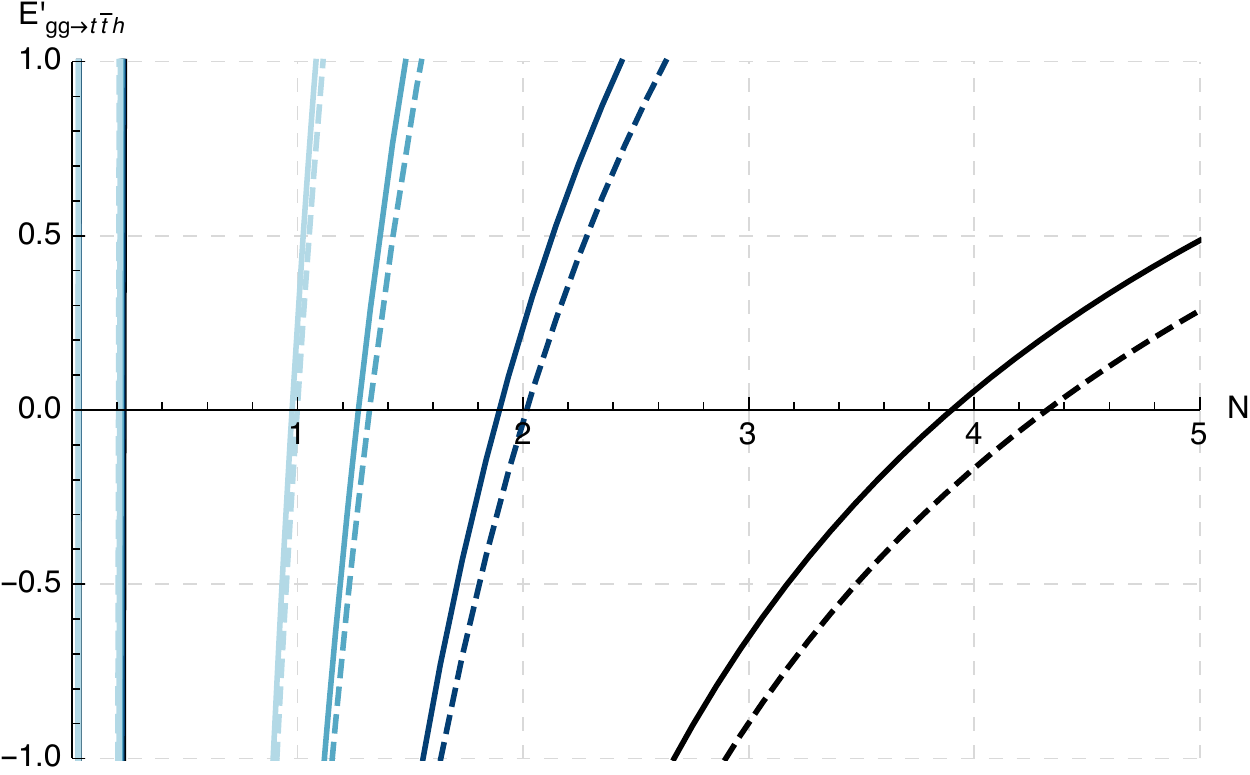}\end{subfigure}
\hspace{0.1cm}
\begin{subfigure}{0.5\textwidth}
\includegraphics[width=\textwidth]{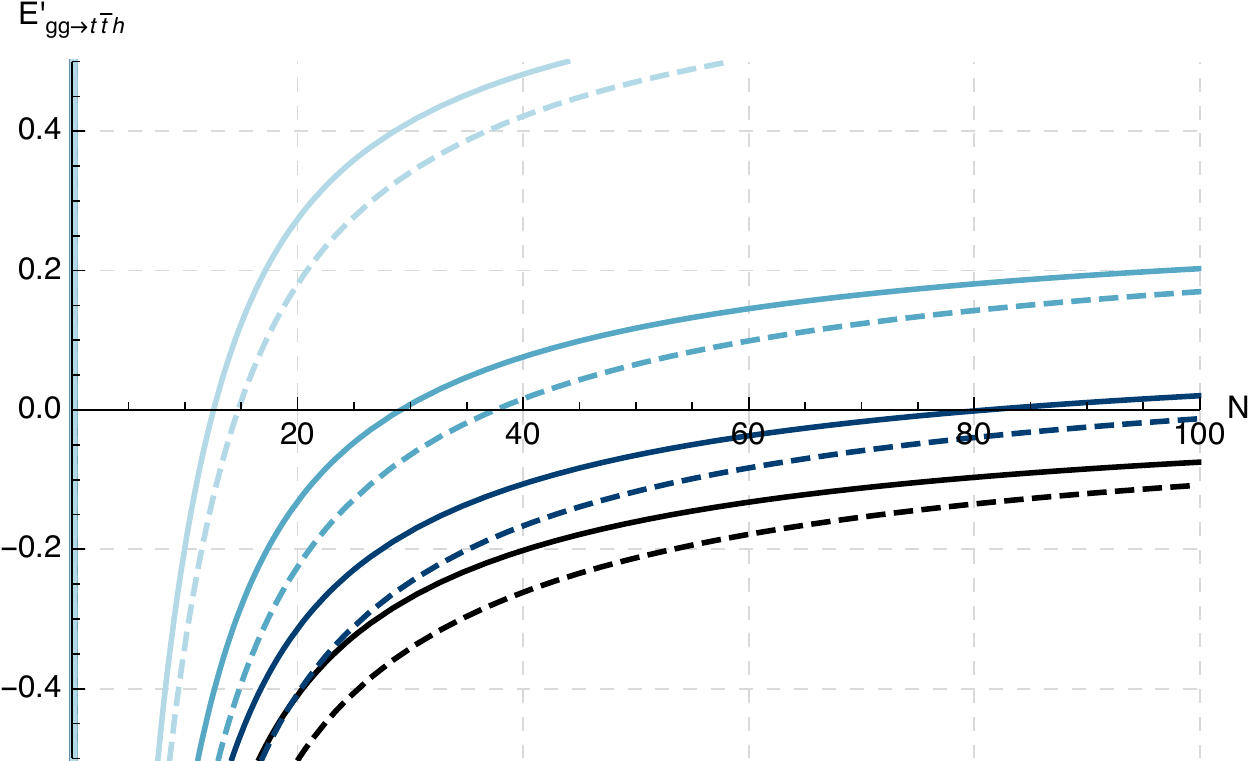}\end{subfigure}}
\vspace{0.5cm}

\mbox{\begin{subfigure}{0.5\textwidth}\includegraphics[width=\textwidth]{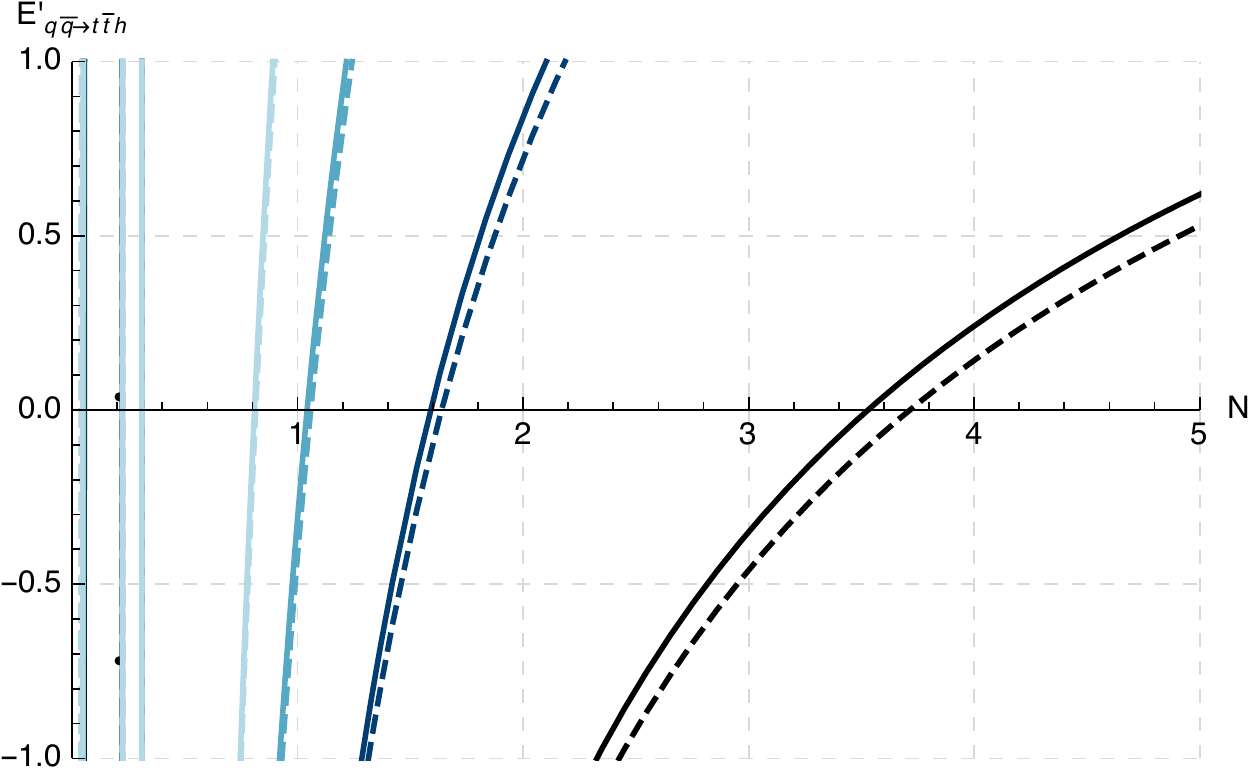}\end{subfigure}
\hspace{0.1cm}
\begin{subfigure}{0.5\textwidth}
\includegraphics[width=\textwidth]{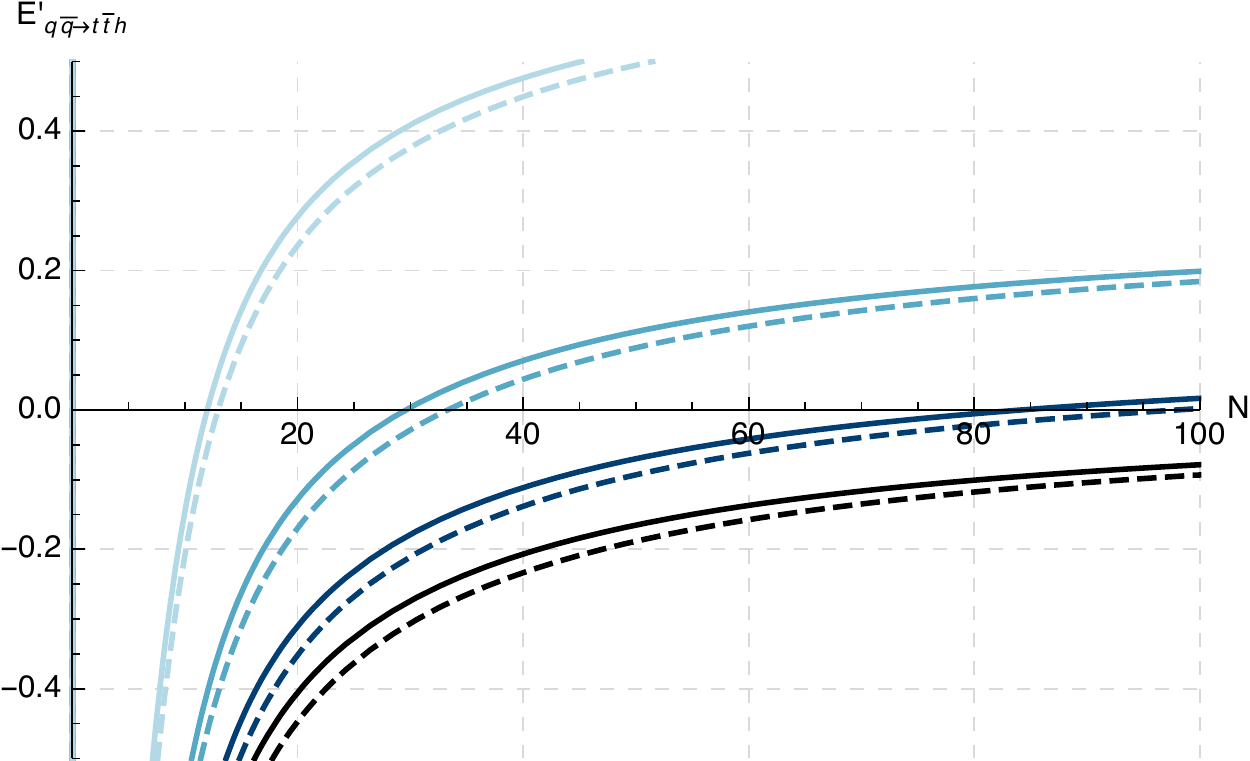}\end{subfigure}}
\caption{$E'_{ij\rightarrow t\bar{t}h}$ with (solid) and without the $g^{(n)}$ contributions (dashed) for the $gg$/$q\bar{q}$ initiated process (top/bottom). We set $\tau/\rho = 0.1$ (black), $0.01$ (dark blue), $0.001$ (blue), $0.0001$ (lightest blue) for the left panel, while for the right panel we use $\tau/\rho = 0.99$ (black), $0.9$ (dark blue), $0.75$ (blue), $0.5$ (lightest blue). }
\label{fig:derivgg}
\end{figure}

\begin{figure}[t]
\centering
\mbox{\begin{subfigure}{0.5\textwidth}\includegraphics[width=\textwidth]{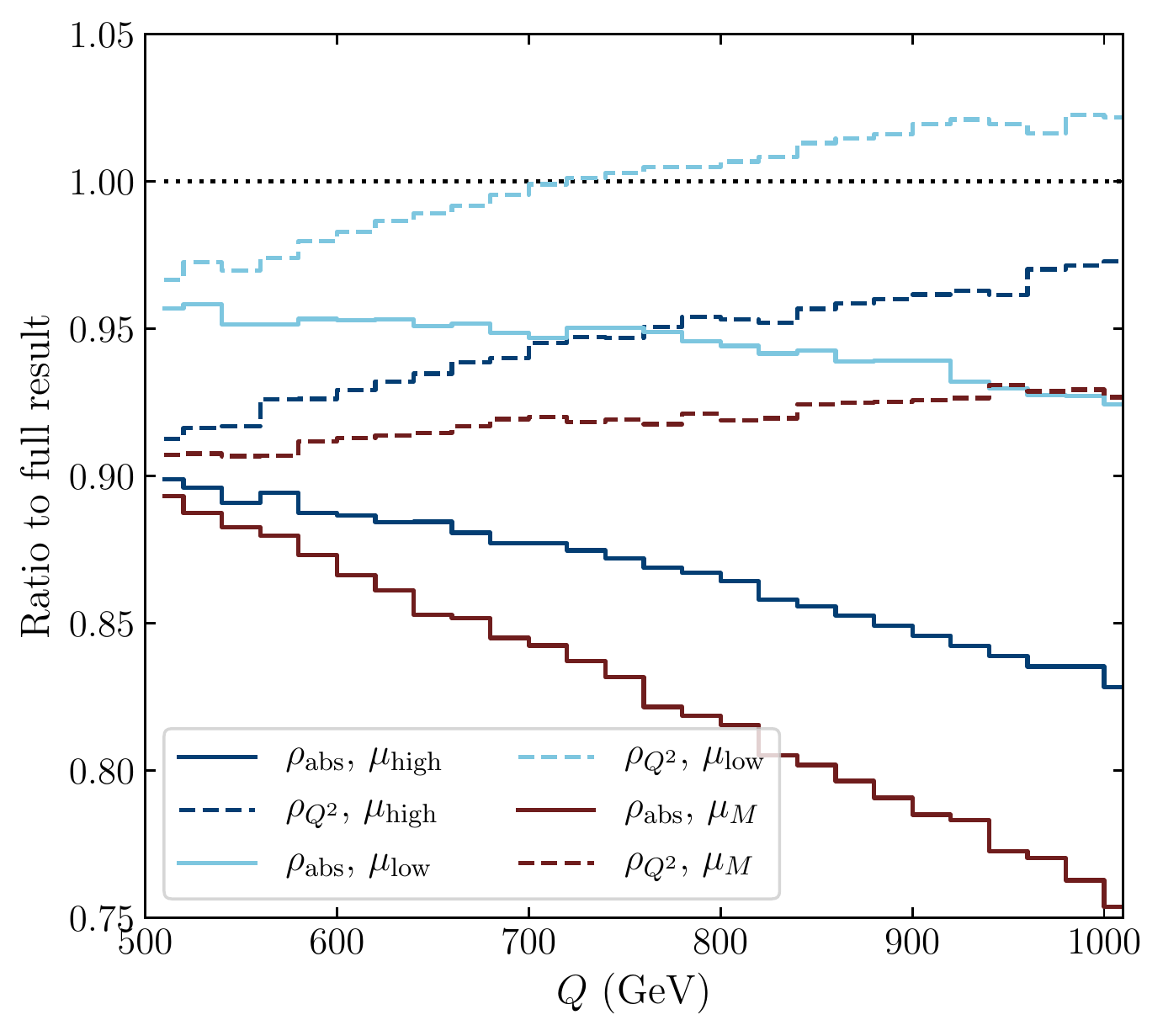}\end{subfigure}
\begin{subfigure}{0.5\textwidth}
\includegraphics[width=\textwidth]{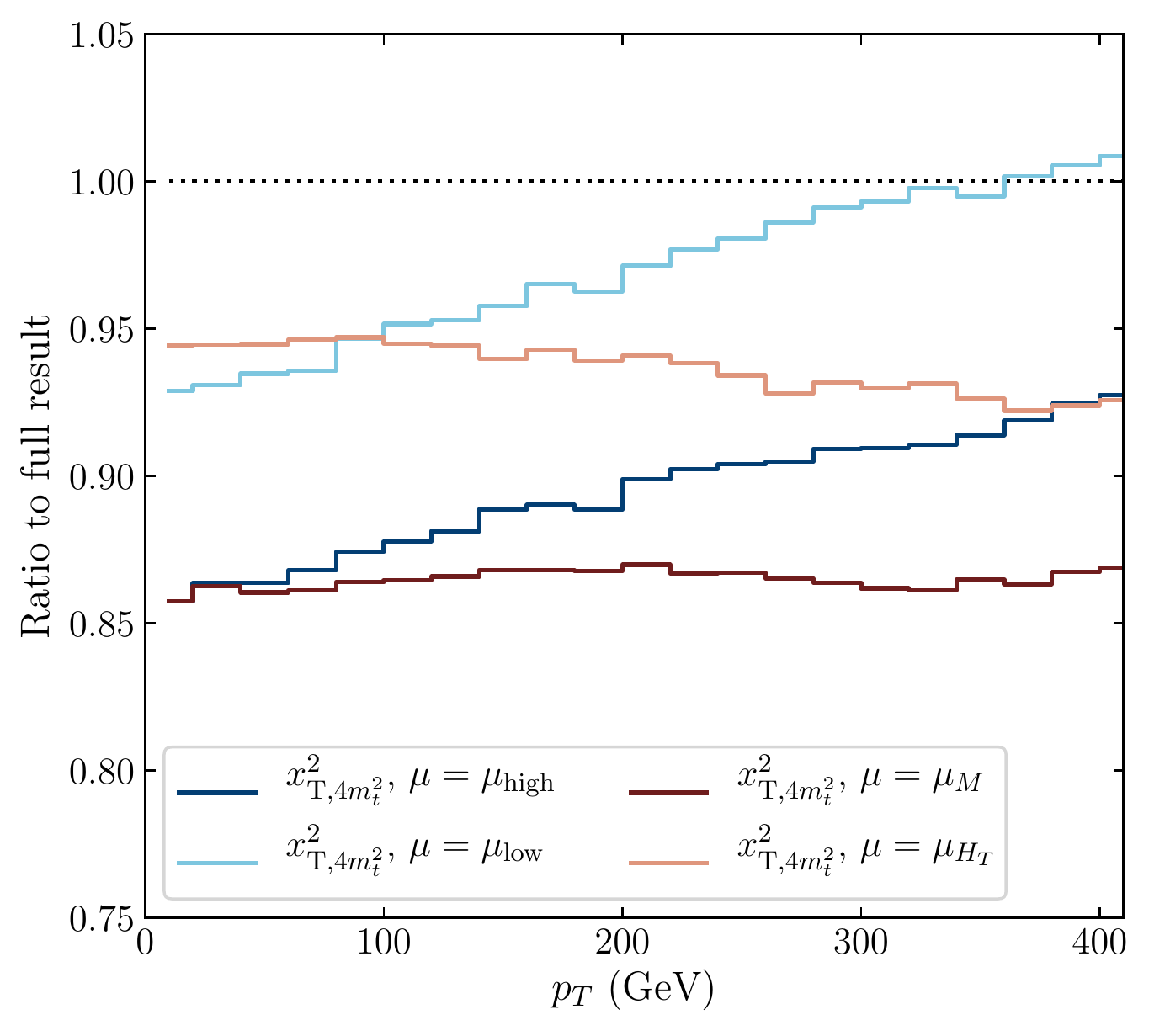}\end{subfigure}}
\caption{Left: The NLL distributions for the ratio between the invariant-mass distribution obtained by setting $\lambda = \lambda_0$ and the full result. The solid (dashed) lines indicate using the threshold definition $\rho_{\rm abs}$ ($\rho_{Q^2}$), while the dark-blue/light-blue/red lines represent the results using $\mu_{\rm high}$/$\mu_{\rm low}$/$\mu_{M}$. Right: The NLL distributions for the ratio between the transverse-mass distribution obtained by setting $\lambda = \lambda_0$ and the full result. We used $x^2_{{\rm T}, 4m_t^2}$ as our threshold variable, and the color indicates the scale choice: $\mu_{\rm high}$ (dark blue), $\mu_{\rm low}$ (light blue), $\mu_{M}$ (red), or $\mu_{H_T}$ (orange).}
\label{fig:N0dym}
\end{figure}

\noindent For $\tau < \rho$, which is the physical domain where $s < S$, we have that $\ln\left(\frac{\tau}{\rho}\right) < 0$, and thus the exponent will grow as $N\rightarrow \infty$. Note that the slope of the (linear) function $-N\ln\left(\frac{\tau}{\rho}\right)$ gets smaller for values of $\rho$ closer to $\tau$ (i.e.~close to the hadronic threshold). For small $N$, the parton luminosity function has a singularity. Combining these two facts, we understand that the function $E_{ij\rightarrow t\bar{t}h}(N)$ has a minimum $(N_0)$, which may also be seen explicitly by plotting $E_{ij\rightarrow t\bar{t}h}(N)$ as we do in Fig.~\ref{fig:diffEtth} (left-hand side, solid lines), and therefore that the dominant contribution to the $N$-integral may be estimated using a saddle-point argument. To this end, we take the derivative with respect to $N$ such that 
\begin{eqnarray}
\label{eq:minimum}
\frac{{\rm d}E_{ij\rightarrow t\bar{t}h}(N)}{{\rm d}N}\Bigg|_{N=N_0} &\equiv& E'_{ij\rightarrow t\bar{t}h}(N_0) \nonumber \\
&=& -\ln\left(\frac{\tau}{\rho}\right)+\frac{{\rm d}\ln(\mathcal{L}_{ij}(N))}{{\rm d}N}\Bigg|_{N=N_0}+ \frac{2}{\alpha_s}\frac{{\rm d}g_i^{(1)}(\lambda)}{{\rm d}N}\Bigg|_{N=N_0} = 0\,.
\end{eqnarray}
The saddle-point approximation of Eq.~\eqref{eq:hihi} then becomes
\begin{eqnarray}
\label{eq:saddle}
{\rm d}\sigma^{\rm LL}_{pp\rightarrow t\bar{t}h} &\simeq& \sum_{i,j} \int_0^1\frac{{\rm d}\rho}{\rho}\,{\rm d}\hat{\sigma}^{\rm LO}_{ij\rightarrow t\bar{t}h}\frac{{\rm e}^{E_{ij\rightarrow t\bar{t}h}(N_0)}}{2\pi i}\int_{c-i\infty}^{c+i\infty}{\rm d}N \, {\rm exp}\left[\frac{E''_{ij\rightarrow t\bar{t}h}(N_0)}{2}(N-N_0)^2\right] \\
&=& \sum_{i,j}\int_0^1\frac{{\rm d}\rho}{\rho}\,{\rm d}\hat{\sigma}^{\rm LO}_{ij\rightarrow t\bar{t}h}\frac{{\rm e}^{E_{ij\rightarrow t\bar{t}h}(N_0)}}{\sqrt{2\pi E''_{t\bar{t}h}(N_0)}}\,, \nonumber
\end{eqnarray}
where we have set $N = N_0 + i t$ and $c = N_0$. This technique is not only useful to estimate the integral, but for our purposes it can be used to determine which value of $N_0$ gives rise to the bulk of the full result. However, we first need to verify the approximation of setting $N\rightarrow N_0$ by comparing Eq.~\eqref{eq:saddle} to the full result, and see that the numerical difference is small, which is what we do in what follows. \\
\noindent We first confirm that the resummation functions minimally influence the location of the minimum of $E_{ij\rightarrow t\bar{t}h}$. This means that instead of using Eq.~\eqref{eq:minimum}, we may use
\begin{eqnarray}
\label{eq:minn01}
-\ln\left(\frac{\tau}{\rho}\right)+\frac{{\rm d}\ln(\mathcal{L}_{ij}(N))}{{\rm d}N}\Bigg|_{N=N_0'} = 0\,,
\end{eqnarray}
to determine $N_0 \simeq N_0'$. The validity of this approximation may be estimated from Fig.~\ref{fig:derivgg} for the $gg$ (top row) and $q\bar{q}$ (bottom row) initiated processes, where we have used $\mu = \mu_{\rm low}$ to set the factorization scale in the luminosity function and the renormalization scale for $\alpha_s$. One can observe that the values of $N_0$ (where the solid lines cross $0$) and $N_0'$ (where the dashed lines cross $0$) are close together for small values of $\tau/\rho$. The difference between the values of $N_0$ and $N_0'$ does increase as $\tau/\rho\rightarrow 1$. However, since the minimum of $E_{ij\rightarrow t\bar{t}h}$ gets less deep for $\tau/\rho \rightarrow 1$ (see Fig.~\ref{fig:diffEtth}), the numerical difference between $E_{ij\rightarrow t\bar{t}h}(N_0)$ and $E_{ij\rightarrow t\bar{t}h}(N_0')$ remains small. 
Secondly, from Fig.~\ref{fig:diffEtth} it can be seen that the approximation of ignoring the $g^{(n)}$ functions to determine the minimum of $E_{ij\rightarrow t\bar{t}h}$ works better when the ratio $M^2/\mu^2$ grows large. On the right-hand side of Fig.~\ref{fig:diffEtth}, we show for $ij = gg$ the difference between the full $E_{gg\rightarrow t\bar{t}h}$ function and the $E_{gg\rightarrow t\bar{t}h}$ function without the $g^{(n)}$ functions for $\mu = \mu_{\rm low}$ and $M = (m_{{\rm T},h}+m_{{\rm T}, 4m_t^2})^2$ for an extreme value for $p_T$ ($p_T = 490$~GeV), where the scale logarithm grows to $\ln(M^2/\mu^2) \simeq 3.1$. One may observe that the effect of including this scale logarithm is negative, and brings the full result closer to the approximated result where the $g^{(n)}$ functions are not considered for $\tau/\rho \rightarrow 1$. The value of $E_{ij\rightarrow t\bar{t}h}$ in the minimum becomes slightly higher for smaller $\tau/\rho$ ratios, and better agreement is found for slightly lower values of $\ln(M^2/\mu^2)$. Keeping these observations in mind, we now proceed and ignore the contribution of the resummation function, thereby determining the minumum value of $N_0$ via Eq.~\eqref{eq:minn01}.  \\
\noindent As a general remark, by comparing the top and bottom panels of Fig.~\ref{fig:derivgg}, we infer that the $q\bar{q}$ luminosity function results in slightly lower values for $N_0$ than the $gg$ luminosity function for small $\tau/\rho$. The difference between using the $q\bar{q}$ and $gg$ luminosity function vanishes for $\tau/\rho \rightarrow 1$. We have verified that the resulting $N_0$ value is minimally influenced by the scale choice, with vanishing dependence as $\tau/\rho \rightarrow 1$. \\
\noindent We have now established that the behavior of the exponent $E_{ij\rightarrow t\bar{t}h}$ is largely determined by the interplay between the luminosity function and $\ln(\tau/\rho)$. Next to dropping the dependence on the resummation function to determine the value of $N_0$, we may make one further approximation: we assume that we may take $\rho$ close to $1$. This approximation is definitely validated for $\rho = \rho_{Q^2}$, but might fail for the other threshold definitions. Next to easing the computation of $N_0$, this approximation has as a further goal that it will estimate the numerical importance of the $\rho\rightarrow 1$ approximation of the soft-anomalous dimension matrices, and the correctness of changing the upper limit of the integration in e.g.~Eq.~\eqref{eq:deltaiint}. Therefore, by using
\begin{eqnarray}
\label{eq:n0minnum}
-\ln\left(\tau\right)+\frac{{\rm d}\ln(\mathcal{L}_{ij}(N))}{{\rm d}N}\Bigg|_{N=N_0} = 0
\end{eqnarray}
to determine the value of $N_0$ for the $q\bar{q}$ and $gg$-initiated channels, and comparing the full resummed result with the approximated one, we may check the validity of setting $\rho = 1$. We then use these values of $N_0$ to set $\lambda_0 = \alpha_s b_0 \ln(\bar{N}_0)$ in the resummation functions. We keep the full $N$-dependence of the PDFs and still perform the inverse Mellin transform numerically. Doing this, we may fully assess the role of the resummation itself, and determine for which value of $N_0$ the resummation functions pick up the correct contribution. The result of this may be seen in the left panel of Fig.~\ref{fig:N0dym} for $\rho_{Q^2}$ and $\rho_{\rm abs}$ (for the invariant-mass distribution) and in the right panel for $x^2_{{\rm T},4m_t^2}$ (for the transverse-momentum distribution), using all scale choices. We do not consider the $\rho_{s_{t\bar{t}}}$ and $x^2_{{\rm T},s_{t\bar{t}}}$ options, since our goal here is to assess the validity of the $\rho_{\rm abs} \rightarrow 1$ and $x^2_{{\rm T},4m_t^2} \rightarrow 1$ approximations.  \\

\vspace{-1em}

\noindent We first consider the invariant-mass distribution. The values of $\bar{N}_0$ that we have found are: $\bar{N}_{0,gg} = 2.4$ and $\bar{N}_{0,q\bar{q}} = 1.7$ for $\tau = (2m_t+m_h)^2/S$, which increase to about $\bar{N}_{0,gg} = 3.1$ and $\bar{N}_{0,q\bar{q}} = 2.2$ at $\tau = Q^2/S$ with $Q = 1$~TeV. Given that these values of $N_0$ are small, we expect that the threshold approximation of the soft-anomalous dimension matrices indeed shows a noticeable effect when compared against the result that uses the full form of the soft-anomalous dimension matrices. \\
With using $\rho_{\rm abs} = (2m_t+m_h)^2/s$, we have a generically lower value for $\tau$ than with using $\rho_{Q^2}$. The highest value for $\tau$ is obtained when using $\rho_{Q^2}$. By choosing $\rho = \rho_{Q^2}$ and $\mu = \mu_{\rm low}$, we introduce a moderate scale logarithm in the $g^{(2)}$ coefficient of about $\ln(M^2/\mu_{\rm low}^2) = 1.4$ at $M = Q \simeq 2\mu_{\rm low}$ and of $\ln(M^2/\mu_{\rm low}^2) = 2.9$ at $M = Q = 1$~TeV. As observed in Fig.~\ref{fig:diffEtth}, a non-zero positive value for this scale logarithm will improve the agreement between $E_{ij\rightarrow t\bar{t}h}$ with the $g^{(n)}$ contributions, and that without it. Since we use Eq.~\eqref{eq:n0minnum} to compute the value of $N_0$, we therefore expect that the agreement between the full result and the $N_0$-approximated result improves as the scale logarithm grows larger. This is precisely the behaviour observed in Fig.~\ref{fig:N0dym}: the best agreement for $\rho = \rho_{Q^2}$ is found by setting $\mu = \mu_{\rm low}$. \\
We now assess the validity of the $\rho \rightarrow 1$ approximation. Since $Q^2 \simeq s$ in the $z\rightarrow 1$ limit, the validity of the approximation of $\rho_{Q^2} \rightarrow 1$ is directly guaranteed. The story is more subtle for $\rho_{\rm abs}$. For $\rho_{\rm abs}$ and $\mu = \mu_{\rm low}$, the role of the luminosities and the value of $\tau$ are unchanged for increasing $Q$, therefore the value of $N_0$ is fixed to the same number for all values of $Q$. The value of the scale logarithm $\ln(M^2/\mu_{\rm low}^2)$ is fixed around $1.4$, so the scale suppression originating from the resummation across all $Q$ values is constant too. If the $\rho_{\rm abs}\rightarrow 1$ limit would be valid across all $Q$ values, we would expect to see a roughly flat line for the ratio between the $N_0$-approximated result and the full result. This is what is observed for the $\rho = \rho_{Q^2}$ and $\mu = \mu_M$ option, where the ratio to the full result is around $-9\%$ for low $Q$ values, and $-7.5\%$ for high $Q$ values. However, we see that the agreement with a roughly flat line gets worse for $\rho_{\rm abs}$ and $\mu = \mu_{\rm low}$ at higher $Q$ values. This is directly caused by the approximation of setting $\rho_{\rm abs} \rightarrow 1$ to calculate $N_0$, which is no longer true at large values of $Q$, as there $Q$ is allowed to deviate significantly from $2m_t+m_h$. Indeed: if one adjusts the value of $\tau$ accordingly by letting it scale as $\tau_{\rm abs} \rightarrow \tau_{\rm abs}/\rho_{\rm abs} \simeq Q^2/S  = \tau_{Q^2}$ to calculate $N_0$ via Eq.~\eqref{eq:n0minnum}, one obtains a deviation from the full result at low $Q$ values of around $-4\%$, while that at large $Q$ values is around $-2.5\%$, following a similar trend as the $\rho_{Q^2}$, $\mu_M$ curve. \\
The same behaviour is reflected in the curve obtained for $\rho = \rho_{\rm abs}$ and $\mu_{\rm high}$. There, the scale logarithm vanishes for all values of $Q$, so we get a worse estimate of $N_0$ by ignoring the $g^{(n)}$ functions. In turn, this results in a generic $10\%$ difference between the full result and the $N_0$-approximated result, as can be observed for low values of $Q$. At large $Q$-values, the $N_0$-approximated result deviates significantly from the full result. If the $\rho_{\rm abs}\rightarrow 1$ limit would be valid at large $Q$-values, one would expect a similar behavior of the $\rho = \rho_{\rm abs}$, $\mu_{\rm high}$ and the $\rho=\rho_{Q^2}$, $\mu = \mu_{M}$ results, as the scale logarithm vanishes in both cases. However, we see that at large $Q$, the two curves deviate significantly, which again shows us the failure of the $\rho_{\rm abs} \rightarrow 1$ approximation in the large-$Q$ limit. Note that the difference between the $\rho_{\rm abs}$, $\mu = \mu_{\rm low}$ and $\rho_{\rm abs}$, $\mu = \mu_{\rm high}$ results is not constant for different values of $Q$. This is caused by the $Q$-dependence of the full NLL (not-matched) result for these two choices, which has different behavior in the tail. \\
Unsurprisingly, the worst agreement between the full result and the $N_0$-approximated result is found with $\rho = \rho_{\rm abs}$ and $\mu = \mu_M$ at large $Q$ values. Firstly, for this choice, the sign of the scale logarithm in  $g^{(2)}$ is reversed, therefore resulting in a larger discrepancy between the full form of $E_{ij\rightarrow t\bar{t}h}$, and the one where the resummation functions are neglected. Secondly, we have now established that the $\rho_{\rm abs} \rightarrow 1$ limit is not valid at large values of $Q$. These two facts combined lead to an underestimation of $N_0$, from which a large deviation of the $N_0$-approximated result from the full result follows. \\
Before concluding, we briefly turn our attention to the transverse-momentum distribution (right-hand side of Fig.~\ref{fig:N0dym}). The values of $\bar{N}_0$ we find are: $\bar{N}_{0,gg} = 2.45$ and $\bar{N}_{0,q\bar{q}} = 1.78$ at $p_T = 10$~GeV, which grows to $\bar{N}_{0,gg} = 3.22$ and  $\bar{N}_{0,q\bar{q}} = 2.35$ at $p_T = 410$~GeV. Firstly we note that the scale logarithm in $g^{(2)}$ cancels by setting $\mu = \mu_M$. The option $\mu = \mu_{H_T}$ leads to a roughly constant value of $\ln(M^2/\mu^2) \simeq 1.4-1.6$. Larger values of  $\ln(M^2/\mu^2)$ are found for smaller values of $p_T$, therefore we expect that the $N_0$-approximation with $\mu = \mu_{H_T}$ works better for small $p_T$, which is indeed what we observe. For the fixed-scale options of $\mu = \mu_{\rm high}$ and $\mu_{\rm low}$, we find that the $N_0$ approximation is better for higher $p_T$ values, which is a direct result of the growing value of $\ln(M^2/\mu^2)$ for those two options. \\
Focusing again on the case where $\mu = \mu_M$, which is the choice where the scale logarithm cancels, we see that the approximation works slightly less well in this case than for the invariant-mass distributions for the option $\rho_{Q^2}$, $\mu_M$. This is directly caused by the invalidity of the $x_{{\rm T},4m_t^2}^2 \rightarrow 1$ approximation. As we have seen in Section~\ref{sec:subsectrans}, another option for the threshold parameter that parameterizes the edge of phase space of the $p_T$ distribution is $x_{{\rm T},s_{\bar{t}t}}^2$. By considering where the ${\rm d}^2\sigma_{t\bar{t}h}/{{\rm d}p_T{\rm d}s_{t\bar{t}}}$ distributions peak (not shown here), one infers that this happens for the generic $s_{\bar{t}t}$ value $s_{\bar{t}t} \simeq 1.5\times 4m_t^2$. This generic $s_{t\bar{t}}$ value does not depend on the value of $p_T$: the ${\rm d}^2\sigma_{t\bar{t}h}/{{\rm d}p_T{\rm d}s_{t\bar{t}}}$ distributions peak roughly at the same value of $s_{t\bar{t}}$ for different $p_T$ values. More than $90\%$ of the ${\rm d}\sigma_{t\bar{t}h}/{{\rm d}p_T}$ distribution is contained within $s_{t\bar{t}} < 4\times 4m_t^2$ for small $p_T$, and $s_{t\bar{t}} < 10\times 4m_t^2$ for large $p_T$. This shows that $x_{{\rm T},4m_t^2}^2$ indeed deviates from $1$, and this gives a non-negligible contribution to the soft-anomalous dimension matrices (as was the case for the $\rho_{\rm abs}$ threshold parameter). \\
To conclude this section, we briefly comment on the consequences of the failure of the $\rho\rightarrow 1$ approximation for either the invariant-mass or transverse-momentum distribution. Firstly, the upper limit of the integral in  e.g.~Eq.~\eqref{eq:deltaiint} may formally not be set to $M^2$, but should remain fixed at $Q^2$ instead. Therefore, it becomes dependent on the partonic threshold variable in the cases of $\rho_{\rm abs}$, $\rho_{s_{t\bar{t}}}$ and $x^2_{{\rm T},4m_t^2}$. Secondly, the kinematics of the soft-anomalous dimension matrices may not be approximated, since the $\rho \rightarrow 1$ limit is simply not obeyed for either large values of $Q$, or for the entire $p_T$ distribution. 

\section{Discussion}
\label{sec:discuss}
We have examined the impact of using different threshold variables in NLL threshold resummation for the $t\bar{t}h$ invariant-mass distribution, the transverse-momentum distribution of the Higgs boson when produced in association with a $t\bar{t}$-pair, and the total $t\bar{t}h$ cross section. These threshold variables differ in the way they include NLP corrections, which in $N$-space show up as $\mathcal{O}(1/N)$ contributions. We also have assessed the role of scale variations. An overview of all options that are considered in this work is given in Table~\ref{tab:summary}. \\
To compute our results, we have introduced a novel numerical method that we call the deformation method. This method stabilizes the computation of the notoriously difficult-to-perform inverse Mellin transform. We show that by using this new method,  $\mathcal{O}(10)$-times less computation time is needed to compute resummed observables, while gaining a factor of $4-5$ in numerical accuracy. We believe that our method is helpful to compute resummed distributions in cases where the Mellin transform of the partonic cross section is too involved to obtain analytically. \\
We show that the resummed distributions and the total cross section are stable under the choice of threshold variable, but that the obtained scale uncertainties do vary with this choice. The smallest scale uncertainties are found by parameterizing the threshold boundary in an absolute sense (i.e.~by setting $\rho = \rho_{\rm abs}$), and not let it depend on the observable of interest (which is what happens for e.g.~$\rho = \rho_{Q^2}$). For the invariant-mass distribution, at the scale choice of $\mu = \mu_{\rm low}$ (see Table~\ref{tab:summary}, page~\pageref{tab:summary} for the definitions), we see that the numerical difference across the different threshold variables is very small. For all definitions we find a decrease of the NLO central contribution for $\mu = \mu_{\rm low}$ between $-1.5\%$ and  $-2.5\%$. The resulting distributions for the other two scale choices $\mu = \mu_{\rm high}$ and $\mu_M$ show a large dependence on the choice of threshold variable: corrections to the tail of the NLO invariant-mass distribution vary from $-1.6\%$ to $+17.9\%$. After averaging over all scale choices, we find that the correction with respect to the averaged NLO invariant-mass distribution is around $+4.7\%$ at the peak of the distribution and $+12.1\%$ in the tails. For the transverse-momentum distribution we find that the scale choice $\mu = \mu_{\rm H_T}$ leads to a very constant result across the two different threshold choices. Upon averaging over all scale choices, we find a correction with respect to the averaged NLO transverse-momentum distribution of $+5.2\%$ at the peak and around $+10.5\%$ in the tail of the distribution. \\
Sometimes, resummed results in direct QCD are obtained using an approximation on the kinematics of the soft-anomalous dimension matrices. This approximation is justified by claiming that the resummed results are only valid in the large-$N$ limit ($N\rightarrow \infty$). However, we show that this large-$N$ limit does not apply to the kinematics of the soft-anomalous dimension matrices. The large-$N$ limit \emph{has} to be used for computing the effects of soft-gluon contributions, since we have no control over subleading $\mathcal{O}(1/N)$ results that originate from for example the emissions of next-to-soft gluons and soft quarks. In computing the effects of soft-gluon contributions, the large-$N$ limit leads to the well-known resummation coefficients $g^{(i)}$. However, the large-$N$ limit may not be used when performing the inverse-Mellin transform, as there all values of $N$ can be probed. Using a saddle-point argument we show that the average value of $N$ in this \emph{inverse-Mellin transform integral} is around $1$ for the $q\bar{q}$-induced channel, and around $2$ for the $gg$-induced channel. This shows that the value of $N$ in the inverse Mellin transform is determined by the interplay between the shapes of the PDFs, and the value of the hadronic threshold variable $\tau$. This means that the $N\rightarrow \infty$ limit does not apply to the factor $\rho^{N}$ present in the Mellin transform of the partonic cross section. It follows that the partonic threshold limit of $\rho \rightarrow 1$ is not obeyed, and that the simplification of the soft-anomalous dimension matrices using this limit is invalid. \\
The numerical effect of this approximation is different for different observables. For the invariant-mass distribution it leads to an overall normalization that is wrong by around $+3\%$. Similarly, the approximation leads to an overestimation of the total cross section of $+3\%$ as well. In contrast, the shape of the transverse-momentum distribution is altered, as the correction obtained after approximating the kinematics of the soft-anomalous dimension matrices is not constant for all $p_T$ values: at the peak of the distribution the overestimation is around $+2.9\%$, while in the tails this grows to about $+6.5\%$. These differences lie within the scale uncertainties of the results. \\
A natural follow-up of our analysis would be to investigate the effects of NNLL threshold resummation. It is known that NNLL threshold resummation further stabilizes the scale-dependence of the results~\cite{Kulesza:2017ukk,Kulesza:2020nfh}, but it does not impact the central values to a great extent. Therefore, we expect that the extension to NNLL does not change our conclusions. Another extension to our work would be to examine the impact of NLP effects on the $t\bar{t}h$ distribution to its full extent. The NLP leading-logarithmic terms would originate from either next-to-soft gluon emissions, or from the emission of soft quarks, and it would certainly be interesting to study their effect once a resummation framework for these contributions exists. 

\section*{Acknowledgments}
This work received support from the Dutch NWO-I program 156, "Higgs as Probe and Portal". MvB also acknowledges support from the Science and Technology Facilities Council (grant number ST/T000864/1). 

\appendix

\section{Useful definitions for NLL resummation}
\label{app:definitions}
We use the following definition of the QCD $\beta$-function
\begin{eqnarray}
\frac{{\rm d}\alpha_s(\mu^2)}{{\rm d}\ln(\mu^2)} \equiv \beta(\alpha_s(\mu^2)) = -\alpha_s^2(\mu^2)\sum_{n=0}^{\infty} b_n\alpha_s^n(\mu^2)\,.
\end{eqnarray}
For NLL resummation we only need $b_0$ and $b_1$, defined by~\cite{Gross:1973id,Politzer:1973fx,Caswell:1974gg,Jones:1974mm,Egorian:1978zx}
\begin{eqnarray}
b_0 &=& \frac{11 C_A - 4 T_R n_f}{12 \pi}\,,\;\;\;\;  \;\;\;\;\;
b_1 \;=\; \frac{17 C_A^2-10 C_A T_R n_f-6 C_F T_R n_f}{24 \pi^2}\,,\;  
\end{eqnarray}
with $T_R = 1/2$, $C_A = 3$ and $C_F = \frac{4}{3}$. The number of active flavors is denoted by $n_f$ and is set equal to $5$ in this work. The resummation functions of Eq.~\eqref{eq:deltai} read
\begin{eqnarray}
g^{(1)}_i(\lambda) &=& \frac{A_i^{(1)}}{2\pi b_0^2}\left(2\lambda+(1-2\lambda)\ln(1-2\lambda)\right), \\
\label{eq:g2} 
g^{(2)}_i(\lambda,M^2/\mu_F^2,M^2/\mu_R^2) &=& \frac{A^{(1)}_i b_1}{2\pi b_0^3}\left[2\lambda+\ln(1-2\lambda)+\frac{1}{2}\ln^2(1-2\lambda)\right]-\frac{A^{(2)}_i}{2\pi^2b_0^2}\left(2\lambda+\ln(1-2\lambda)\right)  \nonumber\\
&& +\frac{A^{(1)}_i}{2\pi b_0}\left(2\lambda+\ln(1-2\lambda)\right)\ln\frac{M^2}{\mu_R^2} - \frac{A^{(1)}_i}{\pi b_0}\lambda\ln\frac{M^2}{\mu_F^2},
\end{eqnarray}
with the coefficients $A^{(1),(2)}_a$ given by \cite{Catani:1989ne}
\begin{align}
\label{eq:37}
  A_a^{(1)} &= C_a\,, \qquad \qquad A_a^{(2)} = 
\frac{C_a}{2} \left[C_A\Bigg(\frac{67}{18}-\zeta(2)\Bigg)-\frac{10}{9}T_R n_f \right].
\end{align}
The $\lambda$ coefficient is defined through $\lambda = \alpha_s b_0 \ln\bar{N}$ with $\alpha_s \equiv \alpha_s(\mu_R^2)$ and $\bar{N} = {\rm e}^{\gamma_E}N$ with $N$ the Mellin moment and $\gamma_E$ the Euler-Mascheroni constant. \\
At lowest order, the soft-anomalous dimension matrix $\mathbf{S}^{(0)}_{ij\rightarrow t\bar{t}h}$ is given by
\begin{eqnarray}
\label{eq:soft0order}
\mathbf{S}^{(0)}_{q\bar{q} \rightarrow t \bar{t} h} = \begin{pmatrix}C_A^2 & 0 \\ 0 & \frac{C_AC_F}{2}\end{pmatrix}, \quad 
\mathbf{S}^{(0)}_{gg \rightarrow t \bar{t} h} = \begin{pmatrix}2C_A^2 C_F & 0 & 0 \\ 0 & C_F(C_A^2-4) & 0 \\ 0 & 0 & C_A^2 C_F \end{pmatrix},
\end{eqnarray}
in the bases $(c_{\mathbf{1}}^q, c_{\mathbf{8}}^q)$ and $(c_{\mathbf{1}}^g,c_{\mathbf{8S}}^g,c_{\mathbf{8A}}^g)$ respectively. Denoting $q(c_i)\bar{q}(c_j) \rightarrow t(c_t)\bar{t}(c_{\bar{t}}) h$ with $c_k$ the color indices belonging to the fundamental representation, we may write the base tensors for the $q\bar{q}$ channel as
\begin{eqnarray}
\label{eq:tensorqq}
c_{\mathbf{1}}^q = \delta_{c_ic_j}\delta_{c_tc_{\bar{t}}}, \quad 
c_{\mathbf{8}}^q = t_{c_jc_i}^et_{c_tc_{\bar{t}}}^e,
\end{eqnarray}
where $t^a_{c_ic_j}$ denotes the generator of ${\rm SU}(3)$ in the fundamental representation, normalized via ${\rm Tr}[t^{a}t^b] = \delta^{ab}/2$. Repeated indices are summed over. For the $g(a_i)g(a_j)$ channel, with $a_k$ indicating a color index in the adjoint representation, we may write 
\begin{eqnarray}
\label{eq:tensorgg}
c_{\mathbf{1}}^g = \delta_{a_ia_j}\delta_{c_tc_{\bar{t}}}, \quad 
c_{\mathbf{8S}}^g = t_{c_tc_{\bar{t}}}^e d^{ea_i a_j}, \quad 
c_{\mathbf{8A}}^g = it_{c_tc_{\bar{t}}}^e f^{ea_i a_j},
\end{eqnarray}
with $f^{abc}$ the structure constants of ${\rm SU}(3)$ defined through $[t^a,t^b]=if^{abc}t^c$, and $d^{abc}$ the symmetric tensor of ${\rm SU}(3)$.
The color tensors of Eq.~\eqref{eq:tensorqq} and~\eqref{eq:tensorgg} are orthogonal, which is the reason why the soft anomalous dimension matrix is diagonal. The color structure of the hard scattering matrix element needs to be projected onto these bases. For the $q\bar{q}$ channel this is straightforward since there is only one color channel ($t_{c_j c_i}^e t_{c_tc_{\bar{t}}}^e$). Denoting the matrix element (stripped of the color tensors) as $\mathcal{M}_{q\bar{q}}$, the lowest-order hard function in the basis  $(c_{\mathbf{1}}^q, c_{\mathbf{8}}^q)$ reads
\begin{eqnarray}
\mathbf{H}_{q\bar{q}\rightarrow t\bar{t}h}^{(0)} = \begin{pmatrix} 0&0\\
0  &  |\mathcal{M}_{q\bar{q}}|^2 \end{pmatrix}.
\end{eqnarray}
For the $gg$ channel this projection is slightly more involved. There are three color structures, given by $t^{a_i}_{c_t k}t^{a_j}_{k c_{\bar{t}}}$, $t^{a_j}_{c_t k}t^{a_i}_{k c_{\bar{t}}}$, $if^{da_ia_j}t^{d}_{c_t c_{\bar{t}}}$. The hard scattering matrix element (stripped of the color tensors) in the basis $\left(t^{a_i}_{c_t k}t^{a_j}_{k c_{\bar{t}}}, t^{a_j}_{c_t k}t^{a_i}_{k c_{\bar{t}}}, if^{da_ia_j}t^{d}_{c_t c_{\bar{t}}}\right)$ is denoted by 
\begin{eqnarray}
\mathcal{M} = \begin{pmatrix}\mathcal{M}_1  \\
 \mathcal{M}_2\\
 \mathcal{M}_3
\end{pmatrix}.
\end{eqnarray}
The hard scattering matrix element $h$ in the basis $(c_{\mathbf{1}}^g,c_{\mathbf{8S}}^g,c_{\mathbf{8A}}^g)$ then becomes 
\begin{eqnarray}
h &=& \left[\mathbf{S}^{(0)}_{gg\rightarrow t \bar{t}h}\right]^{-1} \begin{pmatrix}\delta_{a_ia_j}\delta_{c_tc_{\bar{t}}} \\t_{c_{\bar{t}}c_t}^e d^{a_i a_je} \\ -it_{c_{\bar{t}}c_t}^e f^{a_i a_je} \end{pmatrix}\left(t^{a_i}_{c_t k}t^{a_j}_{k c_{\bar{t}}}\mathcal{M}_1 + t^{a_j}_{c_t k}t^{a_i}_{k c_{\bar{t}}}\mathcal{M}_2 + if^{da_ia_j}t^{d}_{c_t c_{\bar{t}}}\mathcal{M}_3\right) \\
&=& \begin{pmatrix}\frac{1}{2C_A}\left(\mathcal{M}_1+\mathcal{M}_2\right)  \\
\frac{1}{2}\left(\mathcal{M}_1+\mathcal{M}_2\right)\\
\frac{1}{2}\left(\mathcal{M}_1-\mathcal{M}_2+2\mathcal{M}_3\right)
\end{pmatrix}.\nonumber 
\end{eqnarray}
With this, the lowest-order hard function for $gg$ scattering is
\begin{eqnarray}
\label{eq:hsgg}
\mathbf{H}_{gg\rightarrow t\bar{t}h}^{(0)} = h.h^{\dagger} = \begin{pmatrix} \frac{1}{C_A^2}H_{22} &  \frac{1}{C_A}H_{22}  &  \frac{1}{C_A}H_{23} \\ \frac{1}{C_A}H_{22} &  H_{22}  &  H_{23} \\
 \frac{1}{C_A}H_{23}^{
\dagger}& H_{23}^{
\dagger}  &  H_{33} \end{pmatrix},
\end{eqnarray}
with $H_{22} = \frac{1}{4}\left|\mathcal{M}_1+\mathcal{M}_2\right|^2$, $H_{23} =  \frac{1}{4}\left(\mathcal{M}_1+\mathcal{M}_2\right)\left(\mathcal{M}_1^{\dagger}-\mathcal{M}_2^{\dagger}+2\mathcal{M}_3^{\dagger}\right)$, and \\
$H_{33} =\frac{1}{4} \left|\mathcal{M}_1-\mathcal{M}_2+2\mathcal{M}_3\right|^2$. 
\subsection{The soft-anomalous dimension matrices} 
\label{sec:softanomalousdim}
The procedure on how to calculate the soft-anomalous dimension matrix is outlined in Ref.~\cite{KIDONAKIS1996867,Kidonakis:1997gm} for $t\bar{t}$ production and given for $t\bar{t}h$ production in Ref.~\cite{Kulesza:2015vda} with full dependence on the $2\rightarrow 3$ particle kinematics. To write things down in a compact form, it is useful to introduce
\begin{eqnarray*}
L_{t\bar{t}} &=& \frac{1+\beta_{t\bar{t}}^2}{2\beta_{t\bar{t}}}\left[\ln\left(\frac{1-\beta_{t\bar{t}}}{1+\beta_{t\bar{t}}}\right)+i\pi\right],\quad \beta_{t\bar{t}} = \sqrt{1-\frac{4m_t^2}{s_{t\bar{t}}}}\,, \\
\Lambda_{t\bar{t}} &=& \frac{1}{2}\left[T_{1t} + T_{2\bar{t}}+ T_{1\bar{t}} +T_{2t} \right]\,, \\
\Omega_{t\bar{t}} &=& \frac{1}{2}\left[T_{1t} + T_{2\bar{t}} - T_{1\bar{t}} - T_{2t}\right]\,, \\
T_{1t} &=&\ln\left(\frac{m_t^2-t_{1t}}{m_t\sqrt{s}}\right)-\frac{1-i\pi}{2}\,, \quad T_{1\bar{t}} =\ln\left(\frac{m_t^2-t_{1\bar{t}}}{m_t\sqrt{s}}\right)-\frac{1-i\pi}{2}\,, \\T_{2t} &=&\ln\left(\frac{m_t^2-t_{2t}}{m_t\sqrt{s}}\right)-\frac{1-i\pi}{2}\,, \quad T_{2\bar{t}} =\ln\left(\frac{m_t^2-t_{2\bar{t}}}{m_t\sqrt{s}}\right)-\frac{1-i\pi}{2}\,.
\end{eqnarray*}
With these definitions the one-loop soft-anomalous dimension matrix  for the $q\bar{q}$ channel in the basis $(c_{\mathbf{1}}^q,c_{\mathbf{8}}^q)$ reads
\begin{eqnarray}
\label{eq:gammaqq}
\mathbf{\Gamma}^{(1)}_{q\bar{q}\rightarrow t\bar{t}h} = \begin{pmatrix}\Gamma^{q\bar{q}}_{11} & \frac{C_F}{C_A}\Omega_{t\bar{t}} \\
2\Omega_{t\bar{t}} & \Gamma^{q\bar{q}}_{22} \end{pmatrix},
\end{eqnarray}
with 
\begin{eqnarray*}
\Gamma_{11}^{q\bar{q}} &=& -C_F\left(L_{t\bar{t}} + 1\right), \\
\Gamma_{22}^{q\bar{q}} &=& \frac{1}{2}\left[\frac{1}{N_c}\left(L_{t\bar{t}} + 1\right) + \frac{N_c^2-4}{N_c}\Omega_{t\bar{t}} + N_c\Lambda_{t\bar{t}}\right].
\end{eqnarray*}
The one-loop soft-anomalous dimension matrix for the $gg$ channel in the basis $(c_{\mathbf{1}}^g,c_{\mathbf{8S}}^g, c_{\mathbf{8A}}^g)$ is
\begin{eqnarray}
\label{eq:gammagg}
\mathbf{\Gamma}^{(1)}_{gg\rightarrow t\bar{t}h} = \begin{pmatrix}\Gamma^{gg}_{11} & 0 & \Omega_{t\bar{t}} \\
0 & \Gamma^{gg}_{22} & \frac{N_c}{2}\Omega_{t\bar{t}} \\
2\Omega_{t\bar{t}} & \frac{N_c^2-4}{2N_c}\Omega_{t\bar{t}} & \Gamma^{gg}_{33} \end{pmatrix},
\end{eqnarray}
with
\begin{eqnarray*}
\Gamma^{gg}_{11} &=& -C_F\left(L_{t\bar{t}}+1\right),\\
\Gamma^{gg}_{22} &=& \Gamma^{gg}_{33} = \frac{1}{2}\left[\frac{1}{N_c}\left(L_{t\bar{t}}+1\right)+N_c\Lambda_{t\bar{t}}\right].
\end{eqnarray*}
For $s_{t\bar{t}} = 4m_t^2$, we may write
\begin{equation}
L_{t\bar{t}}\Big|_{s_{t\bar{t}}=4m_t^2} = \lim_{\beta_{t\bar{t}}\rightarrow 0}\left[\frac{i\pi}{2\beta_{t\bar{t}}}\right] - 1.
\end{equation}
The function $\Omega_{t\bar{t}}$ vanishes for $s_{t\bar{t}} = 4m_t^2$, while $\Lambda_{t\bar{t}} = \ln\left(1+p_{\rm T}^2/(4m_t^2)\right)-1+i\pi$. Therefore, at $s_{t\bar{t}} = 4m_t^2$, we have a diagonal one-loop soft-anomalous dimension matrix with
\begin{eqnarray*}
\Gamma_{11}^{q\bar{q}}|_{s_{t\bar{t}} = 4m_t^2}  &=& \lim_{\beta_{t\bar{t}}\rightarrow 0}\left(-C_F\left[\frac{i\pi}{2\beta_{t\bar{t}}}\right]\right), \\
\Gamma_{22}^{q\bar{q}}|_{s_{t\bar{t}} = 4m_t^2}  &=&  \lim_{\beta_{t\bar{t}}\rightarrow 0}\left(\frac{N_C}{2}\left[\ln\left(1+p_{\rm T}^2/(4m_t^2)\right)-1+i\pi\right] + \frac{1}{2N_C}\left[\frac{i\pi}{2\beta_{t\bar{t}}}\right]\right),
\end{eqnarray*}
and
\begin{eqnarray*}
\Gamma^{gg}_{11}|_{s_{t\bar{t}} = 4m_t^2} &=& \lim_{\beta_{t\bar{t}}\rightarrow 0}\left(-C_F\left[\frac{i\pi}{2\beta_{t\bar{t}}}\right]\right),\\
\Gamma^{gg}_{22}|_{s_{t\bar{t}} = 4m_t^2} &=& \Gamma^{gg}_{33}|_{s_{t\bar{t}} = 4m_t^2} = \lim_{\beta_{t\bar{t}}\rightarrow 0}\left(\frac{N_C}{2}\left[\ln\left(1+\frac{p_{\rm T}^2}{4m_t^2}\right)-1+i\pi\right]+ \frac{1}{2N_C}\left[\frac{i\pi}{2\beta_{t\bar{t}}}\right]\right).
\end{eqnarray*}
With this, the soft function becomes
\begin{eqnarray}
\label{eq:softfunc}
\mathbf{S}^{(0)}_{q\bar{q}\rightarrow t\bar{t}h} &=& \mathbf{S}^{(0)}_{q\bar{q}\rightarrow t\bar{t}h}
\begin{pmatrix} 1 & 0 \\
0 & {\rm e}^{N_C\left[\ln\left(1+p_{\rm T}^2/(4m_t^2)\right)-1\right] \frac{\ln(1-2\lambda)}{2\pi b_0}}\end{pmatrix}, \\
\mathbf{S}_{gg\rightarrow t\bar{t}h} &=& \mathbf{S}^{(0)}_{gg\rightarrow t\bar{t}h}
\begin{pmatrix} 1 & 0 & 0 \\
0 & {\rm e}^{N_C\left[\ln\left(1+p_{\rm T}^2/(4m_t^2)\right)-1\right] \frac{\ln(1-2\lambda)}{2\pi b_0}}& 0 \\
0 & 0 & {\rm e}^{N_C\left[\ln\left(1+p_{\rm T}^2/(4m_t^2)\right)-1\right] \frac{\ln(1-2\lambda)}{2\pi b_0}} \end{pmatrix}.\nonumber
\end{eqnarray}
This result only holds for $s_{t\bar{t}} = 4m_t^2$. Corrections to the soft function that involve using the full kinematics of the soft anomalous dimension matrices are of NLP. In that case, we also need to diagonalize the one-loop soft-anomalous dimension matrices. We do this using the matrix $\mathbf{R}$ with $\mathbf{R}\mathbf{R}^{-1} = \mathbf{1}$. We may then write (dropping the subscripts)
\begin{eqnarray}
{\rm Tr}\left[\mathbf{H}^{(0)}\mathbf{U}^{\dagger}\mathbf{S}^{(0)}\mathbf{U}\right] &=& {\rm Tr}\left[\mathbf{R}^{-1}\mathbf{H}^{(0)}\left(\mathbf{R}\mathbf{R}^{-1}\right)^{\dagger}\mathbf{U}^{\dagger}\left(\mathbf{R}\mathbf{R}^{-1}\right)^{\dagger}\mathbf{S}^{(0)}\mathbf{R}\mathbf{R}^{-1}\mathbf{U}\mathbf{R}\right] \\
&\equiv& {\rm Tr}\left[\mathbf{H}^{(0)}_{\mathbf{R}}\mathbf{U}^{\dagger}_{\mathbf{R}}\mathbf{S}^{(0)}_{\mathbf{R}}\mathbf{U}_{\mathbf{R}}\right].  \nonumber 
\end{eqnarray}
with
\begin{eqnarray*}
\mathbf{H}^{(0)}_{\mathbf{R}} = \mathbf{R}^{-1}\mathbf{H}^{(0)}\left(\mathbf{R}^{-1}\right)^{\dagger}, &\qquad & 
\mathbf{U}_{\mathbf{R}} = \mathbf{R}^{-1}\mathbf{U}\mathbf{R}\,, \\
\mathbf{S}^{(0)}_{\mathbf{R}} = \mathbf{R}^{\dagger}\mathbf{S}^{(0)}\mathbf{R}, &\qquad & 
\mathbf{U}^{\dagger}_{\mathbf{R}} =  \mathbf{R}^{\dagger}\mathbf{U}^{\dagger}\left(\mathbf{R}^{-1}\right)^{\dagger}.
\end{eqnarray*}
In the $q\bar{q}$ case, the form of $\mathbf{R}$ is simple. The eigenvalues of the one-loop soft-anomalous dimension matrix (denoted by $\lambda_{\pm}^{q\bar{q}}$) are given by
\begin{eqnarray}
\lambda_{\pm}^{q\bar{q}} = \frac{1}{2}\left[\Gamma_{11}^{q\bar{q}}+\Gamma_{22}^{q\bar{q}}\pm\sqrt{(\Gamma_{11}^{q\bar{q}}-\Gamma_{22}^{q\bar{q}})^2+4\Gamma_{12}^{q\bar{q}}\Gamma_{21}^{q\bar{q}}}\right],
\end{eqnarray}
with $\Gamma_{12}^{q\bar{q}} = \frac{C_F}{C_A}\Omega_{t\bar{t}}$ and $\Gamma_{21}^{q\bar{q}} = 2\Omega_{t\bar{t}}$. The eigenvectors are
\begin{eqnarray}
v_{\pm}^{q\bar{q}} = \begin{pmatrix}\frac{\Gamma_{12}^{q\bar{q}}}{\lambda^{q\bar{q}}_{\pm}-\Gamma^{q\bar{q}}_{11}} \\ 1\end{pmatrix},
\end{eqnarray}
such that $\mathbf{R}_{q\bar{q}}$ becomes
\begin{eqnarray}
\mathbf{R}_{q\bar{q}} = \begin{pmatrix} \frac{\Gamma_{12}^{q\bar{q}}}{\lambda^{q\bar{q}}_{+}-\Gamma^{q\bar{q}}_{11}} & \frac{\Gamma_{12}^{q\bar{q}}}{\lambda^{q\bar{q}}_{-}-\Gamma^{q\bar{q}}_{11}}\\
1 & 1 \end{pmatrix}, \quad 
\mathbf{R}_{q\bar{q}}^{-1} = \frac{1}{\frac{\Gamma_{12}^{q\bar{q}}}{\lambda^{q\bar{q}}_{+}-\Gamma^{q\bar{q}}_{11}}-\frac{\Gamma_{12}^{q\bar{q}}}{\lambda^{q\bar{q}}_{-}-\Gamma^{q\bar{q}}_{11}}}\begin{pmatrix} 1 & -\frac{\Gamma_{12}^{q\bar{q}}}{\lambda^{q\bar{q}}_{-}-\Gamma^{q\bar{q}}_{11}} \\
 -1 &\frac{\Gamma_{12}^{q\bar{q}}}{\lambda^{q\bar{q}}_{+}-\Gamma^{q\bar{q}}_{11}} \end{pmatrix}.
\end{eqnarray}
With this, the resummation of the soft function (Eq.~\eqref{eq:softres}) for the $q\bar{q}$ channel becomes
\begin{eqnarray}
\mathbf{S}^{(0)}_{\mathbf{R},q\bar{q}\rightarrow t\bar{t}h} = \mathbf{R}_{q\bar{q}}^{\dagger}\mathbf{S}^{(0)}_{q\bar{q}\rightarrow t\bar{t}h}\mathbf{R}_{q\bar{q}}\begin{pmatrix}{\rm e}^{\left(\lambda^{q\bar{q}}_++\lambda^{q\bar{q},*}_+\right) \frac{\ln(1-2\lambda)}{2\pi b_0}} & 0 \\ 
0 & {\rm e}^{\left(\lambda^{q\bar{q}}_-+\lambda^{q\bar{q},*}_-\right) \frac{\ln(1-2\lambda)}{2\pi b_0}}  \end{pmatrix}.\label{eq:softres2qq}
\end{eqnarray}
The eigenvalues of a $3\times 3$ matrix are harder to write down. The eigenvalue equation that needs to be solved reads
\begin{eqnarray*}
0 &=& (\lambda^{gg}_i)^3+\left[-\Gamma_{11}^{gg}-\Gamma_{22}^{gg}-\Gamma_{33}^{gg}\right](\lambda^{gg}_i)^2+\left[\Gamma^{gg}_{11}\Gamma^{gg}_{22}+\Gamma^{gg}_{11}\Gamma^{gg}_{33}+\Gamma^{gg}_{22}\Gamma^{gg}_{33}-\Gamma^{gg}_{31}\Gamma^{gg}_{13}-\Gamma^{gg}_{32}\Gamma^{gg}_{23}\right]\lambda^{gg}_i \\
&& + \left[\Gamma^{gg}_{11}\Gamma^{gg}_{23}\Gamma^{gg}_{32}+\Gamma^{gg}_{22}\Gamma^{gg}_{13}\Gamma^{gg}_{31}-\Gamma^{gg}_{11}\Gamma^{gg}_{22}\Gamma^{gg}_{33}\right] \\
&\equiv& (\lambda^{gg}_i)^3 + b(\lambda^{gg}_i)^2 + c(\lambda^{gg}_i) + d,
\end{eqnarray*}
with $\Gamma^{gg}_{13} = \Omega_{t\bar{t}}$, $\Gamma^{gg}_{31} = 2\Omega_{t\bar{t}}$, $\Gamma^{gg}_{23} = \frac{N_C}{2}\Omega_{t\bar{t}}$, $\Gamma^{gg}_{32} = \frac{N_C^2-4}{2N_C}\Omega_{t\bar{t}}$. A cubic equation of this kind may be solved by defining
\begin{eqnarray}
\Delta_0 = b^2 - 3 c, \quad \Delta_1 = 2b^3-9bc+27d, \quad C = \left(\frac{\Delta_1 + \sqrt{\Delta_1^2-4\Delta_0^3}}{2}\right)^{1/3}, \quad \xi = \frac{-1+\sqrt{-3}}{2},
\end{eqnarray}
such that the eigenvalues read
\begin{eqnarray}
\lambda^{gg}_i = -\frac{1}{3}\left[b+\xi^kC+\frac{\Delta_0}{\xi^k C}\right],
\end{eqnarray}
with $k = 0, 1, 2$. The eigenvectors follow as
\begin{eqnarray}
v^{gg}_i = \begin{pmatrix}\frac{\Gamma^{gg}_{13}}{\lambda_i^{gg}-\Gamma^{gg}_{11}}\\\frac{\Gamma^{gg}_{23}}{\lambda_i^{gg}-\Gamma^{gg}_{22}}\\1\end{pmatrix}.
\end{eqnarray}
From this we directly obtain the matrix $\mathbf{R}_{gg}$
\begin{eqnarray}
\mathbf{R}_{gg} = \begin{pmatrix} \frac{\Gamma^{gg}_{13}}{\lambda_1^{gg}-\Gamma^{gg}_{11}}& \frac{\Gamma^{gg}_{13}}{\lambda_2^{gg}-\Gamma^{gg}_{11}} & \frac{\Gamma^{gg}_{13}}{\lambda_3^{gg}-\Gamma^{gg}_{11}} \\
\frac{\Gamma^{gg}_{23}}{\lambda_1^{gg}-\Gamma^{gg}_{22}} &
\frac{\Gamma^{gg}_{23}}{\lambda_2^{gg}-\Gamma^{gg}_{22}} &
\frac{\Gamma^{gg}_{23}}{\lambda_3^{gg}-\Gamma^{gg}_{22}} \\
1 & 1 & 1 \end{pmatrix} \equiv  \begin{pmatrix} v_{11}^{gg} & v_{21}^{gg} & v_{31}^{gg} \\ 
v_{12}^{gg} & v_{22}^{gg} & v_{32}^{gg} \\
1 & 1 & 1 \end{pmatrix}, 
\end{eqnarray}
such that
\begin{eqnarray}
\mathbf{R}_{gg}^{-1} = \frac{1}{\mathcal{N}}\begin{pmatrix}  v_{22}^{gg}-v_{32}^{gg} & v_{31}^{gg}-v_{21}^{gg} & v_{21}^{gg}v_{32}^{gg}-v_{22}^{gg}v_{31}^{gg} \\
v_{32}^{gg}-v_{12}^{gg} & v_{11}^{gg}-v_{31}^{gg} & v_{12}^{gg}v_{31}^{gg}-v_{11}^{gg}v_{32}^{gg} \\
v_{12}^{gg}-v_{22}^{gg} & v_{21}^{gg}-v_{11}^{gg} & v_{11}^{gg}v_{22}^{gg}-v_{12}^{gg}v_{21}^{gg} \end{pmatrix},
\end{eqnarray}
with
\begin{eqnarray}
 \mathcal{N} = v^{gg}_{11}(v^{gg}_{22}-v_{32}^{gg})-v_{12}^{gg}(v_{21}^{gg}-v_{31}^{gg})+v_{21}^{gg}v_{32}^{gg}-v_{22}^{gg}v_{31}^{gg}\,.
\end{eqnarray}
With this, the resummation of the soft function for the $gg$ channel becomes
\begin{eqnarray}
\mathbf{S}^{(0)}_{\mathbf{R},gg\rightarrow t\bar{t}h} = \mathbf{R}_{gg}^{\dagger}\mathbf{S}^{(0)}_{gg\rightarrow t\bar{t}h}\mathbf{R}_{gg}\begin{pmatrix}{\rm e}^{\left(\lambda^{gg}_1+\lambda^{gg,*}_1\right) \frac{\ln(1-2\lambda)}{2\pi b_0}} & 0 & 0 \\ 
0 & {\rm e}^{\left(\lambda^{gg}_2+\lambda^{gg,*}_2\right) \frac{\ln(1-2\lambda)}{2\pi b_0}} & 0 \\
0 & 0 & {\rm e}^{\left(\lambda^{gg}_3+\lambda^{gg,*}_3\right) \frac{\ln(1-2\lambda)}{2\pi b_0}}  \\ \end{pmatrix}.\label{eq:softres2gg}
\end{eqnarray}

\section{Kinematics for the top--anti-top--Higgs production process}
\label{app:tth}
Here we briefly describe the connection between the CM frame of the top-anti-top quark pair, and that of the two initial-state partons. In the rest frame of the $t\bar{t}$ state (indicated by the $*$ notation), we may parameterize the momenta of the top and anti-top quarks as follows
\begin{eqnarray}
\label{eq:cmtop}
p_t^* &=& \left(E_t^*,|p_t^*| \sin\theta_t^* \cos\phi_t^*, |p_t^*| \sin\theta_t^* \sin\phi_t^*, |p_t^*| \cos\theta_t^*\right), \\
p_{\bar{t}}^* &=& \left(E_t^*, -|p_t^*| \sin\theta_t^* \cos\phi_t^*, -|p_t^*| \sin\theta_t^* \sin\phi_t^*, -|p_t^*| \cos\theta_t^*\right), \nonumber
\end{eqnarray}
with
\begin{eqnarray}
E_t^* = \frac{\sqrt{s_{t\bar{t}}}}{2}, \qquad |p_t^*| \equiv |\vec{p}_t^{\,*}| =  \frac{\lambda^{1/2}\left(s_{t\bar{t}},m_t^2,m_t^2\right)}{2\sqrt{s_{t\bar{t}}}}.
\end{eqnarray}
The equations for $E_t^*$ and $|p_t^*|$ may directly be derived from the definition of the two-particle phase space that contains the $t\bar{t}$ state
\begin{eqnarray}
{\rm d}\Phi_{t\bar{t}} = \frac{1}{(2\pi)^2}\int {\rm d}^4 p_t\,{\rm d}^4 p_{\bar{t}}\, \delta^{(4)}\left(p_{t\bar{t}} - p_t - p_{\bar{t}}\right)\,\delta^+\left(p_t^2 - m_t^2\right))\,\delta^+\left(p_{\bar{t}}^2 - m_t^2 \right),
\end{eqnarray}
by using that $p_{t\bar{t}}^* = \sqrt{s_{t\bar{t}}}\left(1,0,0,0\right)$ in the CM frame of the $t\bar{t}$ pair. \\
The momenta in Eq.~\eqref{eq:cmtop} need to be boosted to the partonic CM frame. This may be done by using that in the partonic CM frame, $\vec{p}_{t\bar{t}} = -\vec{p}_h$. Therefore, in the partonic CM frame, the four-momentum $p_{t\bar{t}}$ reads
\begin{eqnarray}
p_{t\bar{t}} = \left(E_{t\bar{t}}, 0, -|p_{{\rm T}}|, -m_{{\rm T},h}\sinh\eta \right) = \left(E_{t\bar{t}}, 0, -|p_{h}|\sin\theta_h, -|p_{h}|\cos\theta_h \right),
\end{eqnarray}
with
\begin{eqnarray}
E_{t\bar{t}} = \frac{s+s_{t\bar{t}}-m_h^2}{2\sqrt{s}}, \qquad |p_h| \equiv |\vec{p}_h| = |\vec{p}_{t\bar{t}}| = \frac{\lambda^{1/2}\left(s,s_{t\bar{t}},m_h^2\right)}{2\sqrt{s}}. 
\end{eqnarray}
We have used the azimuthal symmetry of the matrix element to set $\cos\phi_h = 0$ and $\sin\phi_h = 1$. The boost matrix is 
\begin{eqnarray}
 \begin{pmatrix} E_{t\bar{t}} \\ \vec{p}_{t\bar{t}} \end{pmatrix} = \begin{pmatrix}\gamma & -\gamma \vec{\beta} \\ -\gamma \vec{\beta} & \Lambda \end{pmatrix} \begin{pmatrix} E_{t\bar{t}}^*   \\ \vec{0} \end{pmatrix},
\end{eqnarray}
with 
\begin{eqnarray}
 \gamma = \frac{E_{t\bar{t}}}{\sqrt{s_{t\bar{t}}}}, \qquad \vec{\beta} = \frac{\vec{p}_{t\bar{t}}}{E_{t\bar{t}}}.
\end{eqnarray}
The matrix $\Lambda$ is given by
\begin{eqnarray}
 \Lambda = \begin{pmatrix}
 1 + (\gamma-1)\frac{\beta_x^2}{\beta^2} & (\gamma -1)\frac{\beta_x\beta_y}{\beta^2}& (\gamma -1)\frac{\beta_x\beta_z}{\beta^2} \\
 (\gamma -1)\frac{\beta_y\beta_x}{\beta^2}& 1 + (\gamma-1)\frac{\beta_y^2}{\beta^2} &  (\gamma -1)\frac{\beta_y\beta_z}{\beta^2} \\
 (\gamma -1)\frac{\beta_z\beta_x}{\beta^2}&   (\gamma -1)\frac{\beta_z\beta_y}{\beta^2} &1 + (\gamma-1)\frac{\beta_z^2}{\beta^2} \\
 \end{pmatrix}\,,
 \end{eqnarray}
with $\beta_x$ the $x$-component of $\vec{\beta}$, and $
 \beta^2 = \vec{\beta} \cdot \vec{\beta}$. With this, we may write the four-vectors $p_t$ and $p_{\bar{t}}$ in the partonic CM frame.  For the invariants $t_{1t}$, $t_{2t}$, $t_{1\bar{t}}$, and $t_{2\bar{t}}$, we only need $E_t$, $E_{\bar{t}}$, $p_{t,z}$ and $p_{\bar{t},z}$. These read
 \begin{eqnarray*}
 E_t &=& \gamma E_t^* - \gamma\beta_x|p_t^*|\cos\phi_t^*\sin\theta_t^* - \gamma\beta_y|p_t^*|\sin\phi_t^*\sin\theta_t^* - \gamma\beta_z|p_t^*|\cos\theta_t^*\,, \\
 p_{t,z} &=& -\gamma\beta_z E_t^* + (\gamma-1)\frac{\beta_x \beta_z}{\beta^2}|p_t^*|\cos\phi_t^*\sin\theta_t^* \\
 && \hspace{4cm}+ (\gamma-1)\frac{\beta_y\beta_z}{\beta^2}|p_t^*|\sin\phi_t^*\sin\theta_t^* + \left((\gamma-1)\frac{\beta_z^2}{\beta^2}+1\right)|p_t^*|\cos\theta_t^*\,, \\
 E_{\bar{t}} &=& \gamma E_t^* + \gamma\beta_x|p_t^*|\cos\phi_t^*\sin\theta_t^* + \gamma\beta_y|p_t^*|\sin\phi_t^*\sin\theta_t^* + \gamma\beta_z|p_t^*|\cos\theta_t^*\,, \\
 p_{\bar{t},z} &=& -\gamma\beta_z E_t^* - (\gamma-1)\frac{\beta_x \beta_z}{\beta^2}|p_t^*|\cos\phi_t^*\sin\theta_t^* \\
 && \hspace{4cm}- (\gamma-1)\frac{\beta_y\beta_z}{\beta^2}|p_t^*|\sin\phi_t^*\sin\theta_t^* - \left((\gamma-1)\frac{\beta_z^2}{\beta^2}+1\right)|p_t^*|\cos\theta_t^* .
 \end{eqnarray*}
The invariants then become 
\begin{eqnarray*}
t_{1t} = m_t^2 - 2p_1\cdot p_t = m_t^2-\sqrt{s} \left(E_t - p_{t,z}\right), &\qquad&
t_{1\bar{t}} = m_t^2 - 2p_1\cdot p_{\bar{t}} = m_t^2-\sqrt{s} \left(E_{\bar{t}} - p_{\bar{t},z}\right), \\
t_{2t} = m_t^2 - 2p_2\cdot p_t =m_t^2-\sqrt{s}\left(E_t + p_{t,z}\right), &\qquad&
t_{2\bar{t}} = m_t^2 - 2p_2\cdot p_{\bar{t}} = m_t^2-\sqrt{s}\left(E_{\bar{t}} + p_{\bar{t},z}\right).
\end{eqnarray*}
When $x_T^2 = 1$, $p_{z,h} = 0$ as $\cosh\eta = 1$. Therefore, in the threshold limit $\beta_z = 0$. Setting $\beta_x = 0$ by choosing a convenient frame, we then have 
 \begin{eqnarray*}
 E_t = \gamma E_t^* - \gamma\beta_y|p_t^*| \sin\phi_t^*\sin\theta_t^* , &\qquad& 
 E_{\bar{t}} = \gamma E_t^* + \gamma\beta_y|p_t^*|\sin\phi_t^*\sin\theta_t^*, \\
p_{t,z} = |p_t^*|\cos\theta_t^*, &\qquad& p_{\bar{t},z} = -|p_t^*|\cos\theta_t^*.
 \end{eqnarray*}
Moreover, when $s_{t\bar{t}} = 4m_t^2$, $|p_t^*| = 0$, so in that case all invariants reduce to the same number. When $s_{t\bar{t}}$ moves away from $4m_t^2$, none of the invariants reduce to the same number.

\section{The derivative method}
\label{app:deriv}
\noindent The derivative method was introduced in Ref.~\cite{Kulesza:2002rh}, and later extended in Ref.~\cite{Kulesza:2015vda} to also include a second derivative. This method results in one factor of $1/N$ for each PDF and each derivative. This factor suppresses Eq.~\eqref{eq:expgrowth}, and one may hope that then the numerical integration converges quickly enough. Consider
\begin{eqnarray}
\label{eq:deriv1}
N\, f(N+1, \mu_F^2) &=& N \int_{x_{\rm min}}^1{\rm d}x \, x^N f(x, \mu_F^2)\,,
\end{eqnarray}
where we have allowed for a possible $x_{\rm min}\neq 0$ as the lower limit of the integral, resulting from demanding $(x_1 x_2 \rho)/\tau \geq 1 $. Eq.~\eqref{eq:deriv1} may be rewritten to
\begin{eqnarray}
\label{eq:deriv12}
f(N+1, \mu_F^2) &=& \frac{1}{N}\int_{x_{\rm min}}^1{\rm d}x \, \frac{\rm d}{{\rm d}x}\,\Big(x^N\Big)\, x f(x, \mu_F^2) \\
&=& \frac{1}{N}x^N \, x f(x, \mu_F^2)\Big|^1_{x_{\rm min}} -  \frac{1}{N}\int_{x_{\rm min}}^1{\rm d}x \, x^N \frac{\rm d}{{\rm d}x}\Big(x f(x, \mu_F^2)\Big)\,.\nonumber 
\end{eqnarray}
The PDFs vanish at the upper limit of the integral ($x=1$). For $x_{\rm min} = 0$, the integrand also vanishes at the lower limit, and we may neglect the boundary term. The lower bound $x_{\rm min}$ is precisely the value of $x_{1,2}$ where $x_1 x_2 \rho / \tau = 1$. Therefore, for $x_{\rm min} \neq 0$, it is to be expected that the boundary term results in a finite contribution, although it will be suppressed by a factor of $\frac{1}{N}$ (before taking the inverse Mellin transform). The second factor on the second line of Eq.~\eqref{eq:deriv12} also has a suppression factor, but this is compensated by the large-$N$-dependence of the integral. Since we are interested in the $N\rightarrow\infty$ limit, we may hope that its contribution is negligible. \emph{It is important that one checks this assumption explicitly for every process, as it cannot be guaranteed from the outset that this assumption holds true.} If the contribution is indeed negligible, we may replace
\begin{eqnarray}
\label{eq:firstderiv}
    f(N+1,\mu_F^2) \rightarrow -\frac{1}{N}\int_{x_{\rm min}}^1{\rm d}x \,x^{N}\frac{{\rm d}}{{\rm d}x}\left(xf(x,\mu_F^2)\right),
\end{eqnarray}
where we see that we have introduced one suppression factor of $\frac{1}{N}$. We can further manipulate the second factor on the last line of Eq.~\eqref{eq:deriv12} and introduce a second factor of $\frac{1}{N}$ via
\begin{eqnarray}
-\int_{x_{\rm min}}^1{\rm d}x \, x^N \frac{\rm d}{{\rm d}x}\Big(x f(x, \mu_F^2)\Big) &=& -\frac{1}{N}\int_{x_{\rm min}}^1{\rm d}x \frac{\rm d}{{\rm d}x} \,\Big(x^N\Big) x \, \frac{\rm d}{{\rm d}x}\Big(x f(x, \mu_F^2)\Big)\\
&=& -\frac{1}{N}x^{N+1} \frac{\rm d}{{\rm d}x}\left(x f(x, \mu_F^2)\right)\Big|^1_{x_{\rm min}} \nonumber \\
&& \hspace{1.8cm} + \frac{1}{N}\int_{x_{\rm min}}^1{\rm d}x \, x^N \, \frac{\rm d}{{\rm d}x}\Big[x \frac{\rm d}{{\rm d}x}(x f(x, \mu_F^2))\Big]. \nonumber 
\end{eqnarray}
Again, we have no control over what happens to the boundary term. We do not know whether it vanishes at $x_{\rm min}$, but also the derivative of $xf(x,\mu_F^2)$ has to vanish at $x=1$, \emph{and it is not clear that this happens for every PDF.} If we still press on and assume that the boundary term goes to zero, we arrive at
\begin{eqnarray}
\label{eq:secondderiv}
f(N+1, \mu_F^2)  &=& \frac{1}{N^2}\int_{x_{min}}^1 {\rm d}x \, x^N \left[\frac{\rm d}{{\rm d} x}\Big(x f(x, \mu_F^2)\Big) +  x\frac{{\rm d}^2}{{\rm d}x^2}\Big(x f(x, \mu_F^2)\Big)\right] \\
&\equiv& \frac{1}{N^2}\int_{x_{min}}^1{\rm d}x \, x^N\, \mathcal{F}\left(x, \mu_F^2\right) .\nonumber 
\end{eqnarray}
Using this result to perform the inverse Mellin transform as in Eq.~\eqref{eq:inverse}, we introduce an $1/N^4$ suppression factor
\begin{eqnarray}
\label{eq:inverse2}
{\rm d}\sigma(\tau) &=& \sum_{i,j}\frac{1}{2\pi i}\int_{c-i\infty}^{c+i\infty}{\rm d}N \int_{x_{\rm min}}^1 {\rm d}x_1 \int_{x_{\rm min}}^1 {\rm d}x_2 \left(\frac{\tau}{x_1 x_2}\right)^{-N} \int_{\rho_{\rm min}}^1 \, {\rm d}\rho \,\frac{\rho^{N-1}}{N^4} \\
&& \hspace{6cm} \times \mathcal{F}_i(x_1, \mu_F^2)\mathcal{F}_j(x_2, \mu_F^2)  \,{\rm d}\hat{\sigma}_{ij}(\rho), \nonumber
\end{eqnarray}
where $\tau$ and $\rho$ are generic threshold variables, and ${\rm d}\hat{\sigma}_{ij}(\rho)$ (${\rm d}\sigma(\tau)$)  is a generic partonic (hadronic) cross section or differential distribution. The derivative method has the clear advantage that it is flexible in the PDFs that one can use, as one does not need to assume a specific parameterization.  However, considering the assumptions on the boundary terms, there are doubts about whether the method can be trusted. The boundary term is artificially introduced as a method to stabilize the numerical integration (see discussion below Eq.~\eqref{eq:expgrowth}). Therefore, the boundary term should not have been there in the first space as in fact $x_{\rm min} = 0$, and one has to verify numerically that the boundary term results in a negligible contribution to the final result. But even if the contribution of the boundary term is negligible, we still do not have access to the analytic form of the $N$-space PDFs. Therefore, one cannot be certain on what the strip of definition actually is, and consequently, what value of $C_{\rm MP}$ can be chosen.

\bibliographystyle{JHEP}
\bibliography{spires}

\end{document}